\documentclass[reprint,
 superscriptaddress,
nofootinbib,
 amsmath,amssymb,
 aps,
pra,
]{revtex4-2}

\usepackage[utf8]{inputenc} 
\usepackage[T1]{fontenc}    
\usepackage{hyperref}       
\usepackage{url}            
\usepackage{booktabs}       
\usepackage{amsfonts}       
\usepackage{nicefrac}       
\usepackage{microtype}      
\usepackage{graphicx}
\usepackage{xcolor, etoolbox}
\usepackage{amsmath,amsfonts,amsthm} 
\usepackage{mathtools}
\usepackage{ragged2e}
\usepackage{braket}      
\usepackage[english]{babel} 
\usepackage{enumitem}
\usepackage{xspace}
\usepackage{dsfont}
\usepackage{bm}
\usepackage{bbm}
\usepackage{dcolumn}
\usepackage{mathrsfs}

\newcommand{\Tr}{\text{Tr}}

\hypersetup{
    colorlinks,
    linkcolor=blue,
    citecolor=blue,
    urlcolor=blue,
    breaklinks=true 
}

\DeclarePairedDelimiter\abs{\lvert}{\rvert}

\begin{document}

\title{Detecting continuous variable entanglement in phase space with the $Q$-distribution}

\author{Martin G\"{a}rttner}
    \email{martin.gaerttner@uni-jena.de}
    \affiliation{Institut f\"{u}r Theoretische Physik, Universit\"{a}t Heidelberg, Philosophenweg 16, 69120 Heidelberg, Germany}
    \affiliation{Physikalisches Institut, Universit\"{a}t Heidelberg, Im Neuenheimer Feld 226, 69120 Heidelberg, Germany}
    \affiliation{Kirchhoff-Institut f\"{u}r Physik, Universit\"{a}t Heidelberg, Im Neuenheimer Feld 227, 69120 Heidelberg, Germany}
        \affiliation{Institute of Condensed Matter Theory and Optics, Friedrich-Schiller-University Jena, Max-Wien-Platz 1, 07743 Jena, Germany}
        
\author{Tobias Haas}
    \email{tobias.haas@ulb.be}
    \affiliation{Centre for Quantum Information and Communication, École polytechnique de Bruxelles, CP 165, Université libre de Bruxelles, 1050 Brussels, Belgium}
\author{Johannes Noll}
    \email{johannes.noll@stud.uni-heidelberg.de}
    \affiliation{Kirchhoff-Institut f\"{u}r Physik, Universit\"{a}t Heidelberg, Im Neuenheimer Feld 227, 69120 Heidelberg, Germany}

\begin{abstract}
We prove a general class of continuous variable entanglement criteria based on the Husimi $Q$-distribution, which represents a quantum state in canonical phase space, by employing a theorem by Lieb and Solovej. We discuss their generality, which roots in the possibility to optimize over the set of concave functions, from the perspective of continuous majorization theory and show that with this approach families of entropic as well as second moment criteria follow as special cases. All derived criteria are compared to corresponding marginal based criteria and the strength of the phase space approach is demonstrated for a family of prototypical example states where only our criteria flag entanglement. Further, we explore their optimization prospects in two experimentally relevant scenarios characterized by sparse data: finite detector resolution and finite statistics. In both scenarios optimization leads to clear improvements enlarging the class of detected states and the signal-to-noise ratio of the detection, respectively.
\end{abstract}

\maketitle

\section{Introduction}
The quest for efficient methods for the detection of entanglement in continuous variable systems dates back as far as the Einstein-Podolsky-Rosen (EPR) paradox \cite{Einstein1935} and has received renewed interest through the rise of optical quantum technologies for which entanglement is a crucial ingredient \cite{Braunstein2005b,Horodecki2009,Guehne2009,Weedbrook2012}. Over the past two decades, a plethora of entanglement criteria\footnote{We use the terms entanglement criteria and separability criteria interchangeably. Such criteria are usually stated as inequalities that are fulfilled for all separable states, and thus their violation shows that a state is entangled.} emerged. Many of them root in demonstrating the negativity of the partially transposed state \cite{Peres1996,Horodecki1996} by means of violating uncertainty relations of suitable chosen observables \cite{Guehne2009,Horodecki2009,Nha2008}\footnote{This ansatz generically excludes the possibility of detecting bound entanglement \cite{Horodecki1998}.}. 

Widely used are non-local EPR-type operators $\boldsymbol{X}_1 + \boldsymbol{X}_2$ and $\boldsymbol{P}_1 - \boldsymbol{P}_2$, which capture correlations between two systems labelled as $1$ and $2$ \cite{Einstein1935}. For these variables, criteria have been formulated in terms of second moments in seminal works by Duan, Giedke, Cirac and Zoller (DGCZ) \cite{Duan2000} and Mancini, Giovannetti, Vitali and Tombesi (MGVT) \cite{Mancini2002,Giovannetti2003} (see also \cite{Simon2000}). They have been shown to be necessary and sufficient for separability in case of Gaussian states \cite{Braunstein2005b,Weedbrook2012,Serafini2017,Lami2018}, but are known to be rather weak beyond as second moments do only reveal partial information about a measured distribution. Here, a significant upgrade have been entropic criteria, derived by Walborn, Taketani, Salles, Toscane and de Matos Filho (WTSTD) in \cite{Walborn2009} and their generalization to Rényi entropies by Saboia, Toscano and Walborn (STW) \cite{Saboia2011}. Strengthend versions of all these criteria to certify steering have been discussed, too \cite{Reid1989,He2013,Walborn2011,Chowdhury2014,Schneeloch2018}.

Besides other notable approaches utilizing modular variables \cite{Gneiting2011,Carvalho2012}, spin observables with discrete spectra \cite{Agarwal2005}, moments of the bipartite density operator \cite{Shchukin2005,Haas2023}, as well as the quantum Fisher matrix \cite{Gessner2017,Qin2019}, all mentioned methods rely on the detection of marginal distributions of the Wigner $W$-distribution. However, the Husimi $Q$-distribution does also constitute an adequate phase space representation of the quantum state \cite{Husimi1940,Cartwright1976,Lee1995} and can be accessed in a single experimental setting \cite{Collett1987,Welsch1999}. Until recently \cite{Haas2021b,Haas2022a}, the possibility of formulating entanglement criteria based on the Husimi $Q$-distribution has been overlooked. Here, we derive an extremely general class of entanglement criteria from the Husimi $Q$-distribution based on concave functions, generalizing the work in \cite{Haas2022a}.

In the spirit of employing uncertainty relations for entanglement detection, we make use of the most general form of the uncertainty principle in phase space: the Lieb-Solovej theorem \cite{Lieb2014,Bengtsson2017,Schupp2022}. It arose from generalizing a lower bound on the differential entropy --- the so-called Wehrl entropy \cite{Wehrl1978,Wehrl1979} --- of the Husimi $Q$-distribution \cite{Lieb1978,Grabowski1984,Carlen1991,Luo2000} and has been formulated for various algebras \cite{Schupp1999,Lieb2016,Lieb2021,Kulikov2022}. Ultimately, the Lieb-Solovej theorem is a majorization relation stating that the vacuum (or any coherent) distribution majorizes all other distributions (see also \cite{Nielsen1999} for the relation between majorization and entanglement for finite dimensional Hilbert spaces). A similar statement has been conjectured for the Wigner $W$-distribution when restricting to Wigner-positive states \cite{Hertz2019,VanHerstraeten2021a,VanHerstraeten2021b,VanHerstraeten2021c}, but is lacking for marginal distributions. In this sense, the Husimi $Q$-distribution offers the unique opportunity to formulate such general relations for uncertainty and entanglement.

However, in experiments, it is of course not possible to measure a distribution over a continuous space to arbitrary precision. Measuring the Husimi $Q$-distribution means to approximate it based on a finite experimental data set \cite{Leonhardt1995,Braunstein2005b,Weedbrook2012,Serafini2017}. Thus, a relevant question is not only whether our criterion can flag entanglement in the limit of precise knowledge of the distribution, but also how one can, based on a fixed measurement budget and resolution, maximize the statistical significance of the detection. We show in this work that the generality of our witness leads to a significant advantage with regard to this task by optimizing over different choices of the concave function.

A common scheme for measuring the Husimi $Q$-distribution is the application of a coherent displacement followed by the detection of the vacuum projection \cite{Welsch1997,Mancini1997,Shen2016}. The experimental capabilities for realizing this scheme have been demonstrated for microwave photons in cavity QED systems \cite{Kirchmair2013}, including the bipartite setup \cite{Wang2016}, atomic gases in optical cavities \cite{Haas2014,Barontini2015}, and trapped ions \cite{Leibfried1996,Gaerttner2017}.
In this scheme the value of the Husimi $Q$-distribution is obtained on a grid of points in the four-dimensional phase space, where each grid point corresponds to a separate experimental measurement in which the corresponding displacement operation is applied. Obtaining high resolution data is thus challenging in terms of the required experimental resources \cite{Landon2018}, which motivates us to study the prospects of our entanglement criteria in the case where the Husimi $Q$ distribution is only known on a discrete grid with finite resolution.

Another established way of accessing the Husimi $Q$-distribution is heterodyne detection \cite{Collett1987,Mandel2013}, where the system modes are split by sending them on a lossless 50/50 beam splitter and subsequently each output mode is interfered with a so-called local oscillator field in a highly excited coherent state. The phase of the local oscillator controls which field quadrature is measured for each output mode and allows to simultaneously detect the local quadratures with minimal uncertainty. This corresponds to direct sampling from the Husimi $Q$-distribution \cite{Stenholm1992}. This scheme has been realized with optical photons \cite{Noh1991,Noh1992,Leonhardt1993,Mueller2016} and more recently with ultracold atomic gases \cite{Kunkel2019}, including the bipartite setting \cite{Kunkel2021}. In particular for cold atom experiments the experimental repetition rate is rather low, making it challenging to obtain sample data with high statistics. This motivates us to study the potential of our entanglement criteria for optimizing the signal-to-noise ratio of the entanglement detection for a given budget of experimental samples.

Let us remark that in this work we discuss our entanglement criteria in detail from both a theoretical and a practical perspective. An accessible description of the criteria and their application is provided in \cite{PRL}.

\textit{The remainder of this work is structured as follows.} In \autoref{sec:CanonicalPhaseSpace} we introduce the necessary background on canonical phase space, including the formulation of the uncertainty principle in terms of the Husimi $Q$-distribution, and on continuous majorization theory. After proving the general entanglement criteria in \autoref{sec:SeparabilityCriteria} we discuss the effects of the various free parameters entering them. In particular, the meaning of the choice of the concave function $f$ occurring in the criteria is made accessible by means of continuous majorization theory. Subsequently, we examine specific criteria, which includes entropic and second moment criteria. We show that the latter are the strongest possible state independent criteria that can be derived from our general criteria and that they are necessary and sufficient for Gaussian states after optimizing over scaling parameters. In \autoref{sec:Comparison} we provide a comparison of our criteria to well-known criteria based on marginals of the Wigner $W$-distribution. We show that our second moment criteria imply the DGCZ criteria and are neither stronger nor weaker than the MGVT criteria. We relate our criteria to marginal based entropic criteria by WTSTD and STW, establishing a detailed understanding of their strengths and limitations and show an example state for which our criteria detect a class of states that is not detectable by any other of the aforementioned entanglement criteria. The various criteria are generalized to arbitrary coarse-grained measurements in \autoref{sec:CoarseGrainedMeasurements}, which is accompanied with a comparison for the finite-resolution Husimi $Q$-distribution of the two-mode squeezed vacuum state. Thereupon, we exemplify how the signal-to-noise ratio can be improved by optimizing over the concave function in the case of finite statistics in \autoref{sec:sampling_meas}. A discussion of our results and of possible future directions is given in \autoref{sec:conclusions}. 

\textit{Notation}. We employ natural units $\hbar = 1$, write quantum operators with bold letters, e.g. $\boldsymbol{\rho}$, and use a bar, e.g. $\bar{Q}$, to mark vacuum quantities. If not specified differently, integrals run over the Euclidean plane $\mathbb{R}^2$.

\section{Preliminaries}
\label{sec:CanonicalPhaseSpace}
We start with introducing the reader to the Husimi $Q$-distribution and discuss in which way it is constrained by the uncertainty principle from the perspective of continuous majorization theory. Let us stress that \textit{all} presented concepts can be generalized to arbitrary simple Lie groups. Readers familiar with these topics may want to omit this section.

\subsection{Conjugate variables and coherent states}
\label{subsec:ConjugateVariables}
Throughout this work, we are concerned with continuous variable quantum systems. In the monopartite setup these are characterized by an infinite dimensional Hilbert space $\dim \mathcal{H} = \infty$ and canonical commutation relations
\begin{equation}
    [\boldsymbol{X}, \boldsymbol{P}] = i \mathds{1}
    \label{eq:CanonicalCommutationRelations}
\end{equation}
for two hermitian operators $\boldsymbol{X}$ and $\boldsymbol{P}$ with continuous and unbounded spectra. A prime example for this description is the harmonic oscillator, in which case $\boldsymbol{X}$ represents the position while $\boldsymbol{P}$ describes the momentum, but it also applies to the quadratures of an electromagnetic field or suitably chosen spin-components in a spinor Bose-Einstein condensate \cite{Gerving2012,Hamley2012,Kunkel2018,Kunkel2019,Kunkel2021}. We also introduce creation and annihiliation operators
\begin{equation}
    \boldsymbol{a}^{\dagger} = \frac{1}{\sqrt{2}} \left( \boldsymbol{X} - i \boldsymbol{P} \right), \quad \boldsymbol{a} = \frac{1}{\sqrt{2}} \left( \boldsymbol{X} + i \boldsymbol{P} \right), 
\end{equation}
respectively, which fulfill bosonic commutation relations
\begin{equation}
    [\boldsymbol{a}, \boldsymbol{a}^{\dagger} ] = \mathds{1},
    \label{eq:BosonicCommutationRelations}
\end{equation}
and single out an unique vacuum state $\ket{0}$ via $\boldsymbol{a} \ket{0} = 0$.

We construct the set of coherent states axiomatically following the group theoretic approach, which allows to associate a set of coherent states to any simple Lie group, for example $SU(2)$ describing quantum spins \cite{Zhang1990,Radcliffe1971,Gilmore1974}. We start from the Heisenberg-Weyl algebra $\mathbb{H}_4$, which consists out of the four operators $\{ \boldsymbol{a}, \boldsymbol{a}^{\dagger}, \boldsymbol{a}^{\dagger} \boldsymbol{a}, \mathds{1}\}$ and is defined via the commutation relation \eqref{eq:BosonicCommutationRelations}. The subgroup leaving the vacuum state $\ket{0}$ invariant up to a phase is $U(1) \otimes U(1)$ as one might apply rotations in the complex plane generated by $\boldsymbol{a}^{\dagger} \boldsymbol{a}$ or $\mathds{1}$ without changing the vacuum. Then, the displacement operator
\begin{equation}
    \boldsymbol{D}(\alpha) = e^{\alpha \boldsymbol{a}^{\dagger} - \alpha^* \boldsymbol{a}},
    \label{eq:DisplacementOperator}
\end{equation}
with a complex phase conveniently parameterized as
\begin{equation}
    \alpha = \frac{1}{\sqrt{2}} \left(x + i p \right),
    \label{eq:ComplexPhase}
\end{equation}
is nothing but an unitary representation of the coset space $\mathbb{H}_4 / U(1) \otimes U(1)$. Applying this operator to the vacuum generates the set of canonical coherent states
\begin{equation}
    \ket{\alpha} = \boldsymbol{D}(\alpha) \ket{0}.
    \label{eq:CoherentStates}
\end{equation}
Note that for conjugate operators defined via Eq. \eqref{eq:CanonicalCommutationRelations} one may equally introduce coherent states as eigenvalues of the annihilation operator
\begin{equation}
    \boldsymbol{a} \ket{\alpha} = \alpha \ket{\alpha}.
\end{equation}
However, the latter definition does not generalize to systems with degrees of freedom described by other algebras.

Coherent states have three interesting properties which are of importance for our later considerations. First, they are \textit{not} orthogonal 
\begin{equation}
    \abs{\braket{\alpha | \alpha'}}^2 = e^{- \abs{\alpha - \alpha'}^2},
    \label{eq:NonOrthogonality}
\end{equation}
but orthogonality is approximately restored for sufficiently distinct $\alpha$ and $\alpha'$. Second, they span an overcomplete basis
\begin{equation}
    \mathds{1} = \int \frac{\mathrm{d} x \, \mathrm{d} p}{2\pi} \boldsymbol{\ket{\alpha} \bra{\alpha}}
    \label{eq:OvercompleteBasis}
\end{equation}
and third, they minimize \textit{all} uncertainty relations, for example the Heisenberg uncertainty relation formulated in terms of variances
\begin{equation}
    \sigma_x \, \sigma_p = \frac{1}{2}.
    \label{eq:HeisenbergUncertaintyRelation}
\end{equation}

\subsection{Husimi $Q$-distribution}
\label{subsec:HusimiQDistribution}
Eq. \eqref{eq:OvercompleteBasis} shows that pure coherent state projectors $\boldsymbol{E}_{\alpha} = \boldsymbol{\ket{\alpha} \bra{\alpha}}$ constitute a positive operator-valued measure (POVM), which defines the Husimi $Q$-distribution \cite{Husimi1940,Schleich2001,Mandel2013,Schupp2022}
\begin{equation}
    Q(x,p) = \Tr \{ \boldsymbol{\rho} \, \boldsymbol{E}_{\alpha} \} = \braket{\alpha | \boldsymbol{\rho} | \alpha},
\end{equation}
with the parameterization \eqref{eq:ComplexPhase} understood.  It covers the quantum mechanical phase space and corresponds to a joint measurement of position $x$ and momentum $p$ with minimum (but still non-zero) uncertainty.

The Husimi $Q$-distribution comes with several desired properties. Most importantly, it is non-negative and bounded from above by unity
\begin{equation}
    0 \le Q (x,p) \le 1
    \label{eq:HusimiQBounds}
\end{equation}
for all $(x,p) \in \mathbb{R}^2$. This is a profound advantage over the Wigner $W$-distribution, which can become negative and hence measures of localization such as entropies are not defined for all states. The two distributions are related via a Weierstrass transform with respect to the vacuum
\begin{equation}
    \begin{split}
        Q(x,p) &= \left( W * \bar{W} \right)(x,p) \\
        &= 2 \pi \int \mathrm{d}x' \, \mathrm{d} p' \, W (x',p') \, \bar{W}(x-x',p-p'),
    \end{split}
    \label{eq:WignerToHusimiQ}
\end{equation}
with the vacuum distribution
\begin{equation}
    \bar{W}(x,p) = \frac{1}{\pi} \, e^{- \left( x^2 + p^2 \right)},
    \label{eq:WignerWVacuum}
\end{equation}
which removes the negativities of the Wigner $W$-distribution. 

Further, the Husimi $Q$-distribution is normalized to unity in the sense of
\begin{equation}
    1 = \int \frac{\mathrm{d}x \, \mathrm{d} p}{2 \pi} \, Q(x,p),
    \label{eq:HusimiQNormalization}
\end{equation}
following from $\Tr \{ \boldsymbol{\rho} \} = 1$. However, it can \textit{not} be considered a probability density function in a strict sense as the underlying random variables do not constitute mutually exclusive events by non-orthogonality \eqref{eq:NonOrthogonality}. It is therefore convenient to refer to it as a quasi-probability distribution. 

As a final remark, the Husimi $Q$-distribution serves as the quantum mechanical extension of the Boltzmann distribution, to which it converges in the classical limit $\hbar \to 0$ \cite{Wehrl1978,Wehrl1979}.

\subsection{Uncertainty principle}
\label{subsec:UncertaintyPrinciple}
In simple words, the uncertainty principle expresses the fact that non-compatible observables can not be measured simultaneously with arbitrary precision \cite{Heisenberg1927}. Hence, it sets bounds on measures of localization of the distributions obtained when measuring the observables of interest. The simplest example for a measure of localization is the variance \cite{Kennard1927,Schroedinger1930,Robertson1929,Robertson1930}, but over the last decades especially classical entropies became reasonable alternatives \cite{Maassen1988,Bialynicki-Birula1975,Coles2017,Hertz2019,Haas2021a,Haas2021b,Haas2022b}.

The differential entropy associated with the Husimi $Q$-distribution is the Wehrl entropy \cite{Wehrl1978,Wehrl1979}
\begin{equation}
    S (Q) = - \int \frac{\mathrm{d}x \, \mathrm{d} p}{2 \pi} \, Q(x,p) \, \ln Q (x,p),
    \label{eq:WehrlEntropy}
\end{equation}
which is bounded from below by the Wehrl-Lieb inequality
\begin{equation}
    S(Q) \ge 1,
    \label{eq:WehrlLiebInequality}
\end{equation}
with equality if and only if the state under consideration is a pure coherent state \cite{Lieb1978,Carlen1991,Luo2000,Lieb2014,Schupp2022}.

The latter statement has been generalized to arbitrary concave averages by allowing for a concave function $f: [0,1] \to \mathbb{R}$ with $f(0)=0$ in the integrand, which is the Lieb-Solovej theorem\footnote{For generalizations see \cite{Schupp1999,Lieb2014} for $SU(2)$, \cite{Lieb2016} for symmetric $SU(N)$ and \cite{Lieb2021,Kulikov2022} for $SU(1,1)$.} \cite{Lieb2014}
\begin{equation}
    \int \frac{\mathrm{d} x \, \mathrm{d} p}{2 \pi} f (Q) \ge \int \frac{\mathrm{d} x \, \mathrm{d} p}{2 \pi} f (\bar{Q}),
    \label{eq:LiebSolovejTheorem}
\end{equation}
again with equality only for coherent states, containing \eqref{eq:WehrlLiebInequality} as a special case for $f(t)=-t \ln t$. Therein, the vacuum Husimi $Q$-distribution is given by
\begin{equation}
    \bar{Q}(x,p) = e^{- \frac{1}{2} \left( x^2 + p^2 \right)},
    \label{eq:HusimiQVacuum}
\end{equation}
which may be replaced on the right hand side of \eqref{eq:LiebSolovejTheorem} by any coherent state. We remark that the condition $f(0) = 0$ ensures the finiteness of both sides (for $f(0) \neq 0$ both sides diverge to $+ \infty$ or $- \infty$), while the importance of $f$ being concave becomes apparent in the context of continuous majorization theory, see \autoref{subsec:ContinuousMajorizationTheory}.

We note that \eqref{eq:LiebSolovejTheorem} is \textit{not} tight for squeezed coherent states as the unitary squeezing operator $\boldsymbol{\Xi}$ does only reduce to a linear symplectic map $\Xi = \text{diag} (\xi, 1/\xi)$ with $\xi > 0$ in phase space such that $(x,p) \to (\xi x, p / \xi)$ when applied to the Wigner $W$-distribution. More precisely, the correspondence between Gaussian unitaries on states and affine symplectic maps on the Husimi $Q$-distribution is broken by the Weierstrass transformation \eqref{eq:WignerToHusimiQ} in case of squeezing. From a quantum optics perspective, this is due to the vacuum signal in the heterodyne measurement being not equally squeezed with the input signal.
Therefore, one has to introduce a squeezing transformation in \eqref{eq:LiebSolovejTheorem} and minimize with respect to $\xi$ in order to end up with a relation which is tight for \textit{all} pure Gaussian states.

\begin{figure*}[t!]	
    \centering
    \includegraphics[width=0.99\textwidth]{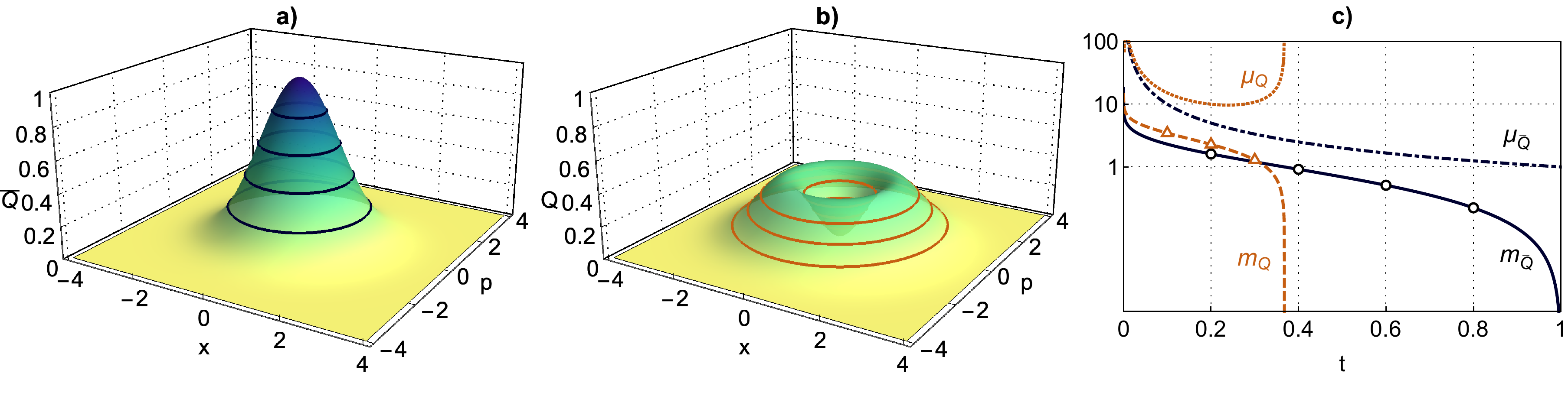}
    \caption{\textbf{a)} and \textbf{b)} show the Husimi $Q$-distributions of the vacuum state \eqref{eq:HusimiQVacuum} and the first Fock state \eqref{eq:HusimiQFock}, respectively. The areas enclosed by the black and dark orange circles (divided by $2 \pi$) give the values for the corresponding level functions at given height, which are depicted as plot markers in \textbf{c)}, together with the level and density-level functions in a logarithmic plot. For $t \le 1/e$ we have $\mu_Q (t) \ge \mu_{\bar{Q}} (t)$, while for $1/e < t \le 1$ we obtain $\mu_{\bar{Q}} (t) \ge \mu_Q (t) = 0$. Irrespective of the concave function $f(t)$ multiplied to the density-level functions, the area below this product for the vacuum state is always smaller than for the Fock state, which illustrates the majorization relation $Q \prec \bar{Q}$.}
    \label{fig:MajorizationTheory}
\end{figure*}

\subsection{Continuous majorization theory}
\label{subsec:ContinuousMajorizationTheory}
Most generally, measures of localization are described within the framework of majorization theory. For continuous distributions, \textit{localization} replaces the notion of \textit{ordering} in case of discrete probability distributions. Some probability density function $\varphi (x,p)$ is said to be majorized by another distribution $\varphi' (x,p)$, written as $\varphi \prec \varphi'$, if and only if 
\begin{equation}
    \int \frac{\mathrm{d} x \, \mathrm{d}p}{2 \pi} \, f(\varphi) \ge \int \frac{\mathrm{d} x \, \mathrm{d}p}{2 \pi} \, f(\varphi')
    \label{eq:ContinuousMajorizationDefinition}
\end{equation}
for all concave functions $f: \mathcal{I} \to \mathbb{R} $ with $f(0) = 0$ and $\mathcal{I} = [0, \max \{ \varphi, \varphi' \}] $ \cite{Marshall2011,VanHerstraeten2021a,VanHerstraeten2021b,VanHerstraeten2021c}. Hence, $\varphi \prec \varphi'$ expresses the intuition that $\varphi'$ is more localized than $\varphi$ in most general terms.

Interestingly, the Lieb-Solovej theorem \eqref{eq:LiebSolovejTheorem} is fundamentally a statement about localization of distributions in phase space and states that no Husimi $Q$-distribution $Q$ is more localized than the vacuum Husimi $Q$-distribution $\bar{Q}$ (or equivalently all coherent ones), i.e. $Q \prec \bar{Q}$. Note that it is sufficient for $Q(x,p)$ to be continuous, non-negative and normalized in order to apply \eqref{eq:ContinuousMajorizationDefinition} and hence the violation of Kolmogorov's third axiom can be safely disregarded.

To make these insights more explicit, we introduce some techniques from continuous majorization theory. We begin with the level function $m_Q (t)$, which is associated with a given Husimi $Q$-distribution $Q(x,p)$ and is defined as the phase space measure of the domain $\mathcal{A}$ for which the value of the distribution exceeds a given threshold $t$, i.e.
\begin{equation}
    m_Q (t) = \int_{\mathcal{A}} \frac{\mathrm{d} x \, \mathrm{d} p}{2 \pi} \,\, \text{with} \,\, \mathcal{A} = \{ (x,p): Q(x,p) \ge t \},
    \label{eq:LevelFunctionDefinition}
\end{equation}
where $t \in \mathcal{I} = [0,1]$ as a result of the boundedness of the Husimi $Q$-distribution \eqref{eq:HusimiQBounds}. The definition \eqref{eq:LevelFunctionDefinition} is conveniently rewritten as
\begin{equation}
    m_Q (t) = \int \frac{\mathrm{d} x \, \mathrm{d} p}{2 \pi} \, \Theta \left[ Q(x,p) - t \right],
    \label{eq:LevelFunctionWithTheta}
\end{equation}
with $\Theta$ denoting the Heaviside $\Theta$-function. The negative derivative of the latter quantity defines the density-level function
\begin{equation}
    \mu_Q (t) = - \frac{\mathrm{d}}{\mathrm{d} t} m_Q (t) = \int \frac{\mathrm{d} x \, \mathrm{d} p}{2 \pi} \, \delta \left[ Q(x,p) - t \right],
    \label{eq:DensityLevelFunction}
\end{equation}
where $\delta$ is the Dirac $\delta$-distribution. 

With \eqref{eq:DensityLevelFunction} at hand we can simplify the integrals appearing in the Lieb-Solovej theorem \eqref{eq:LiebSolovejTheorem} as
\begin{equation}
    \int \frac{\mathrm{d} x \, \mathrm{d} p}{2 \pi} \, f (Q) = \int_\mathcal{I} \mathrm{d} t \, f (t) \, \mu_Q (t).
    \label{eq:FunctionsIntegralIdentity}
\end{equation}
If $f$ is differentiable almost everywhere on $\mathcal{I}$ we can equally write with \eqref{eq:LevelFunctionWithTheta}
\begin{equation}
        \int \frac{\mathrm{d} x \, \mathrm{d} p}{2 \pi} \, f (Q) = \int_\mathcal{I} \mathrm{d} t \, \frac{\mathrm{d} f(t)}{\mathrm{d}t} \, m_Q (t),
    \label{eq:FunctionsIntegralIdentity2}
\end{equation}
where we used $f(0)=0$ and $m_Q (1)=0$. The latter two equations show that continuous majorization relations ultimately boil down to a comparison of density-level or level functions. Note that these functions remain invariant for all transformations with unit determinant, i.e. those which preserve phase space area elements, including symplectic transformations such as rotations and squeezings of the coordinate axes as well as displacements (see \eqref{eq:DensityLevelFunctionGaussianTrafo} for the case of Gaussian distributions).

We illustrate all introduced concepts by comparing the Husimi $Q$-distributions and their localization of the vacuum state \eqref{eq:HusimiQVacuum} with the first excited Fock state $\ket{1}$, for which we obtain
\begin{equation}
    Q(x,p) = \frac{x^2 + p^2}{2} \, e^{- \frac{1}{2} \left( x^2 + p ^2 \right)}.
    \label{eq:HusimiQFock}
\end{equation}
The two distributions are shown in \autoref{fig:MajorizationTheory} \textbf{a)} and \textbf{b)}, respectively. The corresponding level functions are obtained by integrating over the area enclosed by the Husimi $Q$-distribution (and dividing by $2 \pi$) at a given height, which we sketch for a few heights with black and dark orange circles, respectively. The resulting level and density-level functions are plotted in \autoref{fig:MajorizationTheory} \textbf{c)}, where the aforementioned values are represented by plot markers. While the vacuum curves take positive values up to $t=1$, the Fock curves evaluate to zero for $t > 1/e$. If we multiply the density-level functions with a concave function which overweights small $t$, for example $f(t) = t^{1/5}$, the integrated area of the Fock curve enclosed with the $t$-axis will be larger than the area below the vacuum curve. Similarly, if we choose a function pronouncing large $t$, for example $f(t)=-t^5$, the area will again be larger for the Fock curve due to the minus sign. This exemplifies the majorization relation $Q \prec \bar{Q}$ in form of the Lieb-Solovej theorem \eqref{eq:LiebSolovejTheorem} with \eqref{eq:FunctionsIntegralIdentity}  understood for two choices for $f$. The intuition provided here will be useful later when choosing suitable $f$ for certifying entanglement.

\section{Entanglement criteria}
\label{sec:SeparabilityCriteria}
We extend the latter considerations to the bipartite case and derive various classes of entanglement criteria using the Lieb-Solovej theorem \eqref{eq:LiebSolovejTheorem} for general non-local observables.

\subsection{Bipartite setup and non-local operators}
We now investigate two continuous variable quantum systems forming a bipartition, which is described by a bipartite quantum state $\boldsymbol{\rho}_{12}$. The degrees of freedom of the two subsystems are encoded in the local operators $\boldsymbol{X}_j, \boldsymbol{P}_j$, where $j \in \{1,2 \}$ labels the subsystem, which fulfill
\begin{equation}
    \left[ \boldsymbol{X}_j, \boldsymbol{P}_k \right] = i \delta_{j k} \mathds{1}.
    \label{eq:CanonicalCommutationRelationsBipartition}
\end{equation}
For the sake of generality we incorporate the effect of rotations in the local subsystems, which will play an important role in the comparison of our criteria to criteria based on marginal distributions. We describe a rotation around an angle $\vartheta \in [0, 2 \pi)$ by a rotation matrix
\begin{equation}
    \mathcal{T} (\vartheta) = \begin{pmatrix}
    \cos \vartheta & \sin \vartheta \\
    - \sin \vartheta & \cos \vartheta
    \end{pmatrix},
    \label{eq:RotationMatrix}
\end{equation}
leading to rotated operators via the transformation
\begin{equation}
    (\boldsymbol{X}_j, \boldsymbol{P}_j) \to (\boldsymbol{R}_j, \boldsymbol{S}_j) = \mathcal{T} (\vartheta_j) (\boldsymbol{X}_j, \boldsymbol{P}_j).
\end{equation}
As the canonical commutation relations \eqref{eq:CanonicalCommutationRelationsBipartition} are conserved by local rotations, we can associate Husimi $Q$-distributions to the full system as well as to both subsystems along the lines of \autoref{subsec:ConjugateVariables} and \autoref{subsec:HusimiQDistribution}. More precisely, the global Husimi $Q$-distribution is obtained by applying a POVM formed by local coherent states $\boldsymbol{E}_j = \boldsymbol{\ket{\alpha_j} \bra{\alpha_j}}$ to both subsystems
\begin{equation}
    \begin{split}
        Q(r_1, s_1, r_2, s_2) &= \Tr \left\{ \boldsymbol{\rho}_{12} \, \left(\boldsymbol{E}_1 \otimes \boldsymbol{E}_2 \right) \right\} \\
        &= (\bra{\alpha_1} \otimes \bra{\alpha_2}) \, \boldsymbol{\rho}_{12} \, (\ket{\alpha_1} \otimes \ket{\alpha_2}).
    \end{split}
    \label{eq:GlobalHusimiQDefinition}
\end{equation}
The resulting distribution is still bounded in the sense of \eqref{eq:HusimiQBounds}, but now normalized to unity with respect to the four-dimensional phase space measure $\mathrm{d}r_1 \mathrm{d}s_1 \mathrm{d}r_2 \mathrm{d}s_2 / 4 \pi^2$. Local Husimi $Q$-distributions result from integrating out the complementary phase space variables, for example
\begin{equation}
    \begin{split}
        Q (r_1, s_1) &= \Tr \left\{ \boldsymbol{\rho}_{1} \, \boldsymbol{E}_1\right\} \\
        &= \int \frac{\mathrm{d} r_2 \, \mathrm{d}s_2}{2 \pi} \, Q(r_1, s_1, r_2, s_2)
    \end{split}
\end{equation}
for subsystem $1$ and analogously for subsystem $2$.

Following \cite{Einstein1935,Duan2000,Giovannetti2003}, we adopt non-local operators for the study of entanglement by adding and subtracting local operators. Additionally, we allow for relative scalings between the four local operators, such that we end up with
\begin{equation}
    \boldsymbol{R}_{\pm} = a_1 \boldsymbol{R}_1 \pm a_2 \boldsymbol{R}_2, \,\, \boldsymbol{S}_{\pm} = b_1 \boldsymbol{S}_1 \pm b_2 \boldsymbol{S}_2,
    \label{eq:NonLocalOperatorsDefinition}
\end{equation}
with $a_1,b_1,a_2,b_2 \ge 0$ and $a_1 b_1 = a_2 b_2$. These operators fulfill the commutation relations 
\begin{equation}
    [\boldsymbol{R}_{\pm}, \boldsymbol{S}_{\pm}] = i (a_1 b_1 + a_2 b_2) \boldsymbol{\mathds{1}}, \quad [\boldsymbol{R}_{\pm}, \boldsymbol{S}_{\mp}] = 0,
    \label{eq:NonLocalOperatorsCommutationRelations}
\end{equation}
showing that pairs of operators with equal indices represent independent oscillator modes with canonical commutation relations.

To express the global Husimi $Q$-distribution in terms of the non-local variables \eqref{eq:NonLocalOperatorsDefinition}, we employ a variable transformation
\begin{equation}
    \begin{split}
        &Q (r_+,s_+,r_-,s_-) = \frac{1}{4 a_1 b_1 a_2 b_2} \\
        &\quad\times Q \left(\frac{r_+ + r_-}{2a_1}, \frac{s_+ + s_-}{2b_1}, \frac{r_+ - r_-}{2a_2}, \frac{s_+ - s_-}{2b_2} \right),
    \end{split}
    \label{eq:GlobalHusimiQNonLocal}
\end{equation}
where the determinant of the Jacobian matrix evaluates to $1/(4 a_1 b_1 a_2 b_2)$. For entanglement criteria, we are particularly interested in the marginals over the mixed variables pairs $(r_{\pm}, s_{\mp})$ of the latter distribution
\begin{equation}
    Q_{\pm} \equiv Q_{\pm} (r_{\pm}, s_{\mp}) = \int \frac{\mathrm{d} r_{\mp} \, \mathrm{d} s_{\pm}}{2 \pi} \, Q (r_+,s_+,r_-,s_-),
    \label{eq:MarginalHusimiQ}
\end{equation}
which are not constrained by the uncertainty principle as the underlying phase space operators commute \eqref{eq:NonLocalOperatorsCommutationRelations}. 

In contrast, the distributions corresponding to equal signs 
\begin{equation}
    Q (r_{\pm}, s_{\pm}) = \int \frac{\mathrm{d} r_{\mp} \, \mathrm{d} s_{\mp}}{2 \pi} \, Q (r_+,s_+,r_-,s_-)
\end{equation}
constitute true Husimi $Q$-distributions and may equally be defined via coherent state projectors with respect to $(\boldsymbol{R}_{\pm}, \boldsymbol{S}_{\pm})$ after a partial trace over the complementary degrees of freedom. Therefore, they are restrained by the uncertainty principle in the form of the Lieb-Solovej theorem \eqref{eq:LiebSolovejTheorem}. However, the different normalization affects the size of the codomain of the Husimi $Q$-distribution (and hence the domain of the concave function $f$) as well as the form of the vacuum expression \eqref{eq:HusimiQVacuum}, which now reads
\begin{equation}
    \bar{Q} (r_{\pm}, s_{\pm}) = \frac{1}{a_1 b_1 + a_2 b_2} \, e^{- \frac{1}{2} \frac{r^2_{\pm} + s^2_{\pm}}{a_1 b_1 + a_2 b_2}}.
    \label{eq:NonLocalVacuumHusimiQMixed}
\end{equation}
Adapted to this setup, the Lieb-Solovej theorem states
\begin{equation}
    \int \frac{\mathrm{d} r_{\pm} \, \mathrm{d} s_{\pm}}{2 \pi} \, f ( Q ) \ge \int \frac{\mathrm{d} r_{\pm} \, \mathrm{d} s_{\pm}}{2 \pi} \, f \left( \bar{Q} \right),
    \label{eq:LiebSolovejTheoremNonLocal}
\end{equation}
for any concave $f: [0, (a_1 b_1 + a_2 b_2)^{-1}] \to \mathbb{R}$ with $f(0)=0$. As discussed at the end of \autoref{subsec:UncertaintyPrinciple}, we additionally have to allow for an optimization over a squeezing transformation $\Xi$ in the non-local Wigner $W$-distribution $W_{\pm}(r_{\pm}, s_{\mp})$ in order to render the latter inequality tight for all pure Gaussian states. As the tightness of an uncertainty relation is closely related to the detection capabilities of entanglement criteria derived from such a relation, we explicitly allow for this possibility in the following.

\subsection{General criteria}
We now show that the distribution $Q_{\pm}$ is non-trivially constrained for all separable states. A bipartite quantum state $\boldsymbol{\rho}_{12}$ is called separable if it can be written as a convex combination of product states, i.e.
\begin{equation}
    \boldsymbol{\rho}_{12} = \sum_j p_j \left(\boldsymbol{\rho}_1^j \otimes \boldsymbol{\rho}_2^j \right),
    \label{eq:SeparableStateDefinition}
\end{equation}
where $p_j \ge 0$ denotes a discrete probability distribution with $\sum_j p_j = 1$. A widespread method to investigate the separability of a given quantum state is to apply a positive but not completely positive trace-preserving map and check whether the resulting operator is still non-negative, i.e. constitutes a valid density operator \cite{Horodecki2009,Guehne2009}. A prime example for this method is the Peres-Horodecki (PPT) criterion \cite{Peres1996,Horodecki1996} which utilizes the partial transpose $\boldsymbol{T}_2$ (conveniently applied to subsystem two) leading to a necessary, but in general not sufficient, condition for separability: for every separable state $\boldsymbol{\rho}_{12}$ the operator $\boldsymbol{\rho}_{12} \to \boldsymbol{\rho}'_{12} = \left(\boldsymbol{\mathds{1}}_1 \otimes \boldsymbol{T}_2 \right) (\boldsymbol{\rho}_{12})$ is non-negative $\boldsymbol{\rho}'_{12} \ge 0$. In particular, $\boldsymbol{\rho}'_{12}$ is \textit{physical} for separable $\boldsymbol{\rho}_{12}$, and hence the Lieb-Solovej theorem \eqref{eq:LiebSolovejTheoremNonLocal} applies also to Husimi $Q$-distributions of $\boldsymbol{\rho}'_{12}$.

In order to relate these distributions to observable distributions we have to translate the action of the partial transpose $\boldsymbol{T}_2$ on the state $\boldsymbol{\rho}_{12}$ into an action onto the variable pairs $(r_{\pm}, s_{\pm})$ spanning phase space. Is is straightforward to show that the partial transpose $\boldsymbol{T}_2$ flips the sign of the local variable $s_2 \to - s_2$, which holds for all quasi-probability distributions covering phase space \cite{Simon2000}. On the level of the non-local variables \eqref{eq:NonLocalOperatorsDefinition} this is equivalent to $s_{\pm} \to s_{\mp}$ implying that the partial transpose $\boldsymbol{T}_2$ corresponds to the transformation $Q (r_{\pm}, s_{\pm}) \to Q'(r_{\pm}, s_{\pm}) = Q_{\pm} (r_{\pm}, s_{\mp})$.

By the PPT criterion the distribution $Q_{\pm} (r_{\pm}, s_{\mp})$ is thus constrained by the Lieb-Solovej theorem \eqref{eq:LiebSolovejTheoremNonLocal}, where the vacuum expression on the right hand side has to be expressed in the mixed variable pairs
\begin{equation}
    \begin{split}
        \bar{Q}'_{\pm} (r_{\pm}, s_{\mp}) &= \bar{Q} (r_{\pm}, s_{\mp}) \\
        &= \frac{1}{a_1 b_1 + a_2 b_2} \, e^{- \frac{1}{2} \frac{r^2_{\pm} + s^2_{\mp}}{a_1 b_1 + a_2 b_2}}.
    \end{split}
    \label{eq:NonLocalVacuumHusimiQ}
\end{equation}
In order to make this statement more explicit we introduce a witness functional $\mathcal{W}_f$
\begin{equation}
    \mathcal{W}_f = \int \frac{\mathrm{d} r_{\pm} \, \mathrm{d} s_{\mp}}{2 \pi} \left[ f (Q_{\pm})  - f \left(\bar{Q}'_{\pm} \right) \right],
    \label{eq:WitnessDefinition}
\end{equation}
with concave $f: \mathcal{J} \to \mathbb{R}$ fulfilling $f(0) = 0$ defined over the interval $\mathcal{J}=[0,\max \{ \max{Q_{\pm}}, (a_1 b_1 + a_2 b_2)^{-1} \}] \subseteq \mathbb{R}^{+}$. Then, our main result is that $\mathcal{W}_f$ is non-negative for all separable states, i.e.
\begin{equation}
    \boldsymbol{\rho}_{12} \text{ separable} \Rightarrow \mathcal{W}_f \ge 0.
    \label{eq:SeparabilityCriteria}
\end{equation}
If the latter inequality is violated for a given state $\boldsymbol{\rho}_{12}$ entanglement is demonstrated.

Let us stress the generality of our criteria \eqref{eq:SeparabilityCriteria}, as one may optimize over the two local rotation angles $\vartheta_1, \vartheta_2$, the four scaling parameters $a_1, b_1, a_2, b_2$ under the constraint $a_1 b_1 = a_2 b_2$, the squeezing transformation $\Xi$ and the class of concave functions $f$ with $f(0)=0$ to witness entanglement. In the following, we discuss the influences of all quantities in detail to develop a systematic understanding of their effect.

\subsection{Rotation angles and scaling parameters}
\label{subsec:RotationsAndScalings}
We begin our study with the rotation angles $\vartheta_1, \vartheta_2$ and the scaling parameters $a_1, b_1, a_2, b_2$. While we already used a matrix representation for the former in \eqref{eq:RotationMatrix}, we write
\begin{equation}
    \mathcal{U} (a_1, a_2, b_1, b_2) = \begin{pmatrix}
    a_1 & 0 & a_2 & 0 \\
    0 & b_1 & 0 & -b_2 \\
    a_1 & 0 & -a_2 & 0 \\
    0 & b_1 & 0 & b_2
    \end{pmatrix}
\end{equation}
for the matrix mixing the local operators in \eqref{eq:NonLocalOperatorsDefinition}, i.e. $(\boldsymbol{R}_+, \boldsymbol{S}_-, \boldsymbol{R}_-, \boldsymbol{S}_+) = \mathcal{U} (a_1, a_2, b_1, b_2) (\boldsymbol{R}_1, \boldsymbol{S}_1, \boldsymbol{R}_2, \boldsymbol{S}_2)$. Then, the four initial local operators $(\boldsymbol{X}_1, \boldsymbol{P}_1, \boldsymbol{X}_2, \boldsymbol{P}_2)$ transform as
\begin{equation}
    \begin{split}
        (\boldsymbol{X}_1, \boldsymbol{P}_1, \boldsymbol{X}_2, \boldsymbol{P}_2)
        &\to (\boldsymbol{R}_+, \boldsymbol{S}_-, \boldsymbol{R}_-, \boldsymbol{S}_+) \\
        &= \mathcal{G} (\vartheta_1, \vartheta_2, a_1, a_2, b_1, b_2) \\
        &\hspace{0.4cm}(\boldsymbol{X}_1, \boldsymbol{P}_1, \boldsymbol{X}_2, \boldsymbol{P}_2),
    \end{split}
\end{equation}
with the full transformation matrix
\begin{equation}
    \begin{split}
        &\mathcal{G} (\vartheta_1, \vartheta_2, a_1, a_2, b_1, b_2) \\
        &= \mathcal{U} (a_1, a_2, b_1, b_2) \left[ \mathcal{T} (\vartheta_1) \oplus \mathcal{T} (\vartheta_2) \right]
    \end{split}
\end{equation}
describing the effects of our six parameters of interest.

For equal scaling in the local quadratures, i.e. $a_1 = b_1$ and $a_2 = b_2$ (which includes the most relevant case $a_1=b_1=a_2=b_2=1$), the matrix $\mathcal{G}$ can be rewritten as a rotation in subsystem $1$ around an angle $\vartheta$ (subsystem $2$ is left unchanged) and a transformation to non-local variables $(r_{\pm}, s_{\mp})$ followed by a rotation in these variables around $\phi = - \vartheta_2$, i.e.
\begin{equation}
    \begin{split}
        \mathcal{G} (\vartheta_1, \vartheta_2, a_1, a_2, a_1, a_2) = &\left[ \mathcal{T}(\phi) \oplus \mathcal{T} (\phi) \right] \\
        &\mathcal{G} (\vartheta, 0, a_1, a_2, a_1, a_2),
    \end{split}
\end{equation}
when choosing $\vartheta = \arctan (\cos \vartheta_1 \cos \vartheta_2 - \sin \vartheta_1 \sin \vartheta_2, \cos \vartheta_1 \sin \vartheta_2 + \sin \vartheta_1 \cos \vartheta_2)$, where $\arctan (x,y)$ denotes the arcus tangens in the Euclidean plane $(x,y)$. As the final rotations $\mathcal{T}(\phi) \oplus \mathcal{T} (\phi)$ leave our criteria \eqref{eq:SeparabilityCriteria} invariant, we only need to optimize over \textit{one} angle $\vartheta$. In contrast, criteria based on marginal distributions require angle tomography over \textit{both} angles $\vartheta_1, \vartheta_2$, which is substantially more costly in terms of experimental runs.

\begin{figure*}[t!]	
    \centering
    \includegraphics[width=0.99\textwidth]{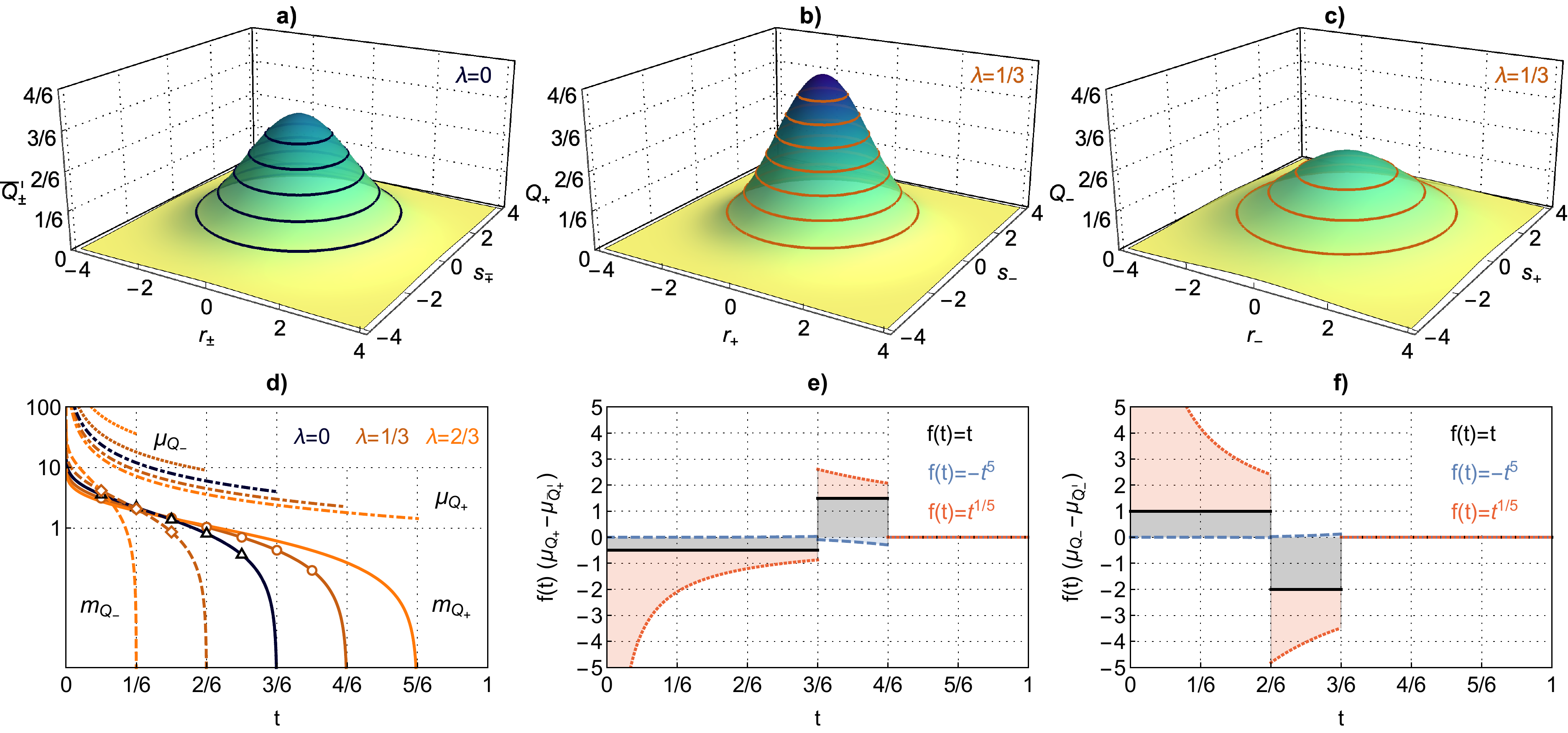}
    \caption{\textbf{a)} and \textbf{b)} / \textbf{c)} show the distributions of the two-mode squeezed vacuum state \eqref{eq:TwoModeSqueezedVacuumState} with $a_1 = b_1 = a_2 = b_2 = 1$, $\vartheta_1 = \vartheta_2 = 0$ and $\xi = 1$ for $\lambda=0$ and $\lambda=1/3$ in $(r_+,s_-)$ / $(r_-,s_+)$ variables, respectively. The areas enclosed by the circles give the values level functions at given height, which are depicted as plot markers in \textbf{d)}, together with level and density-level functions from \eqref{eq:LevelAndDensityLevelFunctionsTMSV} of both variable pairs for selected values of $\lambda$. For $\lambda=1/3$, the integrand of the witness functional \eqref{eq:WitnessDensityLevelFunctions} is shown in \textbf{e)} / \textbf{f)} for $(r_+,s_-)$ / $(r_-,s_+)$ variables.}
    \label{fig:MajorizationTheoryEntanglement}
\end{figure*}

\subsection{Concave function $f$}
\label{subsec:ConcaveFunction}
Following the analysis in \autoref{subsec:ContinuousMajorizationTheory}, our entanglement criteria \eqref{eq:SeparabilityCriteria} state that all distributions $Q_{\pm}$ corresponding to separable states are less localized than the vacuum distribution $\bar{Q}'_{\pm}$. In contrast, some entangled states have sufficiently localized distributions $Q_{\pm}$, such that they are detected by the criteria \eqref{eq:SeparabilityCriteria} for some $f$. The corresponding distribution $Q_{\pm}$ majorizes the vacuum distribution $\bar{Q}'_{\pm}$ if and only if entanglement can be certified for all $f$.

Using the definition of the density-level function \eqref{eq:FunctionsIntegralIdentity}, we can rewrite our criteria as
\begin{equation}
    \mathcal{W}_f = \int_\mathcal{J} \mathrm{d} t \, f (t) \, \left[ \mu_{Q_{\pm}} (t) - \mu_{\bar{Q}'_{\pm}} (t) \right] \ge 0,
    \label{eq:WitnessDensityLevelFunctions}
\end{equation}
showing that also our criteria \eqref{eq:SeparabilityCriteria} reduce to a comparison of the density-level functions $\mu_{Q_{\pm}} (t)$ and $\mu_{\bar{Q}'_{\pm}} (t)$ and an appropriate choice for $f(t)$. Note that the vacuum expression \eqref{eq:NonLocalVacuumHusimiQ} is bounded from above by $(a_1 b_1 + a_2 b_2)^{-1}$, such that the second term in \eqref{eq:WitnessDensityLevelFunctions} is effectively integrated over the interval $[0,(a_1 b_1 + a_2 b_2)^{-1}]$. However, some entangled states have sufficiently localized $Q_{\pm}$, s.t. their density-level functions take positive values for $t > (a_1 b_1 + a_2 b_2)^{-1}$.

As a simple example, we consider the two-mode squeezed vacuum (TMSV) state
\begin{equation}
    \ket{\psi} = \sqrt{1 - \lambda^2} \sum_{n=0}^{\infty} (- \lambda)^n \, \ket{n} \otimes \ket{n},
    \label{eq:TwoModeSqueezedVacuumState}
\end{equation}
with $\lambda \in [0,1]$ being the squeezing parameter, which is entangled for all $\lambda > 0$. For $a_1 = b_1 = a_2 = b_2 = 1$, $\vartheta_1 = \vartheta_2 = 0$ and $\xi = 1$ we obtain a Gaussian distribution with covariance matrix $V_{\pm} = \frac{2}{1 \pm \lambda} \, \mathds{1}$. In \autoref{fig:MajorizationTheoryEntanglement} \textbf{a)} we show the vacuum case $\lambda=0$ (for which $Q_+=Q_-$), while in \textbf{b)} and \textbf{c)} we illustrate $\lambda = 1/3$ for $Q_+$ and $Q_-$, respectively. Entanglement is only detected in the $(r_{+}, s_{-})$ variables as the corresponding distribution is more localized than the vacuum. The rotational symmetry of these distributions permits analytic calculations of the level and density-level functions when working with polar coordinates, which leads to
\begin{equation}
    \begin{split}
        m_{Q_{\pm}} (t) &= -\text{det}^{1/2} \, V_{\pm} \, \ln \left( t \, \text{det}^{1/2} \, V_{\pm} \right), \\
        \mu_{Q_{\pm}} (t) &= \frac{\text{det}^{1/2} \, V_{\pm}}{t},
    \end{split}
    \label{eq:LevelAndDensityLevelFunctionsTMSV}
\end{equation}
for $t \in [0,\text{det}^{-1/2} V_{\pm}]$, respectively. Both are shown for both variable pairs in \autoref{fig:MajorizationTheoryEntanglement} \textbf{d)} for the vacuum (black curves), $\lambda=1/3$ (dark orange curves) and $\lambda = 2/3$ (light orange curves) with solid (dashed) and dot-dashed (dotted) curves for the $+$ ($-$) level and density-level functions, respectively. The values at the plot markers in $t$-steps of $1/12$ correspond to the phase space measures, i.e. areas divided by $2 \pi$, of the circles in \autoref{fig:MajorizationTheoryEntanglement} \textbf{a)}-\textbf{c}). Further, we show the integrand of our witness functional \eqref{eq:WitnessDensityLevelFunctions} for the distributions \textbf{b)} and \textbf{c)} in \textbf{e)} and \textbf{f)}, respectively, for various choices for the concave function $f(t)$, such that the sum of the shaded areas corresponds to the value of \eqref{eq:WitnessDensityLevelFunctions}. The black solid curves express normalization, while the blue dashed and red dotted curves illustrate that the absolute value of the witness \eqref{eq:WitnessDensityLevelFunctions} increases for smaller monomial exponents for the state \eqref{eq:TwoModeSqueezedVacuumState}. However, entanglement is detected in $Q_+$ for all choices of $f$ as the state is Gaussian,  see \autoref{subsubsec:GaussianStates}.

\subsection{Entropic criteria}
Of particular interest are entropic functionals as there exist a variety of criteria formulated in terms of entropies of measurement distributions \cite{Guehne2004,Guehne2004b,Walborn2009,Saboia2011,Walborn2011,Schneeloch2018,Schneeloch2019,Bergh2021a,Bergh2021b,Haas2022a}. In the following, we discuss entropic criteria as special choices for the concave function $f$. In order to arrive at entropic criteria, it is additionally necessary to apply a monotonous function to both terms in the witness functional \eqref{eq:WitnessDefinition}. More precisely, we consider a monotonically increasing function $g: \mathds{R} \to \mathds{R}$ such that the witness functional becomes
\begin{equation}
    \mathcal{W}_{f,g} = g \left[\int \frac{\mathrm{d} r_{\pm} \, \mathrm{d} s_{\mp}}{2 \pi} f (Q_{\pm}) \right] - g \left[ \int \frac{\mathrm{d} r_{\pm} \, \mathrm{d} s_{\mp}}{2 \pi} f (\bar{Q}'_{\pm}) \right],
    \label{eq:WitnessWithg}
\end{equation}
which still fulfills the entanglement criteria \eqref{eq:SeparabilityCriteria}. As the converse statement $\mathcal{W}_f \le 0$ holds for $-f$ (or equivalently for convex $f$), applying a monotonically \textit{decreasing} function $g$ still results in $\mathcal{W}_{f,g} \ge 0$. We note that this procedure does neither strengthen nor weaken the entanglement criteria.

\subsubsection{Rényi-Wehrl entropies}
We start with the family of Rényi-Wehrl entropies
\begin{equation}
    S_{\beta} (Q_{\pm}) = \frac{1}{1 - \beta} \ln \left[ \int \frac{\mathrm{d} r_{\pm} \, \mathrm{d} s_{\mp}}{2 \pi} \, Q^{\beta}_{\pm} (r_{\pm},s_{\mp}) \right],
    \label{eq:RenyiWehrlEntropyDefinition}
\end{equation}
with entropic orders $\beta \in (0,1) \cup (1, \infty)$. Hence, we choose $f(t) = t^{\beta}$, which is concave / convex and a monotonically increasing / decreasing function $g(t) = \frac{1}{1 - \beta} \ln t$ for $\beta < 1$ / $\beta > 1$, in which case \eqref{eq:WitnessWithg} implies
\begin{equation}
    \mathcal{W}_{\beta} = S_{\beta} (Q_{\pm}) - \frac{\ln \beta}{\beta - 1} - \frac{\ln \det \bar{V}'_{\pm}}{2} \ge 0,
    \label{eq:RenyiWehrlCriteria}
\end{equation}
where $\bar{V}'_{\pm} = (a_1 b_1 + a_2 b_2) \mathds{1}$ denotes the covariance matrix of the vacuum $\bar{Q}'_{\pm}$, for all separable states $\boldsymbol{\rho}_{12}$.

\subsubsection{Tsallis-Wehrl entropies}
For Tsallis-Wehrl entropies
\begin{equation}
    S_{\gamma} (Q_{\pm}) = \frac{1}{\gamma -1} \left[1 - \int \frac{\mathrm{d} r_{\pm} \, \mathrm{d} s_{\mp}}{2 \pi} \, Q_{\pm}^{\gamma} (r_{\pm},s_{\mp}) \right],
    \label{eq:TsallisWehrlEntropyDefinition}
\end{equation}
with $\gamma \in (0,1) \cup (1, \infty)$\footnote{For $\gamma < 0$ we have $f(0) \to \infty$ and hence we have to restrict to Tsallis-Wehrl entropies with $\gamma > 0$.} we choose again monomials $f(t) = t^{\gamma}$, but now we take $g(t) = \frac{1 - t}{\gamma - 1}$, which is also monotonically increasing / decreasing for $\gamma < 1$ / $\gamma > 1$, leading to the criteria
\begin{equation}
    \mathcal{W}_{\gamma} = S_{\gamma} (Q_{\pm}) - \frac{1}{\gamma - 1} \left[ 1 - \frac{(\det \bar{V}'_{\pm})^{\frac{1-\gamma}{2}}}{\gamma} \right] \ge 0.
    \label{eq:TsallisWehrlCriteria}
\end{equation}
As the function $f$ agrees for both entropic families, the criteria \eqref{eq:RenyiWehrlCriteria} and \eqref{eq:TsallisWehrlCriteria} are equally strong in the sense that they detect the same set of entangled states. Therefore, an appropriate choice regarding which entropic family is being considered can be based on other decisive factors depending on the application: while Tsallis' entropy was designed to generalize statistical mechanics to situations where an entropy is non-extensive, Rényi's aim was to formulate a family of entropies which includes not only Shannon's, but also Hartley's and other entropies, which are of particular interest in the context of communication protocols.

\subsubsection{Wehrl entropy}
In the limits $\beta, \gamma \to 1$ we obtain from \eqref{eq:RenyiWehrlCriteria} and \eqref{eq:TsallisWehrlCriteria} the entropic witness
\begin{equation}
    \mathcal{W}_{1} = S_1 (Q_{\pm}) - 1 - \frac{\ln \det \bar{V}'_{\pm}}{2} \ge 0,
    \label{eq:WehrlCriteria}
\end{equation}
in terms of the Wehrl entropy
\begin{equation}
    \hspace{-0.05cm}S_1 (Q_{\pm}) = - \int \frac{\mathrm{d} r_{\pm} \, \mathrm{d}s_{\mp}}{2 \pi} \, Q_{\pm} (r_{\pm}, s_{\mp}) \ln Q_{\pm} (r_{\pm}, s_{\mp}).
    \label{eq:WehrlEntropy2}
\end{equation}
Note that this result also follows directly from \eqref{eq:WitnessDefinition} for the choice $f(t) = - t \ln t$. Note also that for $a_1 = b_1 = a_2 = b_2 = 1$ the corresponding criteria reduce to the criteria reported in \cite{Haas2022a}.

\subsection{Second moment criteria}
Another interesting class of witnesses comprises second moments, which are the simplest quantities revealing information about the localization of a distribution. We derive second moment criteria from our general criteria and discuss them in the context of Gaussian states.

\subsubsection{Determinant of the covariance matrix}
We start from the non-local covariance matrix $\gamma_{\pm}$ of an arbitrary Wigner $W$-distribution
\begin{equation}
    \gamma_{\pm} = \begin{pmatrix}
    \sigma^2_{r_{\pm}} & \sigma_{r_{\pm} s_{\mp}} \\
    \sigma_{r_{\pm} s_{\mp}} & \sigma^2_{s_{\mp}}
    \end{pmatrix},
\end{equation}
which contains all three second moments, i.e. the two variances
\begin{equation}
    \begin{split}
        \sigma^2_{r_{\pm}} &= \int \frac{\mathrm{d}r_{\pm} \, \mathrm{d}s_{\mp}}{2 \pi} \, r^2_{\pm} \, W_{\pm} (r_{\pm}, s_{\mp}), \\
        \sigma^2_{s_{\mp}} &= \int \frac{\mathrm{d}r_{\pm} \, \mathrm{d}s_{\mp}}{2 \pi} \, s^2_{\mp} \, W_{\pm} (r_{\pm}, s_{\mp}),
    \end{split}
\end{equation}
as well as the covariance
\begin{equation}
    \sigma_{r_{\pm} s_{\mp}} = \int \frac{\mathrm{d}r_{\pm} \, \mathrm{d}s_{\mp}}{2 \pi} \, r_{\pm} \, s_{\mp} \, W_{\pm} (r_{\pm}, s_{\mp}),
\end{equation}
which characterizes the correlations between $r_{\pm}$ and $s_{\mp}$. Note here that we have assumed zero expectation values without loss of generality as the entanglement criteria \eqref{eq:SeparabilityCriteria} and all marginal criteria we consider in \autoref{sec:Comparison} are invariant under displacements.

Moreover, we write for the covariance matrix $V_{\pm}$ of the Husimi $Q$-distribution
\begin{equation}
    V_{\pm} = \begin{pmatrix}
    \Sigma^2_{r_{\pm}} & \Sigma_{r_{\pm} s_{\mp}} \\
    \Sigma_{r_{\pm} s_{\mp}} & \Sigma^2_{s_{\mp}}
    \end{pmatrix},
\end{equation}
with second moments
\begin{equation}
    \begin{split}
        \Sigma^2_{r_{\pm}} &= \int \frac{\mathrm{d}r_{\pm} \, \mathrm{d}s_{\mp}}{2 \pi} \, r^2_{\pm} \, Q_{\pm} (r_{\pm}, s_{\mp}), \\
        \Sigma^2_{s_{\mp}} &= \int \frac{\mathrm{d}r_{\pm} \, \mathrm{d}s_{\mp}}{2 \pi} \, s^2_{\mp} \, Q_{\pm} (r_{\pm}, s_{\mp}), \\
        \Sigma_{r_{\pm} s_{\mp}} &= \int \frac{\mathrm{d}r_{\pm} \, \mathrm{d}s_{\mp}}{2 \pi} \, r_{\pm} \, s_{\mp} \, Q_{\pm} (r_{\pm}, s_{\mp}).
    \end{split}
\end{equation}
Using \eqref{eq:WignerToHusimiQ} adapted to the non-local variables, the latter covariance matrix is related to the former via
\begin{equation}
    V_{\pm} = \gamma_{\pm} + \bar{\gamma}_{\pm} = \begin{pmatrix}
        \sigma_{r_{\pm}}^2 + \frac{a_1^2 + a_2^2}{2} & \sigma_{r_{\pm} s_{\mp}} \\
        \sigma_{r_{\pm} s_{\mp}} & \sigma_{s_{\mp}}^2 + \frac{b_1^2 + b_2^2}{2}
          \end{pmatrix},
    \label{eq:CovarianceMatrixDefinition}
\end{equation}
which shows that the variances of the Husimi $Q$-distribution are strictly larger than the ones of the Wigner $W$-distribution, while the covariances agree. Note here that $\bar{\gamma}_{\pm}$ is obtained when marginalizing the Wigner $W$-distribution of the global vacuum after transforming to non-local variables.

To derive second moment criteria, we use the fact that the Wehrl entropy $S_1 (Q_{\pm})$ is maximized by a Gaussian Husimi $Q$-distribution $Q_{\pm}$ for a fixed covariance matrix $V_{\pm}$, i.e.
\begin{align}
    S_1 (Q_{\pm}) \le 1 + \frac{1}{2} \ln \det V_{\pm}.
\end{align}
With this upper bound the Wehrl entropic witness functional \eqref{eq:WehrlCriteria} reduces to the second moment witness
\begin{equation}
    \mathcal{W}_{\det V_{\pm}} = \det V_{\pm} - \det \bar{V}'_{\pm} \ge 0,
    \label{eq:SecondMomentCriteria}
\end{equation}
which is based on the determinant of the covariance matrix $V_{\pm}$ and thus contains information about all three second moments, in particular about the correlations between $r_{\pm}$ and $s_{\mp}$.

\subsubsection{Gaussian states}
\label{subsubsec:GaussianStates}
Gaussian states are characterized by Gaussian Husimi $Q$-distributions, which are of the form 
\begin{equation}
    Q_{\pm} (r_{\pm}, s_{\mp}) = \frac{1}{Z} \, e^{-\frac{1}{2} (r_{\pm}, s_{\mp})^T V_{\pm}^{-1} (r_{\pm}, s_{\mp})},
    \label{eq:GaussianHusimiQ}
\end{equation}
with $Z = \det^{1/2} V_{\pm}$ being a normalization constant. We have again set the expectation values to zero without loss of generality. 

Interestingly, for the class of Gaussian distributions \eqref{eq:GaussianHusimiQ}, the general entanglement criteria \eqref{eq:SeparabilityCriteria} are \textit{equivalent} to the second moment criteria $\mathcal{W}_{\det V_{\pm}} \ge 0$ for \textit{all} concave $f$ with $f(0)=0$. To prove this equivalence, we start from Simon's normal form of the bipartite Wigner covariance matrix \cite{Simon2000} (standard form II in \cite{Duan2000})
\begin{equation}
    \gamma_{12} = \begin{pmatrix}
    m_1 & 0 & m_+ & 0 \\
    0 & m_1 & 0 & m_- & \\
    m_+ & 0 & m_2 & 0 \\
    0 & m_- & 0 & m_2
    \end{pmatrix},
    \label{eq:GlobalCovarianceMatrixNormalForm}
\end{equation}
with $m_1, m_2, m_+, m_- \in \mathbb{R}$ (note that these parameters are also constrained by the uncertainty principle formulated for $\gamma_{12}$). Every bipartite covariance matrix $\gamma_{12}$ can be brought into this form by local, single-mode symplectic transformations $\mathcal{S}_1 \otimes \mathcal{S}_2$ with $\mathcal{S}_1, \mathcal{S}_2 \in Sp(2, \mathbb{R})$, which does not alter the separability of the state as such transformations correspond to the class of local operations and classical communications (LOCCs). After adding the global vacuum $\bar{\gamma}_{12}$ to obtain $V_{12}$, transforming to the non-local variables \eqref{eq:NonLocalOperatorsDefinition} and integrating out the mixed variable pairs according to \eqref{eq:MarginalHusimiQ}, we find the diagonal matrix
\begin{widetext}
    \begin{equation}
        V_{\pm} = \frac{1}{2} \, \begin{pmatrix}
        a_1^2 (1 + 2 m_1) + a_2^2 (1 + 2 m_2) \pm 4 a_1 a_2 m_{+} & 0 \\
        0 & b_1^2 (1 + 2 m_1) + b_2^2 (1 + 2 m_2) \mp 4 b_1 b_2 m_{-}
        \end{pmatrix}.
        \label{eq:GeneralCovarianceMatrixpm}
    \end{equation}
\end{widetext}
Starting from the vacuum $\bar{Q}'_{\pm} (r_{\pm},s_{\mp})$, a general Gaussian distribution $Q_{\pm} (\tilde{r}_{\pm}, \tilde{s}_{\mp})$ with a covariance matrix of the form \eqref{eq:GeneralCovarianceMatrixpm} can be obtained by applying an affine linear coordinate transformation in phase space
\begin{equation}
    (r_{\pm},s_{\mp})  \to (\tilde{r}_{\pm},\tilde{s}_{\mp}) = M_{\pm} \, (r_{\pm},s_{\mp}).
    \label{eq:GaussianCoordinateTransformation}
\end{equation}
Comparing \eqref{eq:GeneralCovarianceMatrixpm} with \eqref{eq:NonLocalVacuumHusimiQ} shows that the transformation matrix $M_{\pm}$ is diagonal and reads
\begin{equation}
    M_{\pm} = \sqrt{V_{\pm} \bar{V}_{\pm}^{'-1}},
    \label{eq:CovarianceMatrixGaussianCoordinateTransformation}
\end{equation}
which is always well-defined as all involved components are positive and real by definition. Indeed, with the transformation \eqref{eq:GaussianCoordinateTransformation} we obtain for the quadratic form in the exponent
\begin{equation}
    \begin{split}
        &(r_{\pm},s_{\mp})^T \bar{V}_{\pm}^{-1} (r_{\pm},s_{\mp}) \\
        &\to (\tilde{r}_{\pm},\tilde{s}_{\mp})^T V_{\pm}^{-1} (\tilde{r}_{\pm},\tilde{s}_{\mp}) \\
        &= (r_{\pm},s_{\mp})^T M_{\pm}^{T} V_{\pm}^{-1} M_{\pm} (r_{\pm},s_{\mp}) \\
        &= (r_{\pm},s_{\mp})^T \bar{V}_{\pm}^{-1} (r_{\pm},s_{\mp}).
    \end{split}
\end{equation}
Further, \eqref{eq:CovarianceMatrixGaussianCoordinateTransformation} implies 
\begin{equation}
    \det M_{\pm}^2 = \frac{\det V_{\pm}}{\det \bar{V}'_{\pm}},
    \label{eq:DeterminantGaussianCoordinateTransformation}
\end{equation}
such that we may obtain a relation between $\det V_{\pm}$ and $\det \bar{V}'_{\pm}$ from $\det M_{\pm}^2$. 

Under the coordinate transformation \eqref{eq:GaussianCoordinateTransformation}, the vacuum distribution $\bar{Q}'_{\pm} (r_{\pm},s_{\mp})$ transforms as
\begin{equation}
    \bar{Q}'_{\pm} (r_{\pm},s_{\mp}) \to Q_{\pm}(\tilde{r}_{\pm},\tilde{s}_{\mp}) = \frac{\bar{Q}'_{\pm} (r_{\pm},s_{\mp})}{\det M_{\pm}}.
    \label{eq:HusimiQGaussianCoordinateTransformation}
\end{equation}
Similarly, the level functions transform as
\begin{equation}
    m_{\bar{Q}'_{\pm}} (t) \to m_{Q_{\pm}} (t) = \det M_{\pm} \, m_{\bar{Q}'_{\pm}} (t \, \det M_{\pm}),
\end{equation}
while for the density-level functions we find
\begin{equation}
    \mu_{\bar{Q}'_{\pm}} (t) \to \mu_{Q_{\pm}} (t) = \det M_{\pm}^2 \, \mu_{\bar{Q}'_{\pm}} (t \, \det M_{\pm}).
    \label{eq:DensityLevelFunctionGaussianTrafo}
\end{equation}
This shows that the level as well as the density-level functions of arbitrary Gaussian distributions are scaled versions of each other, which is also evident in \eqref{eq:LevelAndDensityLevelFunctionsTMSV} and \autoref{fig:MajorizationTheoryEntanglement} \textbf{d)}.

Plugging the relation \eqref{eq:HusimiQGaussianCoordinateTransformation} into our witness functional \eqref{eq:WitnessDefinition} for the first or second term yields
\begin{equation}
    \mathcal{W}_f = \int \frac{\mathrm{d} r_{\pm} \, \mathrm{d} s_{\mp}}{2 \pi} \left[ \det M_{\pm} \, f \left( \frac{\bar{Q}'_{\pm}}{\det M_{\pm}} \right)  - f \left(\bar{Q}'_{\pm} \right) \right]
    \label{eq:WitnessGaussianTransformed1}
\end{equation}
or
\begin{equation}
    \mathcal{W}_f = \int \frac{\mathrm{d} \tilde{r}_{\pm} \, \mathrm{d} \tilde{s}_{\mp}}{2 \pi} \left[ f \left(Q_{\pm} \right) - \frac{f \left( Q_{\pm} \det M_{\pm} \right)}{\det M_{\pm}} \right],
    \label{eq:WitnessGaussianTransformed2}
\end{equation}
respectively. As $f$ is concave with $f(0)=0$, it is a subadditive function and hence fulfills $f(\kappa t) \ge \kappa f(t)$ for all real $\kappa \in [0,1]$ and all $t \in \mathcal{J}$. Thus, $\det M_{\pm} \ge 1$ implies $\mathcal{W}_f \ge 0$ from \eqref{eq:WitnessGaussianTransformed1}, while $\det M_{\pm} \le 1$ implies $\mathcal{W}_f \le 0$ from \eqref{eq:WitnessGaussianTransformed2}. As the latter is equivalent to $\mathcal{W}_f \ge 0$ implying $\det M_{\pm} \ge 1$ by contraposition, we can conclude with \eqref{eq:DeterminantGaussianCoordinateTransformation} that the second moment criteria $\mathcal{W}_{\det V_{\pm}} = \det V_{\pm} - \det \bar{V}'_{\pm} \ge 0$ and the general criteria $\mathcal{W}_f \ge 0$ are equivalent for all Gaussian states and for \textit{all} $f$ under our standard assumptions.

\subsubsection{Optimality}
\label{subsubsec:SecondMomentCriteriaFromRenyiWehrl}
The considerations of the preceding section have strong implications regarding the optimality of the second moment witness \eqref{eq:SecondMomentCriteria}: it is optimal in the sense that no stronger state-independent bound on $\det V_{\pm}$ can be implied from the general criteria \eqref{eq:SeparabilityCriteria} as this would be in contradiction with the latter equivalence in case of Gaussian distributions. More precisely, if there existed such a stronger bound, it would have to hold for all states, in particular for Gaussian states. But since the general criteria \eqref{eq:SeparabilityCriteria} are equivalent to the second moment criteria from \eqref{eq:SecondMomentCriteria} for Gaussian states, there can not be any such bound. Therefore, taking $f(t) \neq - t \ln t$ and maximizing the witness functional $\mathcal{W}_f$ over $Q_{\pm}$ for fixed $V_{\pm}$ cannot lead to stronger second moment criteria. 

We illustrate the aforementioned optimality by showing the inferiority of second moment criteria stemming from maximizing Rényi-Wehrl entropies $S_{\beta} (Q_{\pm})$ for fixed $V_{\pm}$. Depending on the entropic order $\beta > \frac{1}{2}$, Rényi-Wehrl entropies are maximized by the distributions \cite{Vignat2006,Johnson2007}
\begin{equation}
    \begin{split}
        &Q_{\pm} (r_{\pm}, s_{\mp}) \\
        &= Z (\beta) \left(1 - \frac{1}{2} \frac{\beta-1}{2\beta - 1} (r_{\pm}, s_{\mp})^T \, V^{-1}_{\pm} \ (r_{\pm}, s_{\mp}) \right)^{\frac{1}{\beta-1}}_+,
    \end{split}
    \label{eq:StudentTRHusimiQ}
\end{equation}
with $x_+ = \max (x, 0)$ as the support for $\beta > 1$ is defined as the compact domain for which $(r_{\pm},s_{\mp})^T V^{-1}_{\pm} (r_{\pm},s_{\mp}) \le \nu$ (for larger values of $(r_{\pm},s_{\mp})$ the distribution $Q_{\pm}$ would become negative), a normalization factor
\begin{equation}
    Z (\beta) = \begin{cases}
    \frac{\Gamma \left(\frac{1}{1-\beta} \right)}{\Gamma \left(\frac{\beta}{1-\beta} \right)} \frac{1-\beta}{2\beta - 1} \, \text{det}^{-1/2} \, V_{\pm} \quad &\frac{1}{2} < \beta < 1, \\
    \frac{\Gamma \left(\frac{\beta}{\beta-1} + 1 \right)}{\Gamma \left(\frac{\beta}{\beta-1} \right)} \frac{\beta-1}{2\beta - 1} \, \text{det}^{-1/2} \, V_{\pm} \quad &\beta > 1,
    \end{cases}
\end{equation}
and $\Gamma (x)$ denoting the $\Gamma$-function. These correspond to two-dimensional Student-t and Student-r distributions, for $\frac{1}{2} < \beta < 1$ with scale matrix $\Sigma_{\pm} = (\nu - 2) \, V_{\pm}$ and $\nu = \frac{2 \beta}{1 - \beta}$ degrees of freedom and for $\beta > 1$ with scale matrix $\sigma_{\pm} = \nu \, V_{\pm}$ and $\nu = 2 \frac{2 \beta - 1}{\beta - 1}$ degrees of freedom, respectively.

\begin{figure}[t!]
    \centering
    \includegraphics[width=0.85\columnwidth]{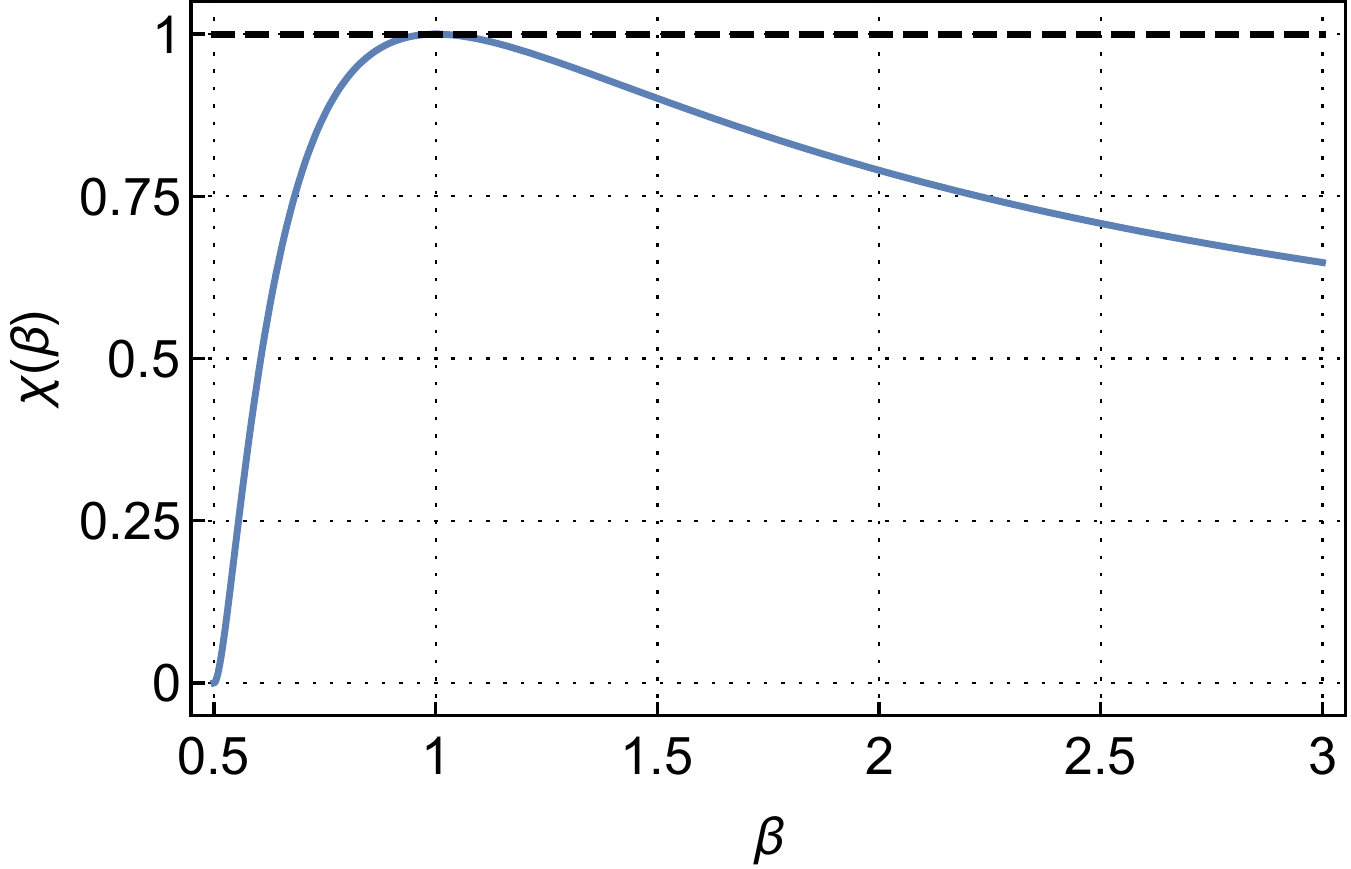}
    \caption{Additional factor $\chi (\beta)$ in the second moment criteria \eqref{eq:SecondMomentCriteriaRenyiWehrl} obtained from maximizing the left hand side of the Rényi-Wehrl criteria \eqref{eq:RenyiWehrlCriteria} for fixed covariance matrix $V_{\pm}$ (blue solid curve). We observe $\chi (\beta) \le 1$ with equality for $\beta \to 1$ (see black dashed line), illustrating the optimality of the main second moment criteria \eqref{eq:SecondMomentCriteria}.}
    \label{fig:SecondMomentCriteriaRenyiWehrl}
\end{figure}

Bounding the left hand side of the Rényi-Wehrl witness \eqref{eq:RenyiWehrlCriteria} from above by the Rényi-Wehrl entropy of the distributions \eqref{eq:StudentTRHusimiQ} and simplifying the result leads to the second moment criteria
\begin{equation}
    \mathcal{\tilde{W}}_{\det V_{\pm}} = \det V_{\pm} - \chi (\beta) \, \det \bar{V}'_{\pm} \ge 0,
    \label{eq:SecondMomentCriteriaRenyiWehrl}
\end{equation}
with the non-negative function
\begin{equation}
    \chi(\beta) = \left(2 - \frac{1}{\beta} \right)^{-\frac{2 \beta}{\beta-1}} \, \beta^{ \frac{2}{\beta - 1}},
\end{equation}
which is shown in \autoref{fig:SecondMomentCriteriaRenyiWehrl}. As $\chi (\beta) \le 1$ for all $\beta \in \left(\frac{1}{2}, 1 \right) \cup (1, \infty)$ with equality in the limit $\beta \to 1$ (in which case the distributions \eqref{eq:StudentTRHusimiQ} converge to Gaussian distributions \eqref{eq:GaussianHusimiQ}), the criteria \eqref{eq:SecondMomentCriteriaRenyiWehrl} are indeed never stronger than \eqref{eq:SecondMomentCriteria}.

\section{Comparison with marginal approach}
\label{sec:Comparison}
Having discussed various families of entanglement criteria following from our general criteria \eqref{eq:SeparabilityCriteria}, we now compare the performance of these criteria to their counterparts based on marginal distributions. 

\subsection{Second moment criteria}
\label{subsec:ComparisonSecondMomentCriteria}
\begin{figure*}[t!]
    \centering
    \includegraphics[width=0.99\textwidth]{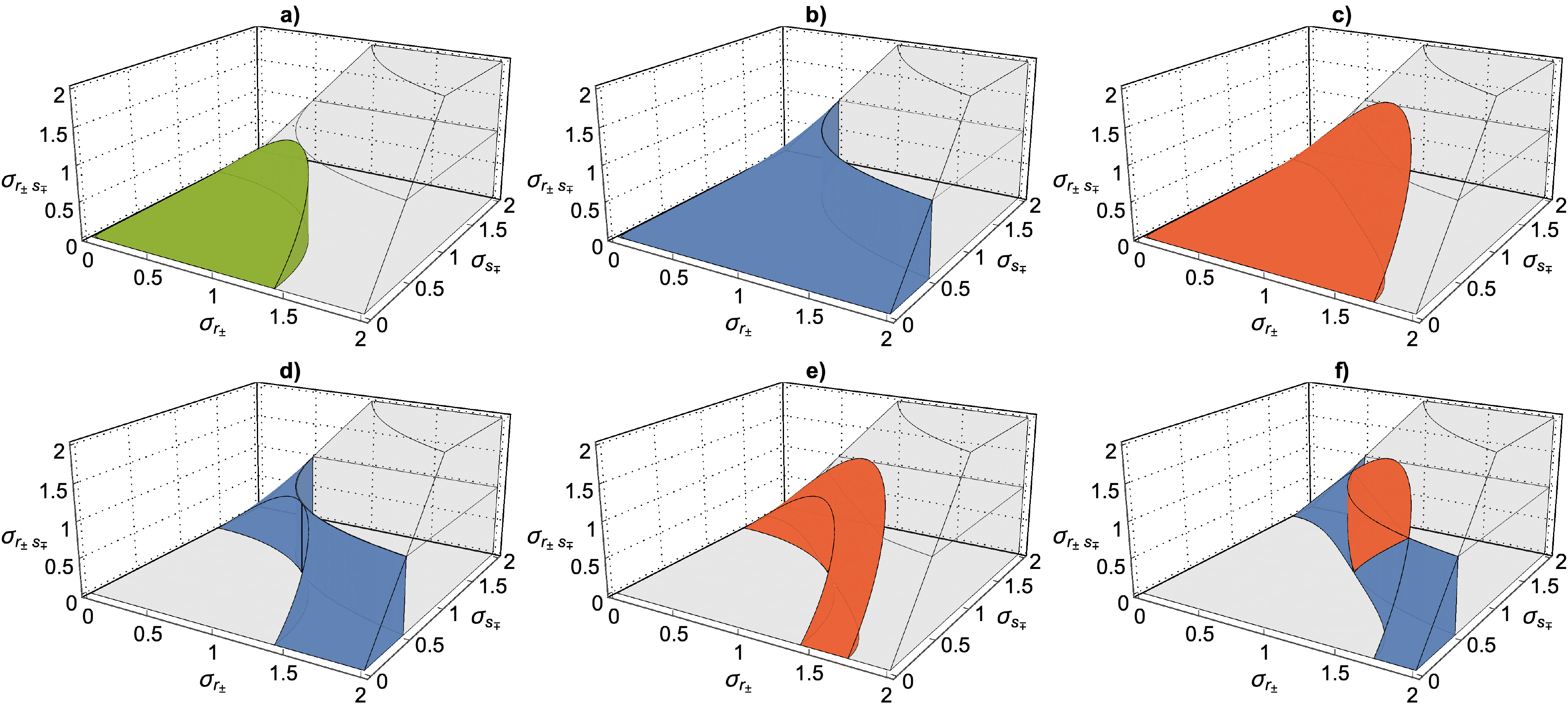}
    \caption{Upper row: Witnessed regions by the DGCZ, MGVT and our second moment criteria are shown in \textbf{a)}, \textbf{b)}, \textbf{c)}, respectively, with the light gray regions indicating the allowed regions from non-negativity of the covariance matrix $\gamma_{\pm}$, i.e. $\sigma_{r_{\pm}} \sigma_{s_{\mp}} \ge \sigma_{r_{\pm} s_{\mp}}$ (overall, we restrict to $\sigma_{r_{\pm} s_{\mp}} \ge 0$ for simplicity). Lower row: Comparisons of witnessed regions between two criteria. \textbf{d)} / \textbf{e)} show the outperformance regions of the MGVT / our criteria over the DGCZ criteria, exemplifying that the former two are strictly stronger conditions than the latter. Our criteria are compared with the MGVT criteria in \textbf{f)}, highlighting their abilities to account for improper angles and squeezings, respectively. A combined figure is provided in \cite{PRL}.}
    \label{fig:SecondMomentCriteriaComparison}
\end{figure*}

There exists a simple relation between our second moment witness \eqref{eq:SecondMomentCriteria} and the MGVT criteria \cite{Mancini2002,Giovannetti2003}. Namely, using that $\det \bar{V}'_{\pm} = 4 \det \bar{\gamma}'_{\pm}$ and multiplying out the determinant with \eqref{eq:CovarianceMatrixDefinition} we find
\begin{equation}
    \begin{split}
        \mathcal{W}_{\det V_{\pm}} &= \mathcal{W}_{\text{MGVT}} + \frac{b_1^2 + b_2^2}{2} \,  \sigma_{r_{\pm}}^2 + \frac{a_1^2 + a_2^2}{2} \, \sigma_{s_{\mp}}^2 \\
        &+\frac{(a_1^2 + a_2^2)(b_1^2 + b_2^2) - 3 (a_1 b_1 + a_2 b_2)^2}{4}\\
        &- \sigma^2_{r_{\pm} s_{\mp}},
    \end{split}
    \label{eq:SecondMomentCriteriaDecomposition}
\end{equation}
where
\begin{equation}
    \mathcal{W}_{\text{MGVT}} = \sigma^2_{r_{\pm}} \, \sigma^2_{s_{\mp}} - \frac{\det \bar{V}'_{\pm}}{4}
    \label{eq:MGVTCriteria}
\end{equation}
is the generalized witness functional for the MGVT criteria. Another interesting set of criteria, which contain a sum rather than a product of variances, are the DGCZ criteria \cite{Duan2000}, expressed through the witness functional
\begin{equation}
    \mathcal{W}_{\text{DGCZ}} = \sigma_{r_{\pm}}^2 + \sigma_{s_{\mp}}^2 - \text{det}^{1/2} \, \bar{V}'_{\pm}.
    \label{eq:DGCZCriteria}
\end{equation}
We show the witnessed regions of the DGCZ, MGVT and our second moment criteria in \autoref{fig:SecondMomentCriteriaComparison} \textbf{a)}, \textbf{b)}, \textbf{c)}, respectively, together with the physically allowed regions by the non-negativity of $\gamma_{\pm}$ (light gray shaded areas) for $a_1 = b_1 = a_2 = b_2 = 1$, $\vartheta_1 = \vartheta_2 = 0$ and $\xi = 1$.

It is well known that the MGVT criteria imply the DGCZ criteria, which follows from $x^2 y^2 \le \frac{1}{4} (x^2 + y^2)^2$ for all $x,y \in \mathbb{R}$, see \autoref{fig:SecondMomentCriteriaComparison} \textbf{d)}. Similarly, our second moment criteria \eqref{eq:SecondMomentCriteria} imply the DGCZ criteria \eqref{eq:DGCZCriteria} for all scaling parameters $a_1, b_1, a_2, b_2$ due to $\sigma^2_{r_{\pm} s_{\mp}} \ge 0$ as well, which we exemplify in \autoref{fig:SecondMomentCriteriaComparison} \textbf{e)}. As the DGCZ criteria are necessary and sufficient for separability in case of Gaussian states when optimized over the scaling parameters $a_1,b_1,a_2,b_2$, the same holds true for the MGVT and our criteria.

Further, our criteria and the MGVT criteria are equivalent after optimizing both criteria over the set of symplectic transformations under which they are \textit{not} invariant together with an optimization over the scaling parameters $a_1, b_1, a_2, b_2$. To that end, we consider symplectic transformations $\mathcal{S}_{\pm} \in Sp(2, \mathbb{R})$ in the non-local phase spaces $(r_{\pm}, s_{\mp})$ transforming the Wigner $W$-distribution as $W_{\pm}(r_{\pm}, s_{\mp}) \to \tilde{W}_{\pm} (\tilde{r}_{\pm}, \tilde{s}_{\mp})$ with $(\tilde{r}_{\pm}, \tilde{s}_{\mp}) = \mathcal{S}_{\pm} \, (r_{\pm}, s_{\mp})$. Then, the non-local covariance matrix transforms as $\gamma_{\pm} \to \tilde{\gamma}_{\pm} = \mathcal{S}_{\pm} \gamma \mathcal{S}_{\pm}^T$ with $\det \mathcal{S}_{\pm} = 1$ \cite{Simon2000}. This implies for our second moment witness functional \eqref{eq:SecondMomentCriteria}
\begin{equation}
    \begin{split}
        \mathcal{W}_{\text{det} V_{\pm}} &\to \tilde{\mathcal{W}}_{\text{det} V_{\pm}} \\
        &= \det \tilde{V}_{\pm} - \det \bar{V}'_{\pm} \\
        &= \det \left( \mathcal{S}_{\pm} \gamma_{\pm} \mathcal{S}_{\pm}^T + \bar{\gamma}_{\pm} \right) - \det \bar{V}'_{\pm} \\
        &= \det \left[ \gamma_{\pm} + \bar{\gamma}_{\pm} (\mathcal{S}_{\pm}^T \mathcal{S}_{\pm})^{-1} \right] - \det \bar{V}'_{\pm},
    \end{split}
    \label{eq:SecondMomentCriteriaSqueezing}
\end{equation}
showing explicitly that the criteria remain invariant under orthogonal transformations $\mathcal{S}_{\pm} \mathcal{S}_{\pm}^T = \mathds{1}$, i.e. rotations, but not under squeezing for which $\mathcal{S}_{\pm} = \Xi = \text{diag} (\xi, 1/\xi)$ with $\xi>0$. In contrast, the MGVT criteria \eqref{eq:MGVTCriteria} are invariant under squeezing only, in which case
\begin{equation}
    \begin{split}
        \mathcal{W}_{\text{MGVT}} \to \tilde{\mathcal{W}}_{\text{MGVT}}
        &= \xi^2 \, \sigma_{r_{\pm}}^2 \, \frac{1}{\xi^2} \, \sigma_{s_{\mp}}^2 - \frac{\det \bar{V}'_{\pm}}{4} \\
        &= \mathcal{W}_{\text{MGVT}}.
    \end{split}
\end{equation}
Note that the DGCZ criteria \eqref{eq:DGCZCriteria} are not invariant under either of the two transformations.

Let us now apply a squeezing transformation to our second moment criteria \eqref{eq:SecondMomentCriteria} in the sense of \eqref{eq:SecondMomentCriteriaSqueezing}, which yields
\begin{equation}
    \begin{split}
        \mathcal{W}_{\text{det} V_{\pm}} &= \left( \xi^2 \, \sigma_{r_{\pm}}^2 + \frac{a_1^2 + a_2^2}{2} \right)\\
        &\hspace{0.05cm} \times \left( \frac{1}{\xi^2} \, \sigma_{s_{\mp}}^2 + \frac{b_1^2 + b_2^2}{2} \right) \\
        &-\sigma_{r_{\pm} s_{\mp}}^2 - \left( a_1 b_1 + a_2 b_2 \right)^2.
    \end{split}
\end{equation}
To optimize over $\xi$, we minimize the latter expression with respect to $\xi$, which gives 
\begin{equation}
    \xi^2 = \frac{\sigma_{s_{\mp}}}{\sigma_{r_{\pm}}} \sqrt{\frac{a_1^2 + a_2^2}{b_1^2 + b_2^2}},
\end{equation}
leading to
\begin{equation}
    \begin{split}
        \mathcal{W}_{\text{det} V_{\pm}} &= \left( \sigma_{r_{\pm}} \sigma_{s_{\mp}} + \frac{\sqrt{(a_1^2 + a_2^2)(b_1^2 + b_2^2)}}{2} \right)^2 \\
        &-\sigma_{r_{\pm} s_{\mp}}^2 - \left( a_1 b_1 + a_2 b_2 \right)^2.
    \end{split}
\end{equation}
The MGVT criteria become optimal for an appropriate choice of the coordinate axes s.t. $\sigma_{r_{\pm} s_{\mp}} = 0$, and then the non-negativity of the latter is equivalent to
\begin{equation}
    \sigma_{r_{\pm}} \sigma_{s_{\mp}} + \frac{\sqrt{(a_1^2 + a_2^2)(b_1^2 + b_2^2)}}{2} - (a_1 b_1 + a_2 b_2) \ge 0.
\end{equation}
Finally, noting that $\sqrt{(a_1^2 + a_2^2)(b_1^2 + b_2^2)} \ge a_1 b_1 + a_2 b_2$ with equality for $b_2 = (b_1 a_2)/a_1$ shows that the statements $\mathcal{W}_{\text{det} V_{\pm}} \ge 0$ and $\mathcal{W}_{\text{MGVT}} \ge 0$ are indeed equivalent after optimizing over the angle $\phi$, the squeezing $\xi$ as well as the scaling parameters $a_1, b_1, a_2, b_2$\footnote{Note that $\mathcal{W}_{\text{det} V_{\pm}}$ and $\mathcal{W}_{\text{MGVT}}$ themselves are still unequal.}.

The different invariances of our criteria \eqref{eq:SecondMomentCriteria} and the MGVT criteria \eqref{eq:MGVTCriteria} cause differences in their performances for fixed $a_1, b_1, a_2, b_2$, depending on whether the considered Husimi $Q$-distribution is sufficiently rotated or squeezed, see \autoref{fig:SecondMomentCriteriaComparison} \textbf{f)}. Further, \eqref{eq:SecondMomentCriteriaDecomposition} provides an intuition for one set of criteria to outperform the other. Roughly speaking, our criteria outperform / are outperformed by the MGVT criteria for $\sigma_{r_{\pm}} \approx \sigma_{s_{\mp}}$ and $\abs{\sigma_{r_{\pm} s_{\mp}}} > 0$ / for $\sigma_{r_{\pm}} \approx 1 / \sigma_{s_{\mp}}$ and $\abs{\sigma_{r_{\pm} s_{\mp}}} \approx 0$.

\subsection{Entropic criteria}
Although there is no direct analog to the relation \eqref{eq:SecondMomentCriteriaDecomposition}, we can derive an inequality between our Wehrl entropic criteria \eqref{eq:WehrlCriteria} and the marginal-based entropic criteria put forward by WTSTD in \cite{Walborn2009}, which will provide an intuition for the later comparison.

We start by decomposing the Wehrl entropy \eqref{eq:WehrlEntropy2} in the sense of a joint entropy
\begin{equation}
    S (Q_{\pm}) = S (F_{\pm}) + S (G_{\mp}) - I (F_{\pm} : G_{\mp}),
    \label{eq:WehrlEntropyDecomposition}
\end{equation}
where the marginals of the Husimi $Q$-distribution read
\begin{equation}
    \begin{split}
        F_{\pm} &\equiv F_{\pm} (r_{\pm}) = \int \frac{\mathrm{d} s_{\mp}}{\sqrt{2 \pi}} \, Q_{\pm}, \\
        G_{\mp} &\equiv G_{\mp} (s_{\mp}) = \int \frac{\mathrm{d} r_{\pm}}{\sqrt{2 \pi}} \, Q_{\pm}.
    \end{split}
\end{equation}
Following \eqref{eq:WignerToHusimiQ}, these distributions correspond to smeared-out versions of the true marginal distributions associated with a measurement of the operators $\boldsymbol{R}_{\pm}$ and $\boldsymbol{S}_{\mp}$, i.e. $f_{\pm}(r_{\pm}) = \braket{r_{\pm} | \boldsymbol{\rho} | r_{\pm}}$ and $g_{\mp}(s_{\mp}) = \braket{s_{\mp} | \boldsymbol{\rho} | s_{\mp}}$, respectively, via
\begin{equation}
    \begin{split}
        F_{\pm} (r_{\pm}) &= \left(f_{\pm} * \bar{f}_{\pm} \right) (r_{\pm}) \\
         &= \sqrt{2 \pi} \int \mathrm{d}r'_{\pm} \, f_{\pm} (r'_{\pm}) \, \bar{f}_{\pm} (r_{\pm}-r'_{\pm}),
    \end{split}
\end{equation}
with $f_{\pm} (r_{\pm})$ being normalized with respect to $\mathrm{d}r_{\pm}/\sqrt{2 \pi}$, and similarly for $g_{\mp}(s_{\mp})$.

Analogous to the covariance $\sigma_{r_{\pm} s_{\mp}}$, we have a term in the decomposition \eqref{eq:WehrlEntropyDecomposition} accounting for the correlations between $r_{\pm}$ and $s_{\mp}$ in \eqref{eq:WehrlEntropyDecomposition}, which is the mutual information
\begin{equation}
    I (F_{\pm} : G_{\mp}) = \int \frac{\mathrm{d}r_{\pm} \, \mathrm{d}s_{\mp}}{2 \pi} \, Q_{\pm} \, \ln \frac{Q_{\pm}}{F_{\pm} \, G_{\mp}}.
    \label{eq:WehrlMutualInformation}
\end{equation}
Note that the mutual information is a non-negative measure for the total correlations which is zero if and only if $r_{\pm}$ and $s_{\mp}$ are uncorrelated. In contrast, a vanishing covariance $\sigma_{r_{\pm} s_{\mp}}$ does not allow to exclude the presence of correlations. 

\begin{figure*}[t!]	
    \centering
    \includegraphics[width=0.99\textwidth]{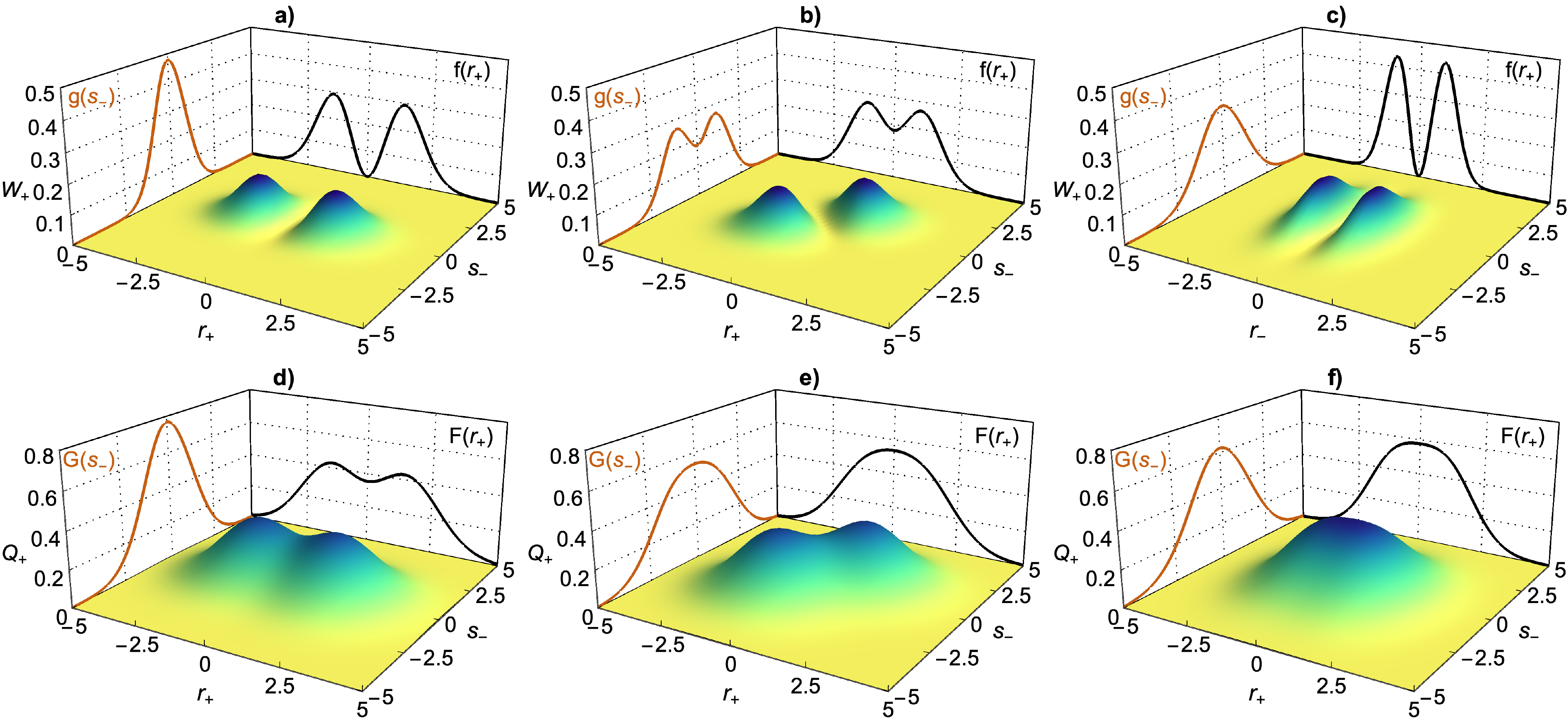}
    \caption{The upper and lower rows show the Wigner $W$- and Husimi $Q$-distributions \eqref{eq:WignerWAndHusimiQExampleState} in the $(r_+,s_-)$ variables, respectively, together with their marginal distributions for $a_1=b_1=a_2=b_2=1, \sigma_+ = 1, \sigma_- = 1.2$ and various choices for the angle $\phi$ and the squeezing parameter $\xi$. \textbf{a}) and \textbf{d)} represent the case $\phi = 0, \xi = 1$. The rotation angle is set to $\phi = \pi/4$ in \textbf{b)} and \textbf{e)} leading to strong correlations between $r_+$ and $s_-$, while in \textbf{c)} and \textbf{f}) the squeezing parameter is changed to $\xi = 3/2$.}
    \label{fig:PhaseSpaceDistributions}
\end{figure*}

In order to relate the differential entropies of $f_{\pm}$ and $F_{\pm}$ we employ the entropy power inequality
\begin{equation}
    e^{2 S(F_{\pm}/\sqrt{2 \pi})} \ge e^{2 S(f_{\pm})} + e^{2 S(\bar{f}_{\pm})},
\end{equation}
use that the vacuum entropy sum is bounded from below \cite{Bialynicki-Birula1975,Coles2017,Hertz2019}
\begin{equation}
    \begin{split}
        &S (\bar{f}_{\pm}) + S (\bar{g}_{\mp}) \\
        &= 1 + \ln \pi + \frac{1}{2} \ln \left[ (a_1^2 + a_2^2) (b_1^2 + b_2^2) \right] \\
        &\ge 1 + \ln \pi + \frac{\ln \det \bar{V}_{\pm}}{2},
    \end{split}
\end{equation}
and utilize midpoint concavity of the logarithm, i.e. that $\ln ((x+y)/2) \ge (\ln x + \ln y)/2$, to arrive at
\begin{equation}
    \mathcal{W}_1 \ge \frac{1}{2} \, \mathcal{W}_{\text{WTSTD}} - I (F_{\pm} : G_{\mp}),
    \label{eq:EntropicCriteriaDecomposition}
\end{equation}
where we introduced the witness functional corresponding to the entropic criteria in \cite{Walborn2009} as
\begin{equation}
    \mathcal{W}_{\text{WTSTD}} = S(f_{\pm}) + S(g_{\mp}) - 1 - \ln \pi - \frac{\ln \det \bar{V}'_{\pm}}{2}.
    \label{eq:WalbornCriteria}
\end{equation}
The inequality \eqref{eq:EntropicCriteriaDecomposition}  can be considered the entropic analog of \eqref{eq:SecondMomentCriteriaDecomposition}. Hence, a necessary (but not sufficient) condition for an entangled state, which is undetected by the WTSTD criteria \eqref{eq:WalbornCriteria}, to be detected by our entropic criteria \eqref{eq:WehrlCriteria} is that sufficiently strong correlations are present, i.e. that $I (F_{\pm} : G_{\mp}) > \mathcal{W}_{\text{WTSTD}} \ge 0$ (see \autoref{subsec:ComparisonExampleState}, especially \autoref{fig:MarginalVsPhaseSpaceCorrelations} b)).

Unfortunately, a similar relation can not be derived for the Rényi-Wehrl criteria \eqref{eq:RenyiWehrlCriteria} and the Rényi entropic criteria by STW, which are given by the witness functional
\begin{equation}
    \begin{split}
        \mathcal{W}_{\text{STW}} &= S_{\alpha}(f_{\pm}) + S_{\beta}(g_{\mp}) + \frac{1}{2 (1 - \alpha)} \ln \frac{\alpha}{\pi} \\
        &+ \frac{1}{2 (1 - \beta)} \ln \frac{\beta}{\pi} - \frac{\ln \det \bar{V}'_{\pm}}{2},
    \end{split}
    \label{eq:SaboiaCriteria}
\end{equation}
with the condition
\begin{equation}
    \frac{1}{\alpha} + \frac{1}{\beta} = 2,
    \label{eq:BabenkoBecknerCondition}
\end{equation}
which can be traced back to the Babenko-Beckner inequality \cite{Beckner1975} appearing in the proof of the corresponding entropic uncertainty relation \cite{Bialynicki-Birula1975,Bialynicki-Birula2011}. A decomposition like \eqref{eq:WehrlEntropyDecomposition} in terms of Rényi-type entropies would require a possibly negative third term, which does not coincide with the Rényi generalization of the mutual information \eqref{eq:WehrlMutualInformation} (an inherently non-negative quantity by definition) in general.

However, there is a crucial difference between the Rényi-type criteria \eqref{eq:RenyiWehrlCriteria} and \eqref{eq:SaboiaCriteria}. While our criteria \eqref{eq:RenyiWehrlCriteria} can be optimized over \textit{all} entropic orders $\beta$, the marginal based criteria \eqref{eq:SaboiaCriteria} are constrained by the condition \eqref{eq:BabenkoBecknerCondition}. More precisely, one marginal distribution is always exponentiated by a real number smaller or equal to one, while the other is always exponentiated by a real number larger or equal to one. As we will see in the following, this is a severe drawback as some entangled states require an optimization where the full phase space distribution is exponentiated with a small or a large number.

\begin{figure*}[t!]	
    \centering
    \includegraphics[width=0.99\textwidth]{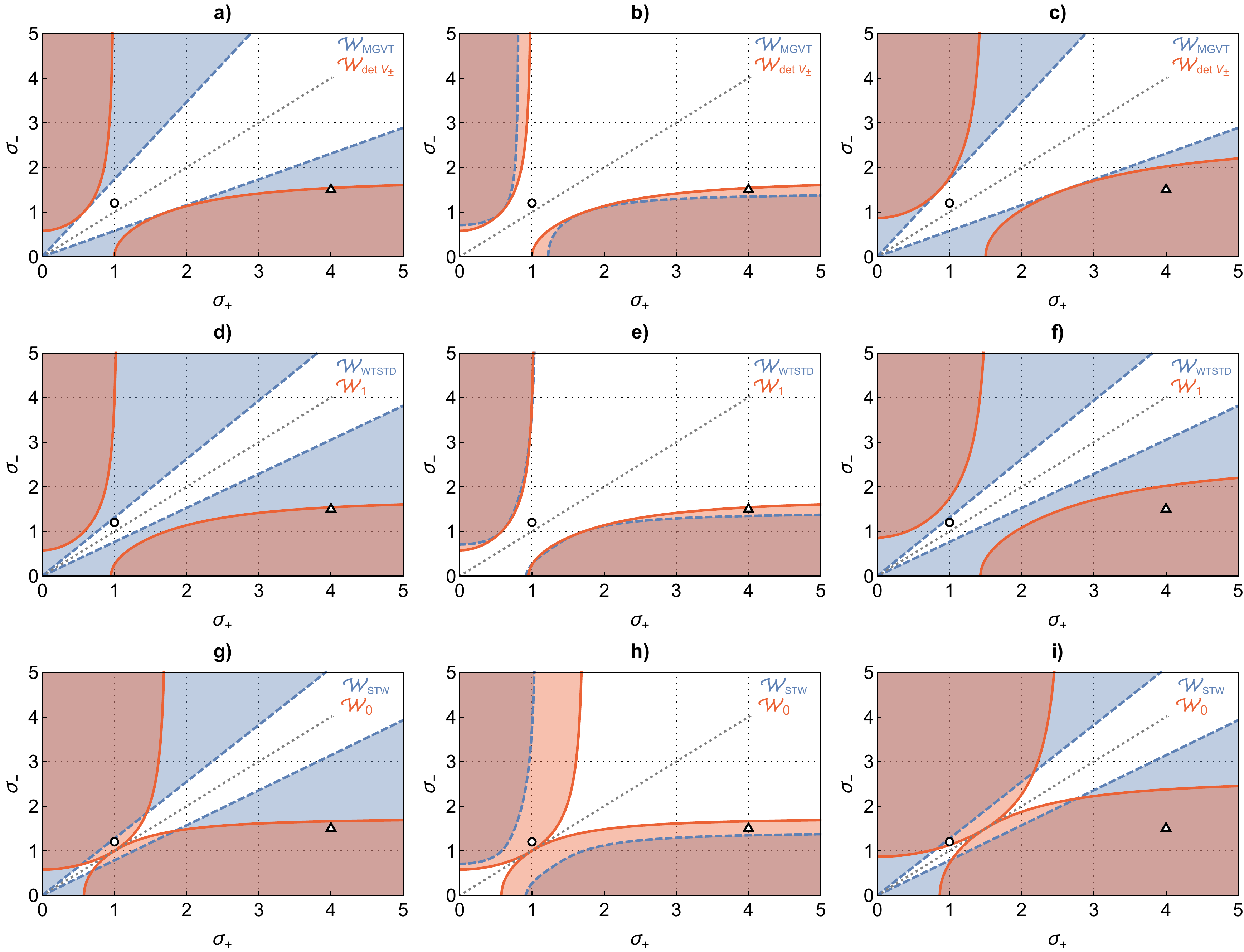}
    \caption{Witnessed regions of various marginal (dashed blue curves) and phase space (solid red curves) criteria for the state \eqref{eq:WignerWAndHusimiQExampleState}. The upper / middle / lower rows show second moment / entropic / optimized Rényi entropic criteria for the choices $\phi = 0, \xi = 1$ (first column), $\phi = \pi/4, \xi = 1$ (second column) and $\phi = 0, \xi = 3/2$ (third column). For $\beta \to 0$ our Rényi-Wehrl criteria \eqref{eq:RenyiWehrlCriteria} certify entanglement for all $\sigma_+ \neq \sigma_-$ after optimizing over $\xi$ as indicated from \textbf{g)} and \textbf{i)}, and hence outperform the STW criteria. In particular, they witness the point $\sigma_+ = 1, \sigma_- = 1.2$ (circle) corresponding to the distributions shown in \autoref{fig:PhaseSpaceDistributions}. For highly correlated variables, i.e. for $\phi = \pi/4$ (middle column), the marginal criteria underperform, see e.g. $\sigma_+ = 4, \sigma_- = 1.5$ (triangle) and also \autoref{fig:MarginalVsPhaseSpaceCorrelations}. A combined figure composed out of the three lower plots is provided in \cite{PRL}.}
    \label{fig:MarginalVsPhaseSpaceCriteria}
\end{figure*}

\subsection{Example state}
\label{subsec:ComparisonExampleState}
Following \cite{Gomes2009a,Gomes2009b,Nogueira2004,Agarwal2005,Saboia2011,Walborn2009}, we consider the family of pure states corresponding to the global wavefunction in position space
\begin{equation}
    \psi (r_1, r_2) = \frac{r_1 + r_2}{\sqrt {\pi \sigma_{-} \sigma^3_{+}}} \, e^{- \frac{1}{4} \left[ \left(\frac{r_1 + r_2}{\sigma_+} \right)^2 + \left(\frac{r_1 - r_2}{\sigma_-}\right)^2 \right]},
    \label{eq:ExampleStateWaveFunction}
\end{equation}
which is entangled for all positive $\sigma_+, \sigma_-$. We generalize this setting by explicitly implementing the effects of squeezings $\Xi$ and rotations in the $(r_{\pm}, s_{\mp})$ variables around an angle $\phi \in [0, 2 \pi)$ (realized by tuning $\vartheta_1$ and $\vartheta_2$ as discussed in \autoref{subsec:RotationsAndScalings}) which leads to the non-local Wigner $W$- and Husimi $Q$-distributions given explicitly for $a_1=b_1=a_2=b_2=1$ in \hyperref[app:Distributions]{Appendix A}. We show both distributions for fixed scalings $a_1=b_1=a_2=b_2=1$, fixed $\sigma_+ = 1, \sigma_- = 1.2$ and the three characteristic choices $(\xi = 1, \phi = 0), (\xi = 1, \phi = \pi/4), (\xi = 3/2, \phi = 0)$ (from left to right) together with their marginal distributions in \autoref{fig:PhaseSpaceDistributions}.

\begin{figure*}[t!]	
    \centering
    \includegraphics[width=0.75\textwidth]{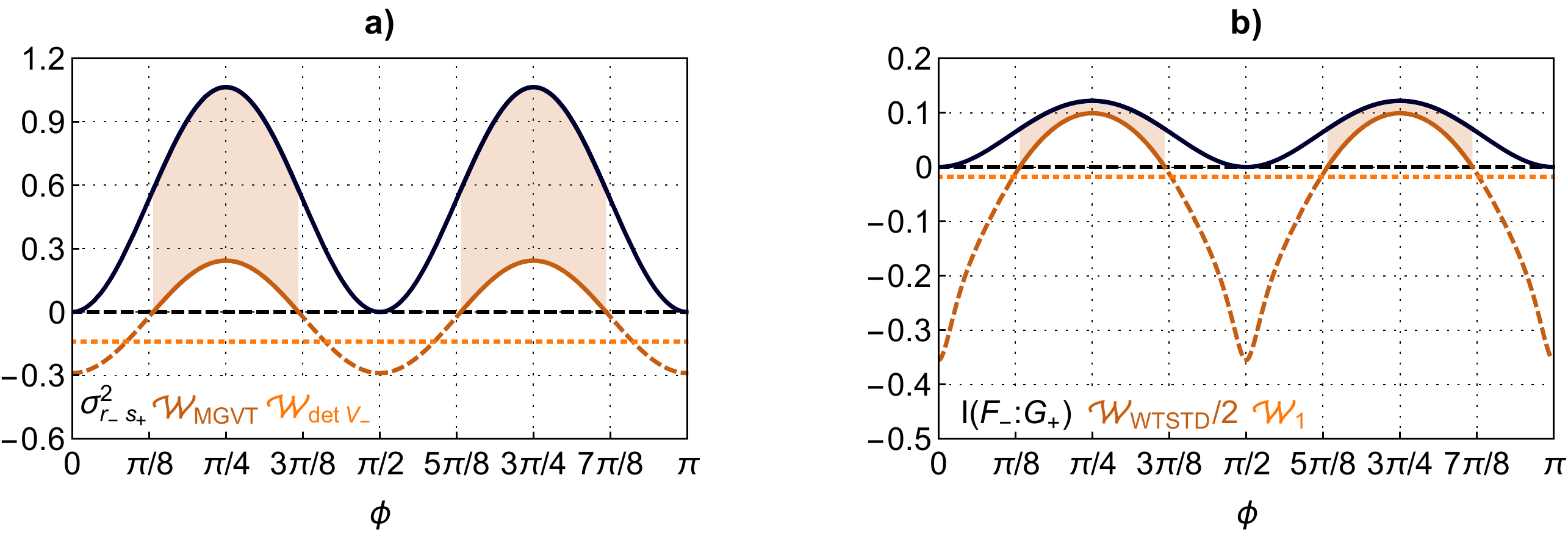}
    \caption{Illustration of all quantities appearing in the second moment / entropic relations \eqref{eq:SecondMomentCriteriaDecomposition} / \eqref{eq:EntropicCriteriaDecomposition} in \textbf{a} / \textbf{b)} for the state \eqref{eq:WignerWAndHusimiQExampleState} with $\sigma_+ = 4, \sigma_- = 1.5$ (triangle in \autoref{fig:MarginalVsPhaseSpaceCriteria}) and $\xi=1$ as functions of $\phi$. For sufficiently correlated $r_{\pm}$ and $s_{\mp}$, i.e. for $\pi/8 \lesssim \phi \lesssim 3 \pi/8$ and $5\pi/8 \lesssim \phi \lesssim \pi$, the marginal criteria (dark orange dashed to solid curves) fail, while the phase space criteria (light orange dotted curves) still witness entanglement. More precisely, the correlation measures (black solid curves) exceed the values of the marginal witnesses in these regimes (orange shaded regions).}
    \label{fig:MarginalVsPhaseSpaceCorrelations}
\end{figure*}

We analyze the performance of all types of criteria considered in \autoref{sec:Comparison} in detail. We evaluate the phase space (solid red curves) and marginal criteria (blue dashed curves) for the triplet of values for $\phi, \xi$ as shown in \autoref{fig:PhaseSpaceDistributions} and compare their performance in terms of $\sigma_+, \sigma_-$. 

We start with our second moment criteria \eqref{eq:SecondMomentCriteria} and the MGVT criteria \eqref{eq:MGVTCriteria} (first row in \autoref{fig:MarginalVsPhaseSpaceCriteria}). For $\phi = 0, \xi = 1$, shown in \textbf{a)}, the MGVT criteria outperform our criteria, while the converse is true for $\phi = \pi/4, \xi = 1$ as depicted in \textbf{b)} as a result of strong correlations being present. When optimizing over $\phi$ and $\xi$, the two become equivalent (see \autoref{subsec:ComparisonSecondMomentCriteria}) as indicated in \textbf{c)} where $\phi = 0, \xi = 3/2$. In all cases, the distributions shown in \autoref{fig:PhaseSpaceDistributions} corresponding to $\sigma_+ = 1, \sigma_- = 1.2$ are not witnessed (circle), while the point $\sigma_+ = 4, \sigma_- = 1.5$ (triangle) is an example for a state which is witnessed by our criteria but not witnessed by the MGVT criteria when the angle has been chosen unfavorably. The angle dependence of our criteria (light orange dotted curve) and the MGVT criteria (dark orange dashed to solid curve) for this case as well as the covariance (black solid curve) is shown in \autoref{fig:MarginalVsPhaseSpaceCorrelations} \textbf{a)}. Following the intuition provided by \eqref{eq:SecondMomentCriteriaDecomposition}, our criteria can outperform the MGVT criteria whenever $\sigma^2_{r_{\pm} s_{\mp}} > \mathcal{W}_{\text{MGVT}}$, which is fulfilled for $\pi/8 \lesssim \phi \lesssim 3 \pi/8$ and $5\pi/8 \lesssim \phi \lesssim \pi$ (orange shaded regions).

The same analysis is carried out for the Wehrl entropic criteria \eqref{eq:WehrlCriteria} and the WTSTD criteria \eqref{eq:WalbornCriteria} in the middle row of \autoref{fig:MarginalVsPhaseSpaceCriteria}. We observe similar effects as for the second moment criteria. The main difference is that no straightforward optimization leads to an equivalence between them. The point $\sigma_+ = 1, \sigma_- = 1.2$ is not witnessed by any of the two and the point $\sigma_+ = 4, \sigma_- = 1.5$, which is always witnessed by the Wehrl entropic criteria, is only witnessed by the marginal criteria provided that the correlations between $r_{\pm}$ and $s_{\mp}$ remain small. We also plot the quantities appearing in the inequality \eqref{eq:EntropicCriteriaDecomposition} in \autoref{fig:MarginalVsPhaseSpaceCorrelations} \textbf{b)} and find the same outperformance regions for $\phi$ as for the second moment criteria.

At last, we compare the Rényi-Wehrl criteria \eqref{eq:RenyiWehrlCriteria} to the STW criteria \eqref{eq:SaboiaCriteria} and optimize over the entropic order for every choice of $\phi, \xi$. For $\phi = 0$, the STW criteria become optimal for $\alpha \to 1/2$ and $\beta \to \infty$ in the variables $(r_+, s_-)$ and conversely for $(r_-, s_+)$, i.e.
\begin{equation}
    \mathcal{W}_{\text{STW}} = S_{1/2} (f_{+}) + S_{\infty} (g_{-}) - \ln 4 \pi, 
\end{equation}
where $S_{\infty} (g_{-}) = - \ln g^{\text{max}}_{-}$ denotes the min-entropy of $g_-$ with maximum value $g_-^{\text{max}}$ and 
\begin{equation}
    \mathcal{W}_{\text{STW}} = S_{\infty} (f_-) + S_{1/2} (g_+) - \ln 4 \pi,
\end{equation}
respectively \cite{Saboia2011}. For $\phi = \pi/4$, they instead become optimal for $\alpha, \beta \to 1$ and reduce to the WTSTD criteria \eqref{eq:WalbornCriteria}. In contrast, the Rényi-Wehrl criteria \eqref{eq:RenyiWehrlCriteria} become optimal in the limit $\beta \to 0$ independent of $\phi, \xi$.

The results are shown in the lower row of \autoref{fig:MarginalVsPhaseSpaceCriteria}. For $\phi = 0, \xi = 1$, our criteria outperform the STW criteria around $\sigma_+ \approx \sigma_-$, see \textbf{g)}. When optimizing over $\xi$, our criteria witness entanglement for all $\sigma_+, \sigma_- \neq \sigma_+$ (gray dashed line) as indicated in \textbf{i)} and hence outperform the STW criteria completely. In particular, the point $\sigma_+ = 1, \sigma_- = 1.2$ corresponding to the distributions in \autoref{fig:PhaseSpaceDistributions} can only be witnessed with the phase space approach after optimization. For $\phi = \pi/4$ the outperformance occurs even without optimizing over $\xi$, see \textbf{h)}.

\section{Coarse-grained measurements}
\label{sec:CoarseGrainedMeasurements}
We now discuss the influence of finite resolution on the entanglement criteria \eqref{eq:WitnessDefinition}. In particular, we analyze the possibilities offered by an optimization over the concave function $f$.

\subsection{Discretization schemes}
\label{subsec:DiscretizationSchemes}
We aim for a description valid for measurements with constant \textit{and} adapative resolution. The latter is relevant when an experimental procedure produces samples of the Husimi $Q$-distribution, which are binned in a post-measurement process ensuring that the underlying distribution is approximated well. A prime example of such schemes is the quadtree method, which has been employed successfully to witness entanglement of a non-Gaussian state using entropic methods in \cite{Schneeloch2019}.

\begin{figure*}[t!]	
    \centering
    \includegraphics[width=0.85\textwidth]{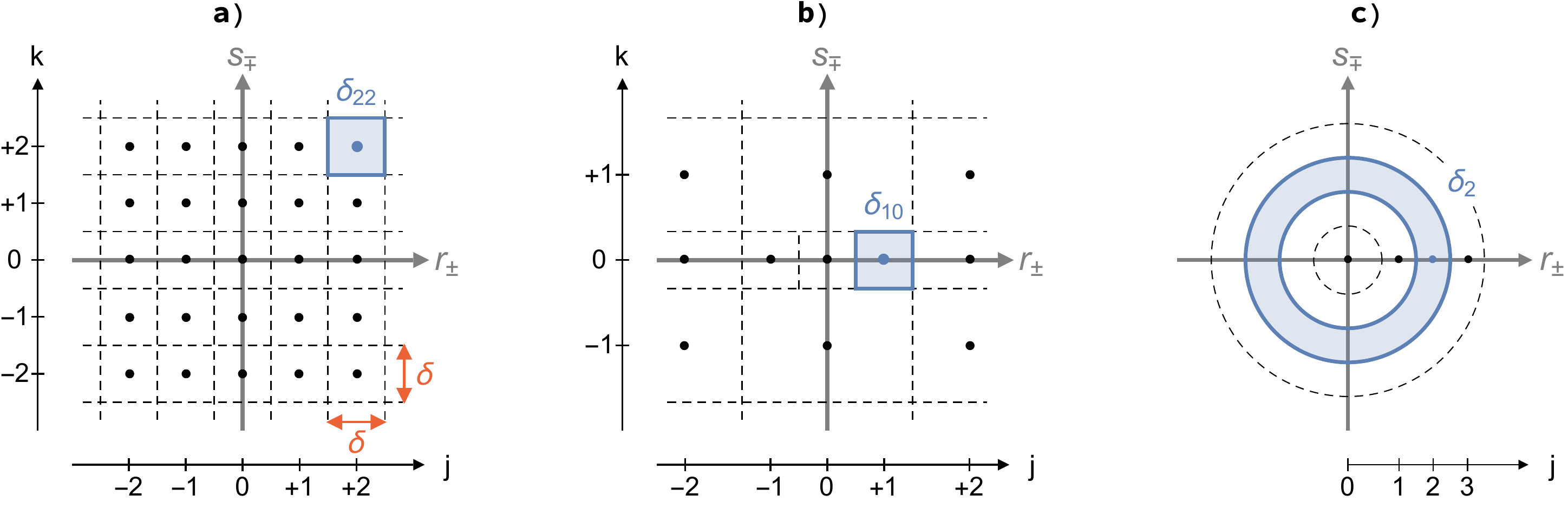}
    \caption{Three typical discretization schemes: \textbf{a)} shows regular quadratic tiles with spacing $\delta$ along both directions, \textbf{b)} an adaptive scheme with smaller tilings closer to the origin and \textbf{c)} a circular tiling around the origin.}
    \label{fig:DiscretizationSchemes}
\end{figure*}

To achieve this generality, we discretize phase space into compact $\delta_{j k}$, where the indices $j,k$ run over integers, but depending on the discretization scheme it may be convenient to draw them from subsets of integers, i.e. $\{j, k \} \in \mathcal{Z} \subseteq \mathbb{Z} \times \mathbb{Z}$. To every tile $\delta_{j k}$ we associate its phase space measure, which we denote by $\Delta_{j k}$, and discrete coordinates $(r^j_{\pm}, s^k_{\mp})$, which are located at the center of $\delta_{j k}$ if it corresponds to a simply connected region.

For regular tilings consisting of rectangles with $\Delta \equiv \Delta_{j k}$ there is a simple relation between the continuous and the discrete sets of coordinates, which reads
\begin{equation}
    r_{\pm} \to r^j_{\pm} = j \, \delta r_{\pm}, \quad s_{\mp} \to s^k_{\mp} = k \, \delta s_{\mp},
    \label{eq:ContinuousToDiscreteCoordinates}
\end{equation}
including the possibility that $\delta r_{\pm} \neq \delta s_{\mp}$. Then, the tile $\delta_{j k}$ is centered at $(r^j_{\pm},s^k_{\mp})$ and hence its domain is given by $\delta_{jk} = [\delta r_{\pm} (j - 1/2), \delta r_{\pm} (j + 1/2)] \times [\delta s_{\mp} (k - 1/2), \delta s_{\mp} (k + 1/2)]$ with constant phase space measure $\Delta = \delta r_{\pm} \, \delta s_{\mp}/(2 \pi)$. In general, such simple relations may only be given recursively or implicitly and hence we work with the discrete indices $j,k$ in the following.

We illustrate three archetypal discretization schemes in \autoref{fig:DiscretizationSchemes}. A regular quadratic tiling with $\delta \equiv \delta r_{\pm} = \delta s_{\mp}$ and $\Delta = \delta^2/(2 \pi)$, most relevant in the context of finite resolution measurements in quantum optics, is shown in \autoref{fig:DiscretizationSchemes} \textbf{a)}. An adaptive scheme with rectangular tiles is sketched in \autoref{fig:DiscretizationSchemes} \textbf{b}), in which case relations like \eqref{eq:ContinuousToDiscreteCoordinates} can only be given recursively. However, such schemes allow to reconstruct the underlying distribution in greater detail in regions where it shows characteristic features, which can be advantageous especially for non-Gaussian distributions. Another adpative scheme, which is based on radially symmetric tilings, is shown in \autoref{fig:DiscretizationSchemes} \textbf{c)}. This method may be employed, for example, when binning sampled data while assuming the underlying distribution to be radially symmetric, such that the discrete coordinates can be labeled with only one non-negative integer-valued index, e.g. $j \in \mathbb{Z}^+_0$.

\subsection{Discretized distributions}
\label{subsec:DiscretizedDistributions}
The discretization procedure leads to a discrete distribution over the coordinate points $(r^j_{\pm},s^k_{\mp})$ by integrating the continuous distribution $Q_{\pm}$ over the corresponding tile $\delta_{j k}$, i.e.
\begin{equation}
    Q^{j k}_{\pm} \equiv Q_{\pm} (r^j_{\pm}, s^k_{\mp}) = \int_{\delta_{jk}} \frac{\mathrm{d} r_{\pm} \, \mathrm{d} s_{\mp}}{2 \pi} \, Q_{\pm}(r_{\pm}, s_{\mp}).
    \label{eq:DiscretizedHusimiQHistogramDefinition}
\end{equation}
This distribution is normalized to unity according to 
\begin{equation}
    \sum_{j,k \in \mathcal{Z}} Q^{j k}_{\pm} = 1.
\end{equation}
Uncertainty relations and entanglement criteria for coarse-grained measurements have been studied extensively in the literature, see e.g. \cite{Bialynicki-Birula2011,Rudnicki2012,Tasca2013,Coles2017}. The most important conclusion from these studies is that both should \textit{not} be formulated in terms of measures of localization with respect to the discrete distribution $Q^{j k}_{\pm}$ as they often underestimate their continuous analogs. For example, consider an experimental procedure with sufficiently low resolution, such that all measurement outcomes lie within a single tile. Then, variances as well as entropies of the resulting discrete distribution evaluate to zero although their continuous analogs are strictly positive.

\begin{figure*}[t!]	
    \centering
    \includegraphics[width=0.99\textwidth]{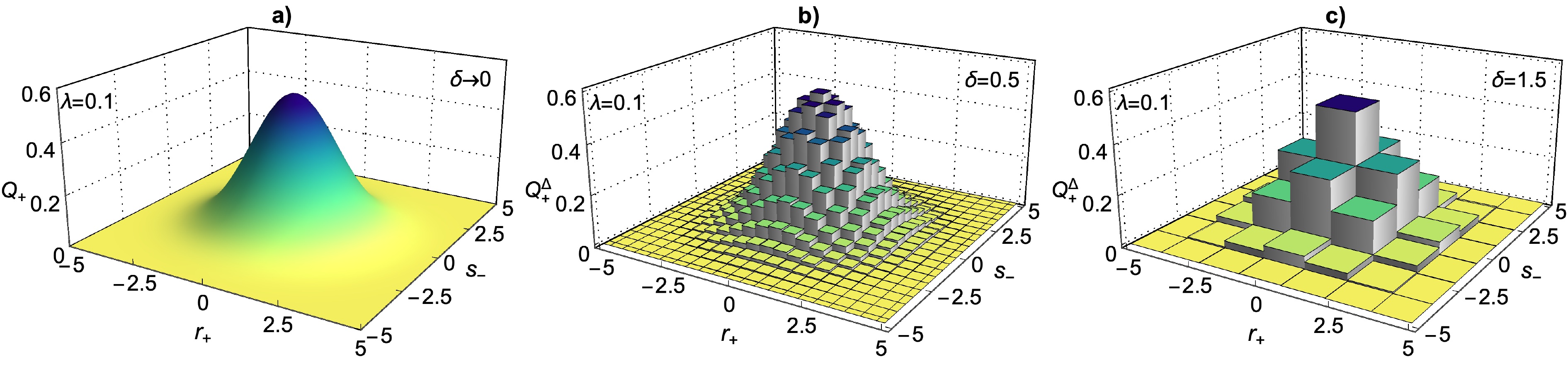}
    \caption{Coarse-graining of the TMSV state \eqref{eq:TwoModeSqueezedVacuumState} with $\lambda=0.1$ in the $(r_+,s_-)$ variables following the regular quadratic discretization scheme shown in \autoref{fig:DiscretizationSchemes} \textbf{a}). Starting from the continuum limit $\delta \to 0$ in \textbf{a)}, the grid spacing $\delta$ increases from \textbf{a)} to \textbf{c)}, resulting in coarser distributions $Q^{\Delta}_{+}$.}
    \label{fig:DiscretizedApproximtation}
\end{figure*}

Therefore, we work with the density of the discrete distribution \eqref{eq:DiscretizedHusimiQHistogramDefinition} over every tile instead. It is defined via
\begin{equation}
    Q^{\Delta}_{\pm} \equiv Q^{\Delta}_{\pm} (r_{\pm}, s_{\mp}) = \sum_{j,k \in \mathcal{Z}}
    \begin{cases}
        \frac{Q^{j k}_{\pm}}{\Delta_{j k}}, &(r_{\pm}, s_{\mp}) \in \delta_{j k}, \\
        0 & \text{else},
    \end{cases}
    \label{eq:DiscretizedHusimiQDefinition}
\end{equation}
and serves as an approximation to the true continuous Husimi $Q$-distribution $Q_{\pm}$ albeit being necessarily discontinuous itself. Thus, it is normalized with respect to the standard phase space measure $\mathrm{d} r_{\pm} \, \mathrm{d} s_{\mp} /(2 \pi)$ and converges to the continuous Husimi $Q$-distribution in the continuum limit, i.e. $Q^{\Delta}_{\pm} (r_{\pm}, s_{\mp}) \to Q_{\pm} (r_{\pm}, s_{\mp})$ for $\Delta_{j k} \to 0$. 

As an example, we consider the effect of coarse-graining for the Husimi $Q$-distribution $Q_{\pm}$ of the TMSV state \eqref{eq:TwoModeSqueezedVacuumState}, which we re-state here for the reader's convenience
\begin{equation}
    Q_{\pm} (r_{\pm}, s_{\mp}) = \frac{1}{Z} \, e^{-\frac{1}{2} (r_{\pm}, s_{\mp})^T V_{\pm}^{-1} (r_{\pm}, s_{\mp})},
\end{equation}
with normalization $Z = \det^{1/2} V_{\pm}$ and covariance matrix $V_{\pm} = \frac{2}{1 \pm \lambda} \mathds{1}$. We show the case $a_1 = b_1 = a_2 = b_2 = 1, \vartheta_1 = \vartheta_2 = 0$ and $\lambda = 0.1$ in \autoref{fig:DiscretizedApproximtation} for a regular quadratic tiling with various spacing $\delta$ along both axes as shown in \autoref{fig:DiscretizationSchemes} \textbf{a}).

\subsection{Discretized criteria}
\label{subsec:DiscretizatizedCriteria}

\subsubsection{General criteria}
In order to derive entanglement criteria which hold for the discretized approximation $Q^{\Delta}_{\pm}$, we apply Jensen's inequality to the integrals appearing in the witness functional \eqref{eq:WitnessDefinition} for every tile \textit{individually}. Adjusted to this setup, i.e. the continuous Husimi $Q$-distribution $Q_{\pm}$ restricted to $\delta_{j k}$ with measure $\Delta_{j k}$ and concave $f$, Jensen's inequality reads
\begin{equation}
    \frac{1}{\Delta_{j k}} \, \int_{\delta_{j k}} \!\!\frac{\mathrm{d}r_{\pm} \, \mathrm{d}s_{\mp}}{2 \pi} \, f (Q_{\pm}) 
    \le f \left( \frac{1}{\Delta_{j k}} \, \int_{\delta_{j k}} \!\!\frac{\mathrm{d}r_{\pm} \, \mathrm{d}s_{\mp}}{2 \pi} \, Q_{\pm} \right).
    \label{eq:JensensInequality}
\end{equation}
Now, discretizing the first term in our witness functional $\mathcal{W}_f$ \eqref{eq:WitnessDefinition} by expanding the phase space integral over all tiles, inserting a multiplicative one and using Jensen's inequality \eqref{eq:JensensInequality} together with the definition of the discrete distribution \eqref{eq:DiscretizedHusimiQHistogramDefinition} leads to
\begin{equation}
    \begin{split}
        \int \frac{\mathrm{d} r_{\pm} \, \mathrm{d} s_{\mp}}{2 \pi} \, f (Q_{\pm}) 
        &= \sum_{j,k \in \mathcal{Z}} \Delta_{j k} \frac{1}{\Delta_{j k}} \int_{\delta_{j k}} \!\!\frac{\mathrm{d} r_{\pm} \, \mathrm{d} s_{\mp}}{2 \pi} \, f (Q_{\pm}) \\
        &\le \sum_{j,k \in \mathcal{Z}} \Delta_{j k} f \left( \frac{1}{\Delta_{j k}} \, Q^{j k}_{\pm} \right).
    \end{split}
    \label{eq:DiscretizedWitnessUpperBound1}
\end{equation}
Now, the right hand side of the latter inequality can be rewritten in terms of the continuous approximation
\begin{equation}
    \sum_{j,k \in \mathcal{Z}} \Delta_{j k} \, f \left( \frac{1}{\Delta_{j k}} \, Q_{\pm}^{j k} \right) = \int \frac{\mathrm{d} r_{\pm} \, \mathrm{d} s_{\mp}}{2 \pi} \, f (Q^{\Delta}_{\pm}),
\end{equation}
which is also a useful relation for computing the discretized witness functional in practice as one might prefer to work with $Q_{\pm}^{jk}$ over $Q^{\Delta}_{\pm}$ to simplify calculations.

Defining a discretized witness functional in terms of $Q^{\Delta}_{\pm}$ instead of $Q_{\pm}$
\begin{equation}
    \mathcal{W}^{\Delta}_f = \int \frac{\mathrm{d} r_{\pm} \, \mathrm{d} s_{\mp}}{2 \pi} \left[ f (Q^{\Delta}_{\pm})  - f \left(\bar{Q}'_{\pm} \right) \right]
    \label{eq:DiscretizedWitnessDefinition}
\end{equation}
allows to conclude $\mathcal{W}_f \le \mathcal{W}^{\Delta}_f$ and therefore our criteria \eqref{eq:WitnessDefinition} imply that all separable states fulfill the discretized criteria  
\begin{equation}
    \boldsymbol{\rho}_{12} \text{ separable} \Rightarrow \mathcal{W}^{\Delta}_f \ge 0,
    \label{eq:DiscretizedSeparabilityCriteria}
\end{equation}
which are weaker in general. We emphasize that the latter inequality is fulfilled for \textit{arbitrary} discretization schemes. Since the discretized witness functional $\mathcal{W}^{\Delta}_f$ is of the same form as its continuous counterpart $\mathcal{W}_f$, we can follow the same arguments as above to derive interesting classes of discrete entanglement criteria. 

\subsubsection{Entropic criteria}
Similar to \eqref{eq:RenyiWehrlEntropyDefinition}, we obtain criteria for Rényi-Wehrl entropies of the discretized approximations
\begin{equation}
    S_{\beta} (Q^{\Delta}_{\pm}) = \frac{1}{1 - \beta} \ln \left[ \int \frac{\mathrm{d} r_{\pm} \, \mathrm{d} s_{\mp}}{2 \pi} \, \left( Q^{\Delta}_{\pm} \right)^{\beta} (r_{\pm},s_{\mp}) \right]
    \label{eq:RenyiWehrlEntropyDiscretized}
\end{equation}
with entropic orders $\beta \in (0,1) \cup (1, \infty)$, by choosing monomials $f(t) = t^{\beta}$ and applying a monotonic function in \eqref{eq:DiscretizedWitnessDefinition}, which yields the discetized witness functional
\begin{equation}
    \mathcal{W}^{\Delta}_{\beta} = S_{\beta} (Q^{\Delta}_{\pm}) - \frac{\ln \beta}{\beta - 1} - \frac{\ln \det \bar{V}'_{\pm}}{2},
    \label{eq:RenyiWehrlCriteriaDicretized}
\end{equation}
analogous to \eqref{eq:RenyiWehrlCriteria}. In general, we can write the Rényi-Wehrl entropies \eqref{eq:RenyiWehrlEntropyDiscretized} in terms of the discretized distribution $Q^{j k}_{\pm}$ as
\begin{equation}
    S_{\beta} (Q^{\Delta}_{\pm}) = \frac{1}{1 - \beta} \, \ln \left[ \sum_{j,k \in \mathcal{Z}} \Delta_{j k}^{1 - \beta} \left( Q^{j k}_{\pm} \right)^{\beta} \right].
\end{equation}
The latter can only be simplified further for regular tilings with $\Delta \equiv \Delta_{j k}$, leading to
\begin{equation}
    S_{\beta} (Q^{\Delta}_{\pm}) = S_{\beta} (Q^{j k}_{\pm}) + \ln \Delta,
\end{equation}
with discrete Rényi-Wehrl entropies
\begin{equation}
    S_{\beta} (Q^{j k}_{\pm}) = \frac{1}{1 - \beta} \, \ln \left[ \sum_{j,k \in \mathcal{Z}} \left( Q^{j k}_{\pm} \right)^\beta \right].
    \label{eq:DiscreteRenyiWehrlEntropyDefinition}
\end{equation}
In the limit $\beta \to 1$, the Rényi-Wehrl entroy \eqref{eq:RenyiWehrlEntropyDiscretized} converges to the Wehrl entropy
\begin{equation}
    \hspace{-0.1cm}S_1 (Q_{\pm}^{\Delta}) = - \int \frac{\mathrm{d}r_{\pm} \, \mathrm{d}s_{\mp}}{2 \pi} \, Q^{\Delta}_{\pm} (r_{\pm}, s_{\mp}) \, \ln Q^{\Delta}_{\pm} (r_{\pm}, s_{\mp}),
    \label{eq:WehrlEntropyDefinition}
\end{equation}
for which the discretized witness functional reads
\begin{equation}
    \mathcal{W}^{\Delta}_{1} = S_{1} (Q^{\Delta}_{\pm}) - 1 - \frac{\ln \det \bar{V}'_{\pm}}{2}.
    \label{eq:WehrlCriteriaDiscretized}
\end{equation}
Irrespective of the discretization scheme we find the relation
\begin{equation}
    S_1 (Q^{\Delta}_{\pm}) = S_1 (Q^{j k}_{\pm}) + \sum_{j, k \in \mathcal{Z}} \ln (\Delta_{j k}) \, Q^{j k}_{\pm}
    \label{eq:DiscreteWehrlEntropiesRelation}
\end{equation}
with the discrete Wehrl entropy being defined as
\begin{equation}
    S_{1} (Q^{j k}_{\pm}) =  - \sum_{j,k \in \mathcal{Z}} Q^{j k}_{\pm} \ln Q^{j k}_{\pm},
\end{equation}
which also follows from \eqref{eq:DiscreteRenyiWehrlEntropyDefinition} in the limit $\beta \to 1$. For regular tilings $\Delta \equiv \Delta_{j k}$ \eqref{eq:DiscreteWehrlEntropiesRelation} reduces to the simple relation
\begin{equation}
    S_{1} (Q^{\Delta}_{\pm}) = S_{1} (Q^{j k}_{\pm}) + \ln \Delta.
\end{equation}

\begin{figure*}[t!]	
    \centering
    \includegraphics[width=0.99\textwidth]{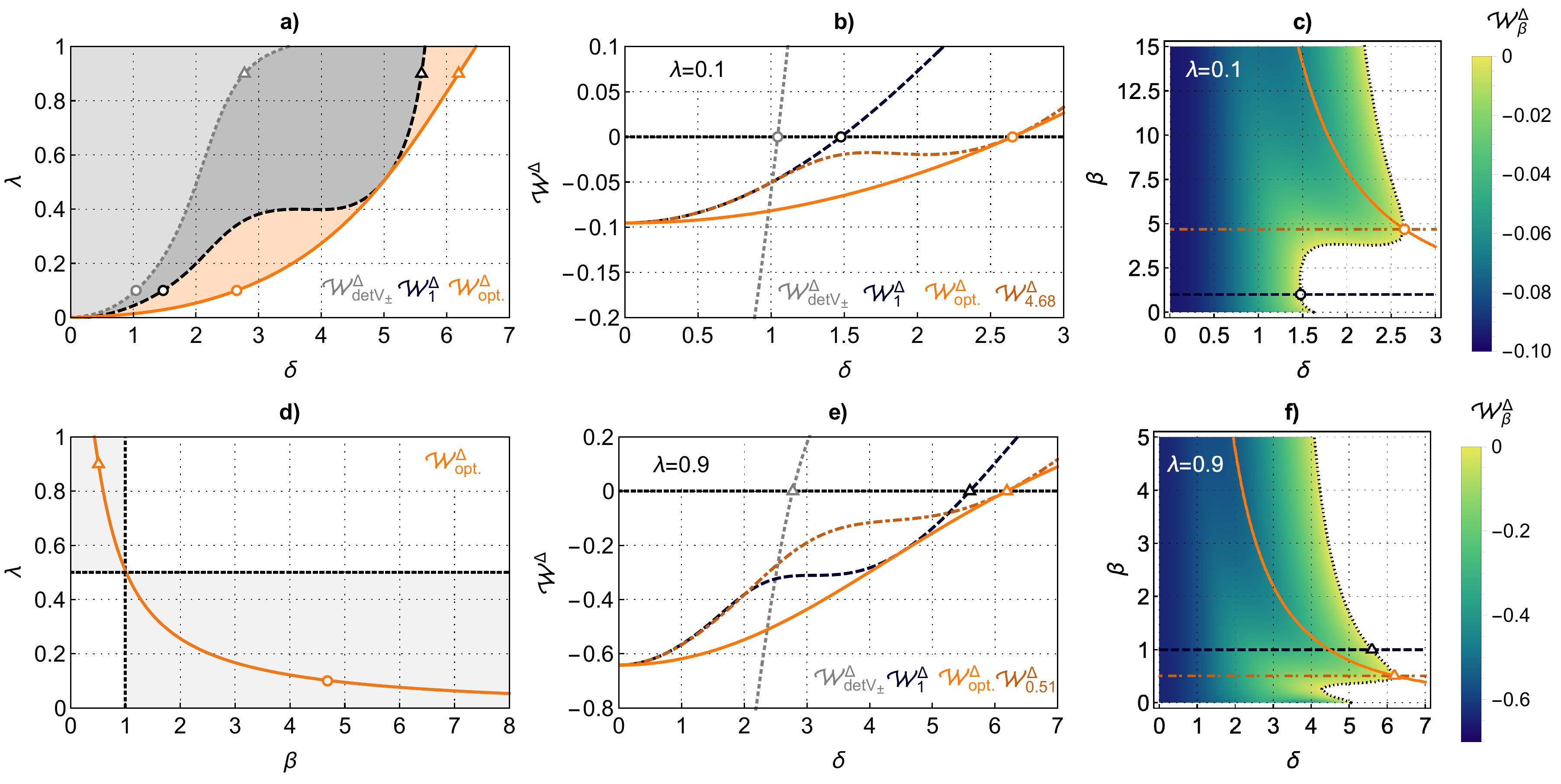}
    \caption{Comparison of discretized entanglement criteria for the TMSV state. As shown in \textbf{a)}, there is a strict hierarchy of second moment (gray dotted curves), Wehrl entropic (black dashed curves) and optimized Rényi-Wehrl entropic (light orange solid curves) criteria in the sense that for a fixed $\lambda$ optimization allows to work with a much coarser grid spacing $\delta$ before the witness breaks down (see also \cite{PRL}). The entanglement analysis of the Rényi-Wehrl criteria as a function of $\beta$ is carried out explicitly in \textbf{c)} and \textbf{f)} for $\lambda = 0.1$ and $\lambda = 0.9$, respectively, where we distinguish between the optimal $\beta$ overall (dark orange dot-dashed curves) and the optimal $\beta$ for every $\delta$ (light orange solid curves). The values of the four witnesses are shown in \textbf{b)} and \textbf{e)} as functions of the grid spacing $\delta$ for the same choices of $\lambda$. The optimal choice for $\beta$ depends on $\lambda$, as plotted in \textbf{d)}, showing that for small entanglement $\lambda \le 0.5$ one should take $\beta \ge 1$ and vice versa.}
    \label{fig:DiscretizedCriteriaTMSV}
\end{figure*}

\subsubsection{Second moment criteria}
Criteria for the second moments of the distribution $Q^{\Delta}_{\pm}$, defined via (we assume vanishing expectation values without loss of generality)
\begin{equation}
    \begin{split}
        \left( \Sigma^{\Delta}_{r_{\pm}} \right)^2 &= \int \frac{\mathrm{d}r_{\pm} \, \mathrm{d}s_{\mp}}{2 \pi} \, r^2_{\pm} \, Q^{\Delta}_{\pm} (r_{\pm}, s_{\mp}), \\
        \left( \Sigma^{\Delta}_{s_{\mp}} \right)^2 &= \int \frac{\mathrm{d}r_{\pm} \, \mathrm{d}s_{\mp}}{2 \pi} \, s^2_{\mp} \, Q^{\Delta}_{\pm} (r_{\pm}, s_{\mp}), \\
        \Sigma^{\Delta}_{r_{\pm} s_{\mp}} &= \int \frac{\mathrm{d}r_{\pm} \, \mathrm{d}s_{\mp}}{2 \pi} \, r_{\pm} \, s_{\mp} \, Q^{\Delta}_{\pm} (r_{\pm}, s_{\mp}),
    \end{split}
\end{equation}
can be formulated in terms of the discretized covariance matrix
\begin{equation}
    V^{\Delta}_{\pm} = \begin{pmatrix}
    \left( \Sigma^{\Delta}_{r_{\pm}} \right)^2 &  \Sigma^{\Delta}_{r_{\pm} s_{\mp}} \\
     \Sigma^{\Delta}_{r_{\pm} s_{\mp}} & \left( \Sigma^{\Delta}_{s_{\mp}} \right)^2
    \end{pmatrix},
\end{equation}
by using that $S_1 (Q_{\pm}^{\Delta}) \le 1 + \frac{1}{2} \ln \det V_{\pm}^{\Delta}$, which leads to the second moment witness functional
\begin{equation}
    \mathcal{W}^{\Delta}_{\det V_{\pm}} = \det V^{\Delta}_{\pm} - \det \bar{V}'_{\pm}.
\end{equation}
One can easily check that the means of the distributions $Q^{\Delta}_{\pm}$ and $Q^{j k}_{\pm}$ agree, i.e. 
\begin{equation}
    (\mu^{\Delta}_{r_{\pm}}, \mu^{\Delta}_{s_{\mp}}) = (\mu_{r^j_{\pm}},\mu_{s^k_{\mp}}), 
\end{equation}
with the continuous means given by
\begin{equation}
    \begin{split}
        \mu^{\Delta}_{r_{\pm}} &= \int \frac{\mathrm{d}r_{\pm} \, \mathrm{d}s_{\mp}}{2 \pi} \, r_{\pm} \, Q^{\Delta}_{\pm} (r_{\pm}, s_{\mp}), \\
        \mu^{\Delta}_{s_{\mp}} &= \int \frac{\mathrm{d}r_{\pm} \, \mathrm{d}s_{\mp}}{2 \pi} \, s_{\mp} \, Q^{\Delta}_{\pm} (r_{\pm}, s_{\mp}),
    \end{split}
\end{equation}
while the discrete means read
\begin{equation}
    \begin{split}
        \mu_{r^j_{\pm}} &= \sum_{j,k \in \mathcal{Z}} r^j_{\pm} \, Q^{j k}_{\pm}, \\
        \mu_{s^k_{\mp}} &= \sum_{j,k \in \mathcal{Z}} s^k_{\mp} \, Q^{j k}_{\pm}.
    \end{split}
\end{equation}
In contrast, the second moments of $Q^{\Delta}_{\pm}$ and $Q^{j k}_{\pm}$ differ. For regular tilings with $\Delta = \delta r_{\pm} \, \delta s_{\mp} / (2 \pi)$ we find (see also \cite{Rudnicki2012,Tasca2013})
\begin{equation}
    \begin{split}
        \left( \Sigma^{\Delta}_{r_{\pm}} \right)^2 &= \Sigma^2_{r^j_{\pm}} + \frac{\left( \delta r_{\pm} \right)^2}{12} \\
        \left( \Sigma^{\Delta}_{s_{\mp}} \right)^2 &= \Sigma^2_{s^k_{\mp}} + \frac{\left( \delta s_{\mp} \right)^2}{12}, \\
        \Sigma^{\Delta}_{r_{\pm} s_{\mp}} &= \Sigma_{r^j_{\pm} s^k_{\mp}},
    \end{split}
    \label{eq:SecondMomentsRelations1}
\end{equation}
with discrete second moments (we assume vanishing expectation values again)
\begin{equation}
    \begin{split}
        \Sigma^2_{r^j_{\pm}} &= \sum_{j,k \in \mathcal{Z}} \left( r^j_{\pm} \right)^2 \, Q^{j k}_{\pm} \\
        \Sigma^2_{s^k_{\mp}} &= \sum_{j,k \in \mathcal{Z}} \left( s^k_{\mp} \right)^2 \, Q^{j k}_{\pm} \\
        \Sigma_{r^j_{\pm} s^k_{\mp}} &= \sum_{j,k \in \mathcal{Z}} r^j_{\pm} \, s^k_{\mp} \, Q^{j k}_{\pm}
    \end{split}
\end{equation}
Eq. \eqref{eq:SecondMomentsRelations1} shows that the discrete second moments $\Sigma^2_{r^j_{\pm}}$ and $\Sigma^2_{s^k_{\mp}}$ indeed underestimate the continuous variances $(\Sigma^{\Delta}_{r_{\pm}})^2$ and $(\Sigma^{\Delta}_{s_{\mp}})^2$, an effect which is cured by taking the variances induced by the finite tile sizes into account.

\subsection{Example state}
To exemplify the advantages offered by an optimization over $f$ in our general discretized criteria \eqref{eq:DiscretizedSeparabilityCriteria} we consider again the Gaussian TMSV state \eqref{eq:TwoModeSqueezedVacuumState} for $a_1 = b_1 = a_2 = b_2 = 1$ and $\vartheta_1 = \vartheta_2 = 0$. The corresponding distribution $Q_{\pm}$ is discretized following a regular quadratic tiling with grid spacing $\delta$ shown in \autoref{fig:DiscretizationSchemes} \textbf{a)}, leading to a discretized distribution $Q^{\Delta}_{\pm}$, which is illustrated for various $\delta$ and $\lambda = 0.1$ in \autoref{fig:DiscretizedApproximtation}. When statistical errors become negligible, which is a justified assumption for many quantum optics setups, the discretized distribution $Q^{\Delta}_{\pm}$ can be measured directly.

Most importantly, the discretization breaks Gaussianity and hence the choice of the function $f$ matters. In \autoref{fig:DiscretizedCriteriaTMSV}~\textbf{a)} we compare the performances of the discretized versions of our second moment criteria (gray dotted curves), the Wehrl entropic criteria (black solid curves) and the optimized Rényi-Wehrl entropic criteria (light orange solid curves) by plotting the witnessed regions for every $\lambda$ as a function of the grid spacing $\delta$. In general, we can say that entropic criteria strongly outperform second moment criteria and that an optimization over the order $\beta$ for every $\lambda$ provides a second substantial improvement, especially for small $\lambda \approx 0.2$. 

We show the regions where the Rényi-Wehrl witness is negative for $\lambda=0.1$ and $\lambda=0.9$ as a function of $\beta$ in \textbf{c)} and \textbf{f)}, respectively, with the black dotted line indicating where the witness evaluates to zero. Note here that the witness functional becomes independent of $\beta$ when $\delta \to 0$ as the underlying state is Gaussian. Therein, we also show the curves obtained for $\beta=1$ (black dashed curves), the value for $\beta$ maximizing entanglement detection (dark orange dot-dashed curves) and the optimal value for $\beta$ for every $\delta$ (light orange solid curves).

The latter curves are drawn in \textbf{b)} and \textbf{e)}, respectively, together with the curves for the second moment criteria (gray). Considering for example $\lambda=0.1$ in \textbf{b)}, the Wehrl entropic witness already breaks down around $\delta \approx 1.5$, which corresponds to the level of discreteness shown in \autoref{fig:DiscretizedApproximtation} \textbf{c)}. In contrast, the optimal Rényi-Wehrl entropic witness certifies entanglement up to $\delta \approx 2.65$.

Further, the optimal choice for $\beta$ is a monotonically decreasing function of $\lambda$, which we plot in \textbf{d)}. Roughly speaking, for large entanglement $\lambda \ge 0.5$ we need small $\beta \le 1$, while for small entanglement $\lambda \le 0.5$ we should choose large $\beta \ge 1$.

\begin{figure*}[t!]	
    \centering
    \includegraphics[width=0.99\textwidth]{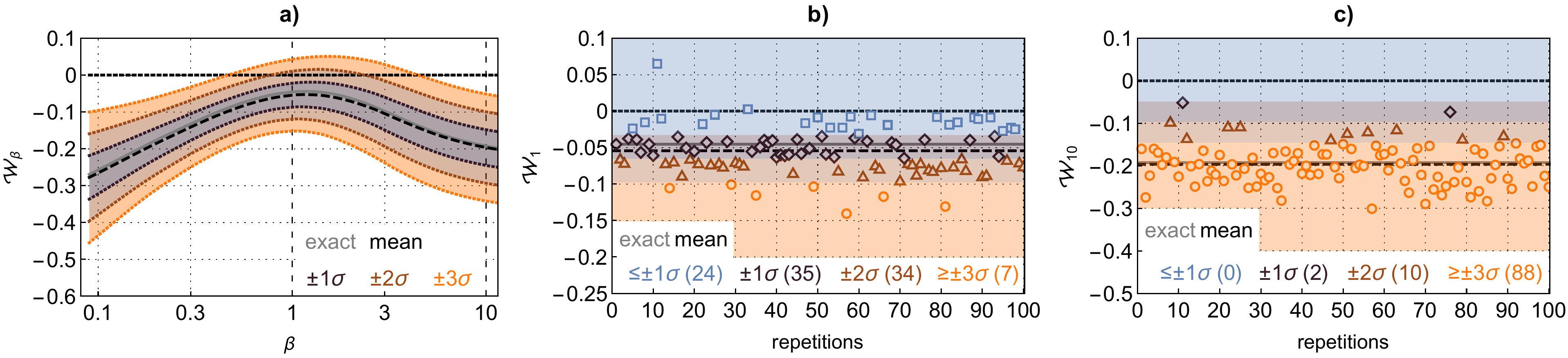}
    \caption{\textbf{a)} Mean (black dashed curve) and confidence intervals (shaded regions) of the Rényi-Wehrl criteria \eqref{eq:RenyiWehrlCriteria} for $10^3$ samples of the state \eqref{eq:HusimiQTMSVMixture} after $10^2$ repetitions of this simulated experiment (see also \cite{PRL}). \textbf{b)} and \textbf{c)} show the computed values for all repetitions for $\beta \to 1$ and $\beta = 10$, respectively. Optimizing over $\beta$, i.e. for $\beta$ on the order of $10^1$, certification of entanglement within large confidence intervals is improved substantially (squares / diamonds / triangles / circles correspond to $\le \pm 1 \sigma, \pm 1 \sigma, \pm 2 \sigma, \ge \pm 3 \sigma$, respectively).}
    \label{fig:RenyiWehrlWitnessSampled}
\end{figure*}

\section{Sampling measurements} \label{sec:sampling_meas}
We investigate our entanglement criteria for the second experimentally relevant scenario: sampling from the Husimi $Q$-distribution with limited statistics.

\subsection{Estimation of functionals of probability densities}
Estimating functionals of a probability density function (PDF), such as the witness functional $\mathcal{W}_f$ \eqref{eq:WitnessDefinition}, from samples is a central problem in statistical data analysis. It generally requires density estimation, the construction of an analytical estimate of the underlying PDF based on observed data. A plethora of methods exists for this including non-parametric approaches like simple data binning or kernel density estimation \cite{Jones1996}, maximum entropy models \cite{Armstrong2019} and parametric deep-learning based approaches \cite{Uria2016, Reed2017}. 

For the specific case of entropies, studied here, also direct estimation techniques have been devised (see \cite{Paninski2003} for a review) including the popular $k$-nearest neighbor method \cite{Kraskov2004}.
Generally, these methods rely on assumptions about the smoothness of the underlying PDF which allow to bound the approximation error. In this respect, the Husimi $Q$-distribution has favorable properties due to the uncertainty principle which leads to an inherently smooth behavior. In contrast to the Wigner $W$-distribution, which can have arbitrarily sharp features in the non-local phase space, the simultaneous detection of two conjugate quadratures leads to a coarse-graining or smoothing of the distribution in the case of Husimi. 

We generate synthetic sample data sets drawn from an analytically known Husimi $Q$-distribution and use a Gaussian mixture model to reconstruct $Q_\pm$ and calculate the witness functional $\mathcal{W}_{\beta}$ \eqref{eq:RenyiWehrlCriteria}.
Using moderate sample set sizes $\mathcal{O}(10^3)$ we are able to reliably estimate the witness functional. Crucially, by using different Rényi-Wehrl entropies parameterized by $\beta$ we find that values of small or large $\beta$ lead to a significantly increased signal-to-noise ratio of the entanglement detection compared to the case of the standard Wehrl entropy ($\beta\to 1$).

\subsection{Example state}
\label{subsec:SamplingMeasurmentsExampleState}
We consider a mixture of two displaced TMSV states with equal squeezing parameters $\lambda$, opposite displacements $\pm r$ with $r \ge 0$ along the $r_{\pm}$ axis and a mixing probability $p \in [0,1]$ s.t. $p=0$ selects the state displaced towards positive values for $r_{\pm}$. Its Husimi $Q$-distribution over a pair of non-local variables is
\begin{equation}
    \begin{split}
        Q_{\pm} (r_{\pm}, s_{\mp}) = (1-p) & \, \frac{1+\lambda}{2} \, e^{- \frac{1 + \lambda}{4} \left[ (r_{\pm} - r)^2 + s^2_{\mp} \right]} \\
        +\, p & \, \frac{1+\lambda}{2} \, e^{- \frac{1 + \lambda}{4} \left[ (r_{\pm} + r)^2 + s^2_{\mp} \right]}.
        \label{eq:HusimiQTMSVMixture}
    \end{split}
\end{equation}
Note that this setup is similar to the non-Gaussian state investigated in \cite{PRL,Walborn2009, Saboia2011}.

We test the performance of the Rényi-Wehrl criteria \eqref{eq:RenyiWehrlCriteria} for the state \eqref{eq:HusimiQTMSVMixture} with $\lambda = 0.8, r = 2, p=0.3$. This state is not witnessed by the second moment criteria. We simulate measurements by drawing $10^3$ samples and evaluate the witness functional using a Gaussian mixture model from the built-in machine-learning density estimation methods in mathematica. This method models the probability density using a mixture of multivariate normal distributions. For details we refer to the documentation of the mathematica function "LearnDensity" and the associated method "GaussianMixture". We gather information on the statistical quantities, in particular mean values and confidence intervals, by repeating this procedure $10^2$ times.

The $\beta$-dependence of the mean value and the three $\sigma$-intervals is shown in \autoref{fig:RenyiWehrlWitnessSampled} \textbf{a)}. As the mean value (dashed black) aligns with the exact result (solid gray), our estimation method is justified a posteriori. We observe that the choice of $\beta$ has a strong influence on the signal-to-noise ratio, which we exemplify for $\beta \to 1$ and $\beta = 10$ in \textbf{b)} and \textbf{c)}, respectively. While the standard Wehrl witness ($\beta \to 1$) is not able to witness entanglement within $\pm 1 \sigma$ for all repetitions, this can be achieved $\beta = 10$. In the latter case entanglement is even certified within $\pm 3 \sigma$ in $88\%$ of all cases (see orange circles in \autoref{fig:RenyiWehrlWitnessSampled} \textbf{c})).

\section{Conclusions and Outlook}
\label{sec:conclusions}
To summarize, using the Husimi $Q$-distribution for constructing entanglement witnesses has several advantages. In contrast to marginal distributions, it contains the full information about the underlying quantum state, which can thus be assessed though measurements in a single experimental setting. Most importantly, the existence of general uncertainty relations permits the derivation of entanglement criteria of much more general form than any known families of marginal based criteria. We showed the resulting strengths in different ways, including derivations of classes of entropic and second moment criteria as well as comparisons with marginal criteria. For a well-known family of states we were able to certify entanglement beyond the capabilities of marginal criteria based on uncertainty relations.

For future work, it would be interesting to investigate whether the requirement $a_1 b_1 = a_2 b_2$ can be relaxed, in which case a tensor product decomposition of the bipartite Hilbert space with respect to the non-local variables becomes impossible. However, we believe that our entanglement criteria do still hold. This conjecture is motivated by the fact that the entropic criteria \eqref{eq:WehrlCriteria} can be derived without the mentioned restriction on the scaling parameters using the entropy power inequality and local uncertainty relations along the lines of \cite{Haas2022a}.

Whether our criteria are the strongest criteria following from the uncertainty principle in phase space and what are the limitations of the Husimi approach, i.e. whether there exist any states which are witnessed by the STW but not by our criteria, are questions of central importance for future investigations. In general, we expect all entanglement criteria based on the linear non-local variables \eqref{eq:NonLocalOperatorsDefinition} to work best whenever the correlations between the local quadratures are mostly linear. In this case, an analysis of fluctuations in the non-local variables resembles the typical statistical analyses with Pearson’s coefficient, i.e. the search for linear correlations between the two subsystems. Therefore, we believe that our criteria become weak for states where entanglement manifests itself in higher-order correlations, which are not accurately captured by linear non-local variables. Such states are often characterized by partly negative Wigner $W$-distributions, consider for example NOON states or the photonic qutrit state. In both cases, we have confirmed hat our witness does \textit{not} flag entanglement (in both cases, optimizing over $\beta$ leads to $\mathcal{W}_{\beta} \to 0$ from above). Nevertheless, we do not see any arguments that would generally exclude detecting entanglement of states with negativities, and we consider finding such a state as an interesting problem for future work.

Moreover, we wonder whether other positive but not completely positive maps beyond the partial transpose have a faithful representation for the Husimi $Q$-distribution, which would allow to formulate yet another different class of entanglement criteria. Here, an approach compensating for the lack of detecting bound entanglement is of special interest. 

As our derivation relies on the group theoretic properties of coherent states, another interesting possibility would be to derive analogous criteria for physical systems described by other algebras, for example spin observables with a $SU(2)$ algebra. Further, it would be desirable to generalize our findings to incorporate entanglement measures in the spirit of \cite{Coles2017,Schneeloch2018,Bergh2021a,Bergh2021b} (see also \cite{dePalma2018c}) to set measurable lower bounds on the amount of entanglement. Finally, other potential directions where the Husimi approach might reveal its strengths encompass criteria for (genuine) multipartite entanglement as well as criteria based on higher-order moments of the Husimi $Q$-distribution in the sense of \cite{Shchukin2005}, which we expect to be of simpler accessibility than those based on higher-order moments of the Wigner $W$-distribution.

Regarding practical applications, we have shown that the optimization prospects of the generalized entanglement criteria based on the Husimi $Q$-distribution lead to clear performance advantages compared to the Wehrl-entropic criteria using the Wehrl entropy in situations with sparse experimental data. In a scenario, where this distribution is only known at a finite number of grid points within phase space, one effectively deals with a step function approximation to the exact continuous distribution. As a result, even for Gaussian states, optimizing the witness functional $\mathcal{W}_f$ over a parameterized family of concave functions $f$ --- in our case monomials $t^\beta$, relating the witness functional to Rényi-Wehrl entropies --- greatly enlarges the range of measurement resolution in which entanglement is detected compared to second moment based or Wehrl entropic criteria.

In a second scenario, where experimental measurements correspond to drawing samples from the Husimi $Q$-distribution, we found that both large ($\gg 1$) and small ($\ll 1$) values of $\beta$ lead to an increased statistical significance of the entanglement detection compared to the standard Wehrl entropy. Intuitively, increasing $\beta$ allows us to reduce the influence of regions of small probability on $\mathcal{W}_f$, i.e. reduce the influence of the tail behavior of the distribution. This can be beneficial because for finite statistics, these regions are very sparsely sampled and thus incur a large statistical error. The good performance for small $\beta$, where the contribution of the tail regions to $\mathcal{W}_f$ are amplified, might be explained by the employed density estimation routine making justified assumptions about the functional form of the tails.

Overall, not only the generality of our entanglement criteria, but also the inherent smoothness of the Husimi $Q$-distribution are found to lead to practical advantages. While one may expect that estimating functionals of a PDF is a much harder task than estimating low moments, it turns out that in the case of the Husimi $Q$-distribution standard density estimation methods work reliably even for rather small sample sizes. Given that the Husimi $Q$-distribution is a complete representation of the state, this leads to advantages for entanglement detection compared to approaches based on second moments or marginals of the Wigner $W$-distribution. 

In the future, one may exploit the prior knowledge about the properties of the Husimi $Q$-distribution --- like smoothness, tail behavior, or symmetries --- more systematically by using tailored density estimation methods \cite{Armstrong2019}. Moreover, for specific choices of $f$ direct ways of extracting $f(Q_\pm)$, or its phase space integral \cite{Paninski2003, Kraskov2004}, from samples exist, circumventing full density estimation, which may lead to even more data-efficient estimates.
Combined with the flexibility of our entanglement criteria to explore larger families of functions $f$, this has the potential to significantly reduce the experimental cost of certifying entanglement in continuous variable systems beyond the results reported here.

\section*{Acknowledgements}
We thank Oliver Stockdale, Stefan Floerchinger and Markus Oberthaler for discussions on the subject and Célia Griffet for comments on the manuscript. M.G. is supported by the Deutsche Forschungsgemeinschaft (DFG, German Research Foundation) under Germany's Excellence Strategy EXC 2181/1 - 390900948 (the Heidelberg STRUCTURES Excellence Cluster) and under SFB 1225 ISOQUANT - 273811115. T.H. acknowledges support by the European Union under project ShoQC within ERA-NET Cofund in Quantum Technologies (QuantERA) program.


\appendix

\section{Phase space distributions}
\label{app:Distributions}
The non-local quasi-probability distributions corresponding to the wave function \eqref{eq:ExampleStateWaveFunction} with arbitrary orientation and squeezing read
\begin{widetext}
    \begin{equation}
        \begin{split}
            W_{\pm} (r_{\pm}, s_{\mp}) &= \frac{1}{2 \pi} \, \exp \left\{ - \frac{\sigma_{\mp}^2}{2} \left[ \xi \sin (\phi) \, r_{\pm} - \frac{\cos (\phi)}{\xi} \, s_{\mp} \right]^2 - \frac{1}{2 \sigma_{\pm}^2} \left[ \xi \cos (\phi) \, r_{\pm} + \frac{\sin (\phi)}{\xi} \, s_{\mp} \right]^2 \right\} \\
            &\times \begin{cases} \frac{\sigma_-}{\sigma_+^3} \, \left[ \xi \cos (\phi) \, r_+ + \frac{\sin (\phi)}{\xi} \, s_- \right]^2 &\text{for  } (r_+, s_-), \\ \frac{\sigma_+^3}{\sigma_-} \, \left[ \xi \sin (\phi) \, r_- - \frac{\cos (\phi)}{\xi} \, s_+ \right]^2 &\text{for  } (r_-, s_+),
            \end{cases} \\
            Q_{\pm} (r_{\pm}, s_{\mp}) &= \frac{1}{\sqrt{\left(\xi^2 + \sigma_-^2 \right) \left(\xi^2 + \sigma_+^2 \right)^5}} \\
            &\times \exp \left\{ - \frac{\sigma_{\mp}^2}{2 \left( \xi^2 + \sigma_{\mp}^2 \right)} \left[\sin (\phi) \, r_{\pm} - \cos (\phi) \, s_{\mp} \right]^2 - \frac{\xi^2}{2 \left( \xi^2 + \sigma_{\pm}^2 \right)} \left[ \cos (\phi) \, r_{\pm} + \sin (\phi) \, s_{\mp} \right]^2 \right\} \\
            &\times \begin{cases} \xi^3 \sigma_- \, \left[ \xi^2 + \sigma_+^2 \left( 1 + \left( \cos (\phi) \, r_+ + \sin (\phi) \, s_- \right)^2 \right) \right] &\text{for  } (r_+, s_-), \\ \xi \sigma_+^3 \, \left[ \sigma_+^2 + \xi^2 \left( 1 + \left( \sin (\phi) \, r_- - \cos (\phi) \, s_+ \right)^2 \right) \right] &\text{for  } (r_-, s_+).
            \end{cases}
        \end{split}
        \label{eq:WignerWAndHusimiQExampleState}
    \end{equation}
\end{widetext}


\bibliography{references.bib}

\begin{thebibliography}{113}%
\makeatletter
\providecommand \@ifxundefined [1]{%
 \@ifx{#1\undefined}
}%
\providecommand \@ifnum [1]{%
 \ifnum #1\expandafter \@firstoftwo
 \else \expandafter \@secondoftwo
 \fi
}%
\providecommand \@ifx [1]{%
 \ifx #1\expandafter \@firstoftwo
 \else \expandafter \@secondoftwo
 \fi
}%
\providecommand \natexlab [1]{#1}%
\providecommand \enquote  [1]{``#1''}%
\providecommand \bibnamefont  [1]{#1}%
\providecommand \bibfnamefont [1]{#1}%
\providecommand \citenamefont [1]{#1}%
\providecommand \href@noop [0]{\@secondoftwo}%
\providecommand \href [0]{\begingroup \@sanitize@url \@href}%
\providecommand \@href[1]{\@@startlink{#1}\@@href}%
\providecommand \@@href[1]{\endgroup#1\@@endlink}%
\providecommand \@sanitize@url [0]{\catcode `\\12\catcode `\$12\catcode
  `\&12\catcode `\#12\catcode `\^12\catcode `\_12\catcode `\%12\relax}%
\providecommand \@@startlink[1]{}%
\providecommand \@@endlink[0]{}%
\providecommand \url  [0]{\begingroup\@sanitize@url \@url }%
\providecommand \@url [1]{\endgroup\@href {#1}{\urlprefix }}%
\providecommand \urlprefix  [0]{URL }%
\providecommand \Eprint [0]{\href }%
\providecommand \doibase [0]{https://doi.org/}%
\providecommand \selectlanguage [0]{\@gobble}%
\providecommand \bibinfo  [0]{\@secondoftwo}%
\providecommand \bibfield  [0]{\@secondoftwo}%
\providecommand \translation [1]{[#1]}%
\providecommand \BibitemOpen [0]{}%
\providecommand \bibitemStop [0]{}%
\providecommand \bibitemNoStop [0]{.\EOS\space}%
\providecommand \EOS [0]{\spacefactor3000\relax}%
\providecommand \BibitemShut  [1]{\csname bibitem#1\endcsname}%
\let\auto@bib@innerbib\@empty
\bibitem [{\citenamefont {Einstein}\ \emph {et~al.}(1935)\citenamefont
  {Einstein}, \citenamefont {Podolsky},\ and\ \citenamefont
  {Rosen}}]{Einstein1935}%
  \BibitemOpen
  \bibfield  {author} {\bibinfo {author} {\bibfnamefont {A.}~\bibnamefont
  {Einstein}}, \bibinfo {author} {\bibfnamefont {B.}~\bibnamefont {Podolsky}},\
  and\ \bibinfo {author} {\bibfnamefont {N.}~\bibnamefont {Rosen}},\ }\bibfield
   {title} {\bibinfo {title} {{Can Quantum-Mechan-ical Description of Physical
  Reality Be Considered Complete?}},\ }\href
  {https://doi.org/10.1103/PhysRev.47.777} {\bibfield  {journal} {\bibinfo
  {journal} {Phys. Rev.}\ }\textbf {\bibinfo {volume} {47}},\ \bibinfo {pages}
  {777} (\bibinfo {year} {1935})}\BibitemShut {NoStop}%
\bibitem [{\citenamefont {Braunstein}\ and\ \citenamefont {van
  Loock}(2005)}]{Braunstein2005b}%
  \BibitemOpen
  \bibfield  {author} {\bibinfo {author} {\bibfnamefont {S.~L.}\ \bibnamefont
  {Braunstein}}\ and\ \bibinfo {author} {\bibfnamefont {P.}~\bibnamefont {van
  Loock}},\ }\bibfield  {title} {\bibinfo {title} {{Quantum information with
  continuous variables}},\ }\href {https://doi.org/10.1103/RevModPhys.77.513}
  {\bibfield  {journal} {\bibinfo  {journal} {Rev. Mod. Phys.}\ }\textbf
  {\bibinfo {volume} {77}},\ \bibinfo {pages} {513} (\bibinfo {year}
  {2005})}\BibitemShut {NoStop}%
\bibitem [{\citenamefont {Horodecki}\ \emph {et~al.}(2009)\citenamefont
  {Horodecki}, \citenamefont {Horodecki}, \citenamefont {Horodecki},\ and\
  \citenamefont {Horodecki}}]{Horodecki2009}%
  \BibitemOpen
  \bibfield  {author} {\bibinfo {author} {\bibfnamefont {R.}~\bibnamefont
  {Horodecki}}, \bibinfo {author} {\bibfnamefont {P.}~\bibnamefont
  {Horodecki}}, \bibinfo {author} {\bibfnamefont {M.}~\bibnamefont
  {Horodecki}},\ and\ \bibinfo {author} {\bibfnamefont {K.}~\bibnamefont
  {Horodecki}},\ }\bibfield  {title} {\bibinfo {title} {{Quantum
  entanglement}},\ }\href {https://doi.org/10.1103/RevModPhys.81.865}
  {\bibfield  {journal} {\bibinfo  {journal} {Rev. Mod. Phys.}\ }\textbf
  {\bibinfo {volume} {81}},\ \bibinfo {pages} {865} (\bibinfo {year}
  {2009})}\BibitemShut {NoStop}%
\bibitem [{\citenamefont {G{\"u}hne}\ and\ \citenamefont
  {T{\'o}th}(2009)}]{Guehne2009}%
  \BibitemOpen
  \bibfield  {author} {\bibinfo {author} {\bibfnamefont {O.}~\bibnamefont
  {G{\"u}hne}}\ and\ \bibinfo {author} {\bibfnamefont {G.}~\bibnamefont
  {T{\'o}th}},\ }\bibfield  {title} {\bibinfo {title} {Entanglement
  detection},\ }\href {https://doi.org/10.1016/j.physrep.2009.02.004}
  {\bibfield  {journal} {\bibinfo  {journal} {Phys. Rep.}\ }\textbf {\bibinfo
  {volume} {474}},\ \bibinfo {pages} {1} (\bibinfo {year} {2009})}\BibitemShut
  {NoStop}%
\bibitem [{\citenamefont {Weedbrook}\ \emph {et~al.}(2012)\citenamefont
  {Weedbrook}, \citenamefont {Pirandola}, \citenamefont
  {Garc\'{i}a-Patr\'{o}n}, \citenamefont {Cerf}, \citenamefont {Ralph},
  \citenamefont {Shapiro},\ and\ \citenamefont {Lloyd}}]{Weedbrook2012}%
  \BibitemOpen
  \bibfield  {author} {\bibinfo {author} {\bibfnamefont {C.}~\bibnamefont
  {Weedbrook}}, \bibinfo {author} {\bibfnamefont {S.}~\bibnamefont
  {Pirandola}}, \bibinfo {author} {\bibfnamefont {R.}~\bibnamefont
  {Garc\'{i}a-Patr\'{o}n}}, \bibinfo {author} {\bibfnamefont {N.~J.}\
  \bibnamefont {Cerf}}, \bibinfo {author} {\bibfnamefont {T.~C.}\ \bibnamefont
  {Ralph}}, \bibinfo {author} {\bibfnamefont {J.~H.}\ \bibnamefont {Shapiro}},\
  and\ \bibinfo {author} {\bibfnamefont {S.}~\bibnamefont {Lloyd}},\ }\bibfield
   {title} {\bibinfo {title} {{Gaussian quantum information}},\ }\href
  {https://doi.org/10.1103/RevModPhys.84.621} {\bibfield  {journal} {\bibinfo
  {journal} {Rev. Mod. Phys.}\ }\textbf {\bibinfo {volume} {84}},\ \bibinfo
  {pages} {621} (\bibinfo {year} {2012})}\BibitemShut {NoStop}%
\bibitem [{\citenamefont {Peres}(1996)}]{Peres1996}%
  \BibitemOpen
  \bibfield  {author} {\bibinfo {author} {\bibfnamefont {A.}~\bibnamefont
  {Peres}},\ }\bibfield  {title} {\bibinfo {title} {{Separability Criterion for
  Density Matrices}},\ }\href {https://doi.org/10.1103/PhysRevLett.77.1413}
  {\bibfield  {journal} {\bibinfo  {journal} {Phys. Rev. Lett.}\ }\textbf
  {\bibinfo {volume} {77}},\ \bibinfo {pages} {1413} (\bibinfo {year}
  {1996})}\BibitemShut {NoStop}%
\bibitem [{\citenamefont {Horodecki}\ \emph {et~al.}(1996)\citenamefont
  {Horodecki}, \citenamefont {Horodecki},\ and\ \citenamefont
  {Horodecki}}]{Horodecki1996}%
  \BibitemOpen
  \bibfield  {author} {\bibinfo {author} {\bibfnamefont {M.}~\bibnamefont
  {Horodecki}}, \bibinfo {author} {\bibfnamefont {P.}~\bibnamefont
  {Horodecki}},\ and\ \bibinfo {author} {\bibfnamefont {R.}~\bibnamefont
  {Horodecki}},\ }\bibfield  {title} {\bibinfo {title} {{Separability of mixed
  states: necessary and sufficient conditions}},\ }\href
  {https://doi.org/https://doi.org/10.1016/S0375-9601(96)00706-2} {\bibfield
  {journal} {\bibinfo  {journal} {Phys. Lett. A}\ }\textbf {\bibinfo {volume}
  {223}},\ \bibinfo {pages} {1} (\bibinfo {year} {1996})}\BibitemShut {NoStop}%
\bibitem [{\citenamefont {Nha}\ and\ \citenamefont {Zubairy}(2008)}]{Nha2008}%
  \BibitemOpen
  \bibfield  {author} {\bibinfo {author} {\bibfnamefont {H.}~\bibnamefont
  {Nha}}\ and\ \bibinfo {author} {\bibfnamefont {M.~S.}\ \bibnamefont
  {Zubairy}},\ }\bibfield  {title} {\bibinfo {title} {{Uncertainty Inequalities
  as Entanglement Criteria for Negative Partial-Transpose States}},\ }\href
  {https://doi.org/10.1103/PhysRevLett.101.130402} {\bibfield  {journal}
  {\bibinfo  {journal} {Phys. Rev. Lett.}\ }\textbf {\bibinfo {volume} {101}},\
  \bibinfo {pages} {130402} (\bibinfo {year} {2008})}\BibitemShut {NoStop}%
\bibitem [{\citenamefont {Horodecki}\ \emph {et~al.}(1998)\citenamefont
  {Horodecki}, \citenamefont {Horodecki},\ and\ \citenamefont
  {Horodecki}}]{Horodecki1998}%
  \BibitemOpen
  \bibfield  {author} {\bibinfo {author} {\bibfnamefont {M.}~\bibnamefont
  {Horodecki}}, \bibinfo {author} {\bibfnamefont {P.}~\bibnamefont
  {Horodecki}},\ and\ \bibinfo {author} {\bibfnamefont {R.}~\bibnamefont
  {Horodecki}},\ }\bibfield  {title} {\bibinfo {title} {{Mixed-State
  Entanglement and Distillation: Is there a ``Bound'' Entanglement in
  Nature?}},\ }\href {https://doi.org/10.1103/PhysRevLett.80.5239} {\bibfield
  {journal} {\bibinfo  {journal} {Phys. Rev. Lett.}\ }\textbf {\bibinfo
  {volume} {80}},\ \bibinfo {pages} {5239} (\bibinfo {year}
  {1998})}\BibitemShut {NoStop}%
\bibitem [{\citenamefont {Duan}\ \emph {et~al.}(2000)\citenamefont {Duan},
  \citenamefont {Giedke}, \citenamefont {Cirac},\ and\ \citenamefont
  {Zoller}}]{Duan2000}%
  \BibitemOpen
  \bibfield  {author} {\bibinfo {author} {\bibfnamefont {L.-M.}\ \bibnamefont
  {Duan}}, \bibinfo {author} {\bibfnamefont {G.}~\bibnamefont {Giedke}},
  \bibinfo {author} {\bibfnamefont {J.~I.}\ \bibnamefont {Cirac}},\ and\
  \bibinfo {author} {\bibfnamefont {P.}~\bibnamefont {Zoller}},\ }\bibfield
  {title} {\bibinfo {title} {{Inseparability Criterion for Continuous Variable
  Systems}},\ }\href {https://doi.org/10.1103/PhysRevLett.84.2722} {\bibfield
  {journal} {\bibinfo  {journal} {Phys. Rev. Lett.}\ }\textbf {\bibinfo
  {volume} {84}},\ \bibinfo {pages} {2722} (\bibinfo {year}
  {2000})}\BibitemShut {NoStop}%
\bibitem [{\citenamefont {Mancini}\ \emph {et~al.}(2002)\citenamefont
  {Mancini}, \citenamefont {Giovannetti}, \citenamefont {Vitali},\ and\
  \citenamefont {Tombesi}}]{Mancini2002}%
  \BibitemOpen
  \bibfield  {author} {\bibinfo {author} {\bibfnamefont {S.}~\bibnamefont
  {Mancini}}, \bibinfo {author} {\bibfnamefont {V.}~\bibnamefont
  {Giovannetti}}, \bibinfo {author} {\bibfnamefont {D.}~\bibnamefont
  {Vitali}},\ and\ \bibinfo {author} {\bibfnamefont {P.}~\bibnamefont
  {Tombesi}},\ }\bibfield  {title} {\bibinfo {title} {{Entangling Macroscopic
  Oscillators Exploiting Radiation Pressure}},\ }\href
  {https://doi.org/10.1103/PhysRevLett.88.120401} {\bibfield  {journal}
  {\bibinfo  {journal} {Phys. Rev. Lett.}\ }\textbf {\bibinfo {volume} {88}},\
  \bibinfo {pages} {120401} (\bibinfo {year} {2002})}\BibitemShut {NoStop}%
\bibitem [{\citenamefont {Giovannetti}\ \emph {et~al.}(2003)\citenamefont
  {Giovannetti}, \citenamefont {Mancini}, \citenamefont {Vitali},\ and\
  \citenamefont {Tombesi}}]{Giovannetti2003}%
  \BibitemOpen
  \bibfield  {author} {\bibinfo {author} {\bibfnamefont {V.}~\bibnamefont
  {Giovannetti}}, \bibinfo {author} {\bibfnamefont {S.}~\bibnamefont
  {Mancini}}, \bibinfo {author} {\bibfnamefont {D.}~\bibnamefont {Vitali}},\
  and\ \bibinfo {author} {\bibfnamefont {P.}~\bibnamefont {Tombesi}},\
  }\bibfield  {title} {\bibinfo {title} {{Characterizing the entanglement of
  bipartite quantum systems}},\ }\href
  {https://doi.org/10.1103/PhysRevA.67.022320} {\bibfield  {journal} {\bibinfo
  {journal} {Phys. Rev. A}\ }\textbf {\bibinfo {volume} {67}},\ \bibinfo
  {pages} {022320} (\bibinfo {year} {2003})}\BibitemShut {NoStop}%
\bibitem [{\citenamefont {Simon}(2000)}]{Simon2000}%
  \BibitemOpen
  \bibfield  {author} {\bibinfo {author} {\bibfnamefont {R.}~\bibnamefont
  {Simon}},\ }\bibfield  {title} {\bibinfo {title} {{Peres-Horodecki
  Separability Criterion for Continuous Variable Systems}},\ }\href
  {https://doi.org/10.1103/PhysRevLett.84.2726} {\bibfield  {journal} {\bibinfo
   {journal} {Phys. Rev. Lett.}\ }\textbf {\bibinfo {volume} {84}},\ \bibinfo
  {pages} {2726} (\bibinfo {year} {2000})}\BibitemShut {NoStop}%
\bibitem [{\citenamefont {Serafini}(2017)}]{Serafini2017}%
  \BibitemOpen
  \bibfield  {author} {\bibinfo {author} {\bibfnamefont {A.}~\bibnamefont
  {Serafini}},\ }\href {https://doi.org/10.1201/9781315118727} {\emph {\bibinfo
  {title} {{Quantum Continuous Variables}}}}\ (\bibinfo  {publisher} {CRC
  Press},\ \bibinfo {year} {2017})\BibitemShut {NoStop}%
\bibitem [{\citenamefont {Lami}\ \emph {et~al.}(2018)\citenamefont {Lami},
  \citenamefont {Serafini},\ and\ \citenamefont {Adesso}}]{Lami2018}%
  \BibitemOpen
  \bibfield  {author} {\bibinfo {author} {\bibfnamefont {L.}~\bibnamefont
  {Lami}}, \bibinfo {author} {\bibfnamefont {A.}~\bibnamefont {Serafini}},\
  and\ \bibinfo {author} {\bibfnamefont {G.}~\bibnamefont {Adesso}},\
  }\bibfield  {title} {\bibinfo {title} {{Gaussian entanglement revisited}},\
  }\href {https://doi.org/10.1088/1367-2630/aaa654} {\bibfield  {journal}
  {\bibinfo  {journal} {New J. Phys.}\ }\textbf {\bibinfo {volume} {20}},\
  \bibinfo {pages} {023030} (\bibinfo {year} {2018})}\BibitemShut {NoStop}%
\bibitem [{\citenamefont {Walborn}\ \emph {et~al.}(2009)\citenamefont
  {Walborn}, \citenamefont {Taketani}, \citenamefont {Salles}, \citenamefont
  {Toscano},\ and\ \citenamefont {de~Matos~Filho}}]{Walborn2009}%
  \BibitemOpen
  \bibfield  {author} {\bibinfo {author} {\bibfnamefont {S.~P.}\ \bibnamefont
  {Walborn}}, \bibinfo {author} {\bibfnamefont {B.~G.}\ \bibnamefont
  {Taketani}}, \bibinfo {author} {\bibfnamefont {A.}~\bibnamefont {Salles}},
  \bibinfo {author} {\bibfnamefont {F.}~\bibnamefont {Toscano}},\ and\ \bibinfo
  {author} {\bibfnamefont {R.~L.}\ \bibnamefont {de~Matos~Filho}},\ }\bibfield
  {title} {\bibinfo {title} {{Entropic Entanglement Criteria for Continuous
  Variables}},\ }\href {https://doi.org/10.1103/PhysRevLett.103.160505}
  {\bibfield  {journal} {\bibinfo  {journal} {Phys. Rev. Lett.}\ }\textbf
  {\bibinfo {volume} {103}},\ \bibinfo {pages} {160505} (\bibinfo {year}
  {2009})}\BibitemShut {NoStop}%
\bibitem [{\citenamefont {Saboia}\ \emph {et~al.}(2011)\citenamefont {Saboia},
  \citenamefont {Toscano},\ and\ \citenamefont {Walborn}}]{Saboia2011}%
  \BibitemOpen
  \bibfield  {author} {\bibinfo {author} {\bibfnamefont {A.}~\bibnamefont
  {Saboia}}, \bibinfo {author} {\bibfnamefont {F.}~\bibnamefont {Toscano}},\
  and\ \bibinfo {author} {\bibfnamefont {S.~P.}\ \bibnamefont {Walborn}},\
  }\bibfield  {title} {\bibinfo {title} {{Family of continuous-variable
  entanglement criteria using general entropy functions}},\ }\href
  {https://doi.org/10.1103/PhysRevA.83.032307} {\bibfield  {journal} {\bibinfo
  {journal} {Phys. Rev. A}\ }\textbf {\bibinfo {volume} {83}},\ \bibinfo
  {pages} {032307} (\bibinfo {year} {2011})}\BibitemShut {NoStop}%
\bibitem [{\citenamefont {Reid}(1989)}]{Reid1989}%
  \BibitemOpen
  \bibfield  {author} {\bibinfo {author} {\bibfnamefont {M.~D.}\ \bibnamefont
  {Reid}},\ }\bibfield  {title} {\bibinfo {title} {{Demonstration of the
  Einstein-Podolsky-Rosen paradox using nondegenerate parametric
  amplification}},\ }\href {https://doi.org/10.1103/PhysRevA.40.913} {\bibfield
   {journal} {\bibinfo  {journal} {Phys. Rev. A}\ }\textbf {\bibinfo {volume}
  {40}},\ \bibinfo {pages} {913} (\bibinfo {year} {1989})}\BibitemShut
  {NoStop}%
\bibitem [{\citenamefont {He}\ and\ \citenamefont {Reid}(2013)}]{He2013}%
  \BibitemOpen
  \bibfield  {author} {\bibinfo {author} {\bibfnamefont {Q.~Y.}\ \bibnamefont
  {He}}\ and\ \bibinfo {author} {\bibfnamefont {M.~D.}\ \bibnamefont {Reid}},\
  }\bibfield  {title} {\bibinfo {title} {{Genuine Multipartite
  Einstein-Podolsky-Rosen Steering}},\ }\href
  {https://doi.org/10.1103/PhysRevLett.111.250403} {\bibfield  {journal}
  {\bibinfo  {journal} {Phys. Rev. Lett.}\ }\textbf {\bibinfo {volume} {111}},\
  \bibinfo {pages} {250403} (\bibinfo {year} {2013})}\BibitemShut {NoStop}%
\bibitem [{\citenamefont {Walborn}\ \emph {et~al.}(2011)\citenamefont
  {Walborn}, \citenamefont {Salles}, \citenamefont {Gomes}, \citenamefont
  {Toscano},\ and\ \citenamefont {Ribeiro}}]{Walborn2011}%
  \BibitemOpen
  \bibfield  {author} {\bibinfo {author} {\bibfnamefont {S.~P.}\ \bibnamefont
  {Walborn}}, \bibinfo {author} {\bibfnamefont {A.}~\bibnamefont {Salles}},
  \bibinfo {author} {\bibfnamefont {R.~M.}\ \bibnamefont {Gomes}}, \bibinfo
  {author} {\bibfnamefont {F.}~\bibnamefont {Toscano}},\ and\ \bibinfo {author}
  {\bibfnamefont {P.~H.~S.}\ \bibnamefont {Ribeiro}},\ }\bibfield  {title}
  {\bibinfo {title} {{Revealing Hidden Einstein-Podolsky-Rosen Nonlocality}},\
  }\href {https://doi.org/10.1103/PhysRevLett.106.130402} {\bibfield  {journal}
  {\bibinfo  {journal} {Phys. Rev. Lett.}\ }\textbf {\bibinfo {volume} {106}},\
  \bibinfo {pages} {130402} (\bibinfo {year} {2011})}\BibitemShut {NoStop}%
\bibitem [{\citenamefont {Chowdhury}\ \emph {et~al.}(2014)\citenamefont
  {Chowdhury}, \citenamefont {Pramanik}, \citenamefont {Majumdar},\ and\
  \citenamefont {Agarwal}}]{Chowdhury2014}%
  \BibitemOpen
  \bibfield  {author} {\bibinfo {author} {\bibfnamefont {P.}~\bibnamefont
  {Chowdhury}}, \bibinfo {author} {\bibfnamefont {T.}~\bibnamefont {Pramanik}},
  \bibinfo {author} {\bibfnamefont {A.~S.}\ \bibnamefont {Majumdar}},\ and\
  \bibinfo {author} {\bibfnamefont {G.~S.}\ \bibnamefont {Agarwal}},\
  }\bibfield  {title} {\bibinfo {title} {{Einstein-Podolsky-Rosen steering
  using quantum correlations in non-Gaussian entangled states}},\ }\href
  {https://doi.org/10.1103/PhysRevA.89.012104} {\bibfield  {journal} {\bibinfo
  {journal} {Phys. Rev. A}\ }\textbf {\bibinfo {volume} {89}},\ \bibinfo
  {pages} {012104} (\bibinfo {year} {2014})}\BibitemShut {NoStop}%
\bibitem [{\citenamefont {Schneeloch}\ and\ \citenamefont
  {Howland}(2018)}]{Schneeloch2018}%
  \BibitemOpen
  \bibfield  {author} {\bibinfo {author} {\bibfnamefont {J.}~\bibnamefont
  {Schneeloch}}\ and\ \bibinfo {author} {\bibfnamefont {G.~A.}\ \bibnamefont
  {Howland}},\ }\bibfield  {title} {\bibinfo {title} {{Quantifying
  high-dimensional entanglement with Einstein-Podolsky-Rosen correlations}},\
  }\href {https://doi.org/10.1103/PhysRevA.97.042338} {\bibfield  {journal}
  {\bibinfo  {journal} {Phys. Rev. A}\ }\textbf {\bibinfo {volume} {97}},\
  \bibinfo {pages} {042338} (\bibinfo {year} {2018})}\BibitemShut {NoStop}%
\bibitem [{\citenamefont {Gneiting}\ and\ \citenamefont
  {Hornberger}(2011)}]{Gneiting2011}%
  \BibitemOpen
  \bibfield  {author} {\bibinfo {author} {\bibfnamefont {C.}~\bibnamefont
  {Gneiting}}\ and\ \bibinfo {author} {\bibfnamefont {K.}~\bibnamefont
  {Hornberger}},\ }\bibfield  {title} {\bibinfo {title} {{Detecting
  Entanglement in Spatial Interference}},\ }\href
  {https://doi.org/10.1103/PhysRevLett.106.210501} {\bibfield  {journal}
  {\bibinfo  {journal} {Phys. Rev. Lett.}\ }\textbf {\bibinfo {volume} {106}},\
  \bibinfo {pages} {210501} (\bibinfo {year} {2011})}\BibitemShut {NoStop}%
\bibitem [{\citenamefont {Carvalho}\ \emph {et~al.}(2012)\citenamefont
  {Carvalho}, \citenamefont {Ferraz}, \citenamefont {Borges}, \citenamefont
  {de~Assis}, \citenamefont {P\'adua},\ and\ \citenamefont
  {Walborn}}]{Carvalho2012}%
  \BibitemOpen
  \bibfield  {author} {\bibinfo {author} {\bibfnamefont {M.~A.~D.}\
  \bibnamefont {Carvalho}}, \bibinfo {author} {\bibfnamefont {J.}~\bibnamefont
  {Ferraz}}, \bibinfo {author} {\bibfnamefont {G.~F.}\ \bibnamefont {Borges}},
  \bibinfo {author} {\bibfnamefont {P.-L.}\ \bibnamefont {de~Assis}}, \bibinfo
  {author} {\bibfnamefont {S.}~\bibnamefont {P\'adua}},\ and\ \bibinfo {author}
  {\bibfnamefont {S.~P.}\ \bibnamefont {Walborn}},\ }\bibfield  {title}
  {\bibinfo {title} {{Experimental observation of quantum correlations in
  modular variables}},\ }\href {https://doi.org/10.1103/PhysRevA.86.032332}
  {\bibfield  {journal} {\bibinfo  {journal} {Phys. Rev. A}\ }\textbf {\bibinfo
  {volume} {86}},\ \bibinfo {pages} {032332} (\bibinfo {year}
  {2012})}\BibitemShut {NoStop}%
\bibitem [{\citenamefont {Agarwal}\ and\ \citenamefont
  {Biswas}(2005)}]{Agarwal2005}%
  \BibitemOpen
  \bibfield  {author} {\bibinfo {author} {\bibfnamefont {G.~S.}\ \bibnamefont
  {Agarwal}}\ and\ \bibinfo {author} {\bibfnamefont {A.}~\bibnamefont
  {Biswas}},\ }\bibfield  {title} {\bibinfo {title} {{Inseparability
  inequalities for higher order moments for bipartite systems}},\ }\href
  {https://doi.org/10.1088/1367-2630/7/1/211} {\bibfield  {journal} {\bibinfo
  {journal} {New J. Phys.}\ }\textbf {\bibinfo {volume} {7}},\ \bibinfo {pages}
  {211} (\bibinfo {year} {2005})}\BibitemShut {NoStop}%
\bibitem [{\citenamefont {Shchukin}\ and\ \citenamefont
  {Vogel}(2005)}]{Shchukin2005}%
  \BibitemOpen
  \bibfield  {author} {\bibinfo {author} {\bibfnamefont {E.}~\bibnamefont
  {Shchukin}}\ and\ \bibinfo {author} {\bibfnamefont {W.}~\bibnamefont
  {Vogel}},\ }\bibfield  {title} {\bibinfo {title} {{Inseparability Criteria
  for Continuous Bipartite Quantum States}},\ }\href
  {https://doi.org/10.1103/PhysRevLett.95.230502} {\bibfield  {journal}
  {\bibinfo  {journal} {Phys. Rev. Lett.}\ }\textbf {\bibinfo {volume} {95}},\
  \bibinfo {pages} {230502} (\bibinfo {year} {2005})}\BibitemShut {NoStop}%
\bibitem [{\citenamefont {Griffet}\ \emph {et~al.}(2023)\citenamefont
  {Griffet}, \citenamefont {Haas},\ and\ \citenamefont {Cerf}}]{Haas2023}%
  \BibitemOpen
  \bibfield  {author} {\bibinfo {author} {\bibfnamefont {C.}~\bibnamefont
  {Griffet}}, \bibinfo {author} {\bibfnamefont {T.}~\bibnamefont {Haas}},\ and\
  \bibinfo {author} {\bibfnamefont {N.~J.}\ \bibnamefont {Cerf}},\ }\bibfield
  {title} {\bibinfo {title} {{Accessing continuous-variable entanglement
  witnesses with multimode spin observables}},\ }\href
  {https://doi.org/10.1103/PhysRevA.108.022421} {\bibfield  {journal} {\bibinfo
   {journal} {Phys. Rev. A}\ }\textbf {\bibinfo {volume} {108}},\ \bibinfo
  {pages} {022421} (\bibinfo {year} {2023})}\BibitemShut {NoStop}%
\bibitem [{\citenamefont {Gessner}\ \emph {et~al.}(2017)\citenamefont
  {Gessner}, \citenamefont {Pezz{\`{e}}},\ and\ \citenamefont
  {Smerzi}}]{Gessner2017}%
  \BibitemOpen
  \bibfield  {author} {\bibinfo {author} {\bibfnamefont {M.}~\bibnamefont
  {Gessner}}, \bibinfo {author} {\bibfnamefont {L.}~\bibnamefont
  {Pezz{\`{e}}}},\ and\ \bibinfo {author} {\bibfnamefont {A.}~\bibnamefont
  {Smerzi}},\ }\bibfield  {title} {\bibinfo {title} {{Entanglement and
  squeezing in continuous-variable systems}},\ }\href
  {https://doi.org/10.22331/q-2017-07-14-17} {\bibfield  {journal} {\bibinfo
  {journal} {{Quantum}}\ }\textbf {\bibinfo {volume} {1}},\ \bibinfo {pages}
  {17} (\bibinfo {year} {2017})}\BibitemShut {NoStop}%
\bibitem [{\citenamefont {Qin}\ \emph {et~al.}(2019)\citenamefont {Qin},
  \citenamefont {Gessner}, \citenamefont {Ren}, \citenamefont {Deng},
  \citenamefont {Han}, \citenamefont {Li}, \citenamefont {Su}, \citenamefont
  {Smerzi},\ and\ \citenamefont {Peng}}]{Qin2019}%
  \BibitemOpen
  \bibfield  {author} {\bibinfo {author} {\bibfnamefont {Z.}~\bibnamefont
  {Qin}}, \bibinfo {author} {\bibfnamefont {M.}~\bibnamefont {Gessner}},
  \bibinfo {author} {\bibfnamefont {Z.}~\bibnamefont {Ren}}, \bibinfo {author}
  {\bibfnamefont {X.}~\bibnamefont {Deng}}, \bibinfo {author} {\bibfnamefont
  {D.}~\bibnamefont {Han}}, \bibinfo {author} {\bibfnamefont {W.}~\bibnamefont
  {Li}}, \bibinfo {author} {\bibfnamefont {X.}~\bibnamefont {Su}}, \bibinfo
  {author} {\bibfnamefont {A.}~\bibnamefont {Smerzi}},\ and\ \bibinfo {author}
  {\bibfnamefont {K.}~\bibnamefont {Peng}},\ }\bibfield  {title} {\bibinfo
  {title} {{Characterizing the multipartite continuous-variable entanglement
  structure from squeezing coefficients and the Fisher information}},\ }\href
  {https://doi.org/10.1038/s41534-018-0119-6} {\bibfield  {journal} {\bibinfo
  {journal} {npj Quantum Inf.}\ }\textbf {\bibinfo {volume} {5}},\ \bibinfo
  {pages} {3} (\bibinfo {year} {2019})}\BibitemShut {NoStop}%
\bibitem [{\citenamefont {Husimi}(1940)}]{Husimi1940}%
  \BibitemOpen
  \bibfield  {author} {\bibinfo {author} {\bibfnamefont {K.}~\bibnamefont
  {Husimi}},\ }\bibfield  {title} {\bibinfo {title} {{Some formal properties of
  the density matrix}},\ }\href {https://doi.org/10.11429/ppmsj1919.22.4_264}
  {\bibfield  {journal} {\bibinfo  {journal} {Proc. Phys.-Math. Soc. Jap. 3rd
  Ser.}\ }\textbf {\bibinfo {volume} {22}},\ \bibinfo {pages} {264} (\bibinfo
  {year} {1940})}\BibitemShut {NoStop}%
\bibitem [{\citenamefont {Cartwright}(1976)}]{Cartwright1976}%
  \BibitemOpen
  \bibfield  {author} {\bibinfo {author} {\bibfnamefont {N.~D.}\ \bibnamefont
  {Cartwright}},\ }\bibfield  {title} {\bibinfo {title} {{A non-negative
  Wigner-type distribution}},\ }\href
  {https://doi.org/10.1016/0378-4371(76)90145-X} {\bibfield  {journal}
  {\bibinfo  {journal} {Phys. A}\ }\textbf {\bibinfo {volume} {83}},\ \bibinfo
  {pages} {210} (\bibinfo {year} {1976})}\BibitemShut {NoStop}%
\bibitem [{\citenamefont {Lee}(1995)}]{Lee1995}%
  \BibitemOpen
  \bibfield  {author} {\bibinfo {author} {\bibfnamefont {H.-W.}\ \bibnamefont
  {Lee}},\ }\bibfield  {title} {\bibinfo {title} {{Theory and application of
  the quantum phase-space distribution functions}},\ }\href
  {https://doi.org/https://doi.org/10.1016/0370-1573(95)00007-4} {\bibfield
  {journal} {\bibinfo  {journal} {Phys. Rep.}\ }\textbf {\bibinfo {volume}
  {259}},\ \bibinfo {pages} {147} (\bibinfo {year} {1995})}\BibitemShut
  {NoStop}%
\bibitem [{\citenamefont {Collett}\ \emph {et~al.}(1987)\citenamefont
  {Collett}, \citenamefont {Loudon},\ and\ \citenamefont
  {Gardiner}}]{Collett1987}%
  \BibitemOpen
  \bibfield  {author} {\bibinfo {author} {\bibfnamefont {M.~J.}\ \bibnamefont
  {Collett}}, \bibinfo {author} {\bibfnamefont {R.}~\bibnamefont {Loudon}},\
  and\ \bibinfo {author} {\bibfnamefont {C.~W.}\ \bibnamefont {Gardiner}},\
  }\bibfield  {title} {\bibinfo {title} {{Quantum Theory of Optical Homodyne
  and Heterodyne Detection}},\ }\href
  {https://doi.org/10.1080/09500348714550811} {\bibfield  {journal} {\bibinfo
  {journal} {J. Mod. Opt.}\ }\textbf {\bibinfo {volume} {34}},\ \bibinfo
  {pages} {881} (\bibinfo {year} {1987})}\BibitemShut {NoStop}%
\bibitem [{\citenamefont {Welsch}\ \emph {et~al.}(1999)\citenamefont {Welsch},
  \citenamefont {Vogel},\ and\ \citenamefont {Opatrný}}]{Welsch1999}%
  \BibitemOpen
  \bibfield  {author} {\bibinfo {author} {\bibfnamefont {D.-G.}\ \bibnamefont
  {Welsch}}, \bibinfo {author} {\bibfnamefont {W.}~\bibnamefont {Vogel}},\ and\
  \bibinfo {author} {\bibfnamefont {T.}~\bibnamefont {Opatrný}},\ }\href
  {https://doi.org/10.1016/S0079-6638(08)70389-5} {\emph {\bibinfo {title} {{II
  Homodyne Detection and Quantum-State Reconstruction}}}},\ \bibinfo {series}
  {Progress in Optics}, Vol.~\bibinfo {volume} {39}\ (\bibinfo  {publisher}
  {Elsevier},\ \bibinfo {year} {1999})\ pp.\ \bibinfo {pages}
  {63--211}\BibitemShut {NoStop}%
\bibitem [{\citenamefont {Floerchinger}\ \emph
  {et~al.}(2021{\natexlab{a}})\citenamefont {Floerchinger}, \citenamefont
  {Haas},\ and\ \citenamefont {M\"uller-Groeling}}]{Haas2021b}%
  \BibitemOpen
  \bibfield  {author} {\bibinfo {author} {\bibfnamefont {S.}~\bibnamefont
  {Floerchinger}}, \bibinfo {author} {\bibfnamefont {T.}~\bibnamefont {Haas}},\
  and\ \bibinfo {author} {\bibfnamefont {H.}~\bibnamefont
  {M\"uller-Groeling}},\ }\bibfield  {title} {\bibinfo {title} {{Wehrl entropy,
  entropic uncertainty relations, and entanglement}},\ }\href
  {https://doi.org/10.1103/PhysRevA.103.062222} {\bibfield  {journal} {\bibinfo
   {journal} {Phys. Rev. A}\ }\textbf {\bibinfo {volume} {103}},\ \bibinfo
  {pages} {062222} (\bibinfo {year} {2021}{\natexlab{a}})}\BibitemShut
  {NoStop}%
\bibitem [{\citenamefont {Floerchinger}\ \emph
  {et~al.}(2022{\natexlab{a}})\citenamefont {Floerchinger}, \citenamefont
  {G\"arttner}, \citenamefont {Haas},\ and\ \citenamefont
  {Stockdale}}]{Haas2022a}%
  \BibitemOpen
  \bibfield  {author} {\bibinfo {author} {\bibfnamefont {S.}~\bibnamefont
  {Floerchinger}}, \bibinfo {author} {\bibfnamefont {M.}~\bibnamefont
  {G\"arttner}}, \bibinfo {author} {\bibfnamefont {T.}~\bibnamefont {Haas}},\
  and\ \bibinfo {author} {\bibfnamefont {O.~R.}\ \bibnamefont {Stockdale}},\
  }\bibfield  {title} {\bibinfo {title} {{Entropic entanglement criteria in
  phase space}},\ }\href {https://doi.org/10.1103/PhysRevA.105.012409}
  {\bibfield  {journal} {\bibinfo  {journal} {Phys. Rev. A}\ }\textbf {\bibinfo
  {volume} {105}},\ \bibinfo {pages} {012409} (\bibinfo {year}
  {2022}{\natexlab{a}})}\BibitemShut {NoStop}%
\bibitem [{\citenamefont {Lieb}\ and\ \citenamefont
  {Solovej}(2014)}]{Lieb2014}%
  \BibitemOpen
  \bibfield  {author} {\bibinfo {author} {\bibfnamefont {E.~H.}\ \bibnamefont
  {Lieb}}\ and\ \bibinfo {author} {\bibfnamefont {J.~P.}\ \bibnamefont
  {Solovej}},\ }\bibfield  {title} {\bibinfo {title} {{Proof of an entropy
  conjecture for Bloch coherent spin states and its generalizations}},\ }\href
  {https://doi.org/10.1007/s11511-014-0113-6} {\bibfield  {journal} {\bibinfo
  {journal} {Acta Math.}\ }\textbf {\bibinfo {volume} {212}},\ \bibinfo {pages}
  {379} (\bibinfo {year} {2014})}\BibitemShut {NoStop}%
\bibitem [{\citenamefont {Bengtsson}\ and\ \citenamefont
  {Zyczkowski}(2017)}]{Bengtsson2017}%
  \BibitemOpen
  \bibfield  {author} {\bibinfo {author} {\bibfnamefont {I.}~\bibnamefont
  {Bengtsson}}\ and\ \bibinfo {author} {\bibfnamefont {K.}~\bibnamefont
  {Zyczkowski}},\ }\href {https://doi.org/10.1017/9781139207010} {\emph
  {\bibinfo {title} {{Geometry of Quantum States, 2nd ed.}}}}\ (\bibinfo
  {publisher} {John Wiley and Sons},\ \bibinfo {year} {2017})\BibitemShut
  {NoStop}%
\bibitem [{\citenamefont {Schupp}(2022)}]{Schupp2022}%
  \BibitemOpen
  \bibfield  {author} {\bibinfo {author} {\bibfnamefont {P.}~\bibnamefont
  {Schupp}},\ }\bibinfo {title} {{Wehrl entropy, coherent states and quantum
  channels}},\ in\ \href {https://doi.org/10.4171/90-2/42} {\emph {\bibinfo
  {booktitle} {{The Physics and Mathematics of Elliott Lieb}}}}\ (\bibinfo
  {publisher} {EMS Press},\ \bibinfo {year} {2022})\ pp.\ \bibinfo {pages}
  {329--344}\BibitemShut {NoStop}%
\bibitem [{\citenamefont {Wehrl}(1978)}]{Wehrl1978}%
  \BibitemOpen
  \bibfield  {author} {\bibinfo {author} {\bibfnamefont {A.}~\bibnamefont
  {Wehrl}},\ }\bibfield  {title} {\bibinfo {title} {{General properties of
  entropy}},\ }\href {https://doi.org/10.1103/RevModPhys.50.221} {\bibfield
  {journal} {\bibinfo  {journal} {Rev. Mod. Phys.}\ }\textbf {\bibinfo {volume}
  {50}},\ \bibinfo {pages} {221} (\bibinfo {year} {1978})}\BibitemShut
  {NoStop}%
\bibitem [{\citenamefont {Wehrl}(1979)}]{Wehrl1979}%
  \BibitemOpen
  \bibfield  {author} {\bibinfo {author} {\bibfnamefont {A.}~\bibnamefont
  {Wehrl}},\ }\bibfield  {title} {\bibinfo {title} {{On the relation between
  classical and quantum-mechanical entropy}},\ }\href
  {https://doi.org/10.1016/0034-4877(79)90070-3} {\bibfield  {journal}
  {\bibinfo  {journal} {Rep. Math. Phys.}\ }\textbf {\bibinfo {volume} {16}},\
  \bibinfo {pages} {353} (\bibinfo {year} {1979})}\BibitemShut {NoStop}%
\bibitem [{\citenamefont {Lieb}(1978)}]{Lieb1978}%
  \BibitemOpen
  \bibfield  {author} {\bibinfo {author} {\bibfnamefont {E.~H.}\ \bibnamefont
  {Lieb}},\ }\bibfield  {title} {\bibinfo {title} {{Proof of an entropy
  conjecture of Wehrl}},\ }\href {https://doi.org/10.1007/BF01940328}
  {\bibfield  {journal} {\bibinfo  {journal} {Commun. Math. Phys.}\ }\textbf
  {\bibinfo {volume} {62}},\ \bibinfo {pages} {35} (\bibinfo {year}
  {1978})}\BibitemShut {NoStop}%
\bibitem [{\citenamefont {Grabowski}(1984)}]{Grabowski1984}%
  \BibitemOpen
  \bibfield  {author} {\bibinfo {author} {\bibfnamefont {M.}~\bibnamefont
  {Grabowski}},\ }\bibfield  {title} {\bibinfo {title} {{Wehrl-Lieb's
  inequality for entropy and the uncertainty relation}},\ }\href
  {https://doi.org/10.1016/0034-4877(84)90029-6} {\bibfield  {journal}
  {\bibinfo  {journal} {Rep. Math. Phys.}\ }\textbf {\bibinfo {volume} {20}},\
  \bibinfo {pages} {153} (\bibinfo {year} {1984})}\BibitemShut {NoStop}%
\bibitem [{\citenamefont {Carlen}(1991)}]{Carlen1991}%
  \BibitemOpen
  \bibfield  {author} {\bibinfo {author} {\bibfnamefont {E.~A.}\ \bibnamefont
  {Carlen}},\ }\bibfield  {title} {\bibinfo {title} {{Some integral identities
  and inequalities for entire functions and their application to the coherent
  state transform}},\ }\href {https://doi.org/10.1016/0022-1236(91)90022-W}
  {\bibfield  {journal} {\bibinfo  {journal} {J. Funct. Anal.}\ }\textbf
  {\bibinfo {volume} {97}},\ \bibinfo {pages} {231} (\bibinfo {year}
  {1991})}\BibitemShut {NoStop}%
\bibitem [{\citenamefont {Luo}(2000)}]{Luo2000}%
  \BibitemOpen
  \bibfield  {author} {\bibinfo {author} {\bibfnamefont {S.}~\bibnamefont
  {Luo}},\ }\bibfield  {title} {\bibinfo {title} {{A simple proof of Wehrl's
  conjecture on entropy}},\ }\href
  {https://doi.org/10.1088/0305-4470/33/16/303} {\bibfield  {journal} {\bibinfo
   {journal} {J. Phys. A Math. Theor.}\ }\textbf {\bibinfo {volume} {33}},\
  \bibinfo {pages} {3093} (\bibinfo {year} {2000})}\BibitemShut {NoStop}%
\bibitem [{\citenamefont {Schupp}(1999)}]{Schupp1999}%
  \BibitemOpen
  \bibfield  {author} {\bibinfo {author} {\bibfnamefont {P.}~\bibnamefont
  {Schupp}},\ }\bibfield  {title} {\bibinfo {title} {{On Lieb's Conjecture for
  the Wehrl Entropy of Bloch Coherent States}},\ }\href
  {https://doi.org/10.1007/s002200050734} {\bibfield  {journal} {\bibinfo
  {journal} {Commun. Math. Phys.}\ }\textbf {\bibinfo {volume} {207}},\
  \bibinfo {pages} {481} (\bibinfo {year} {1999})}\BibitemShut {NoStop}%
\bibitem [{\citenamefont {Lieb}\ and\ \citenamefont
  {Solovej}(2016)}]{Lieb2016}%
  \BibitemOpen
  \bibfield  {author} {\bibinfo {author} {\bibfnamefont {E.~H.}\ \bibnamefont
  {Lieb}}\ and\ \bibinfo {author} {\bibfnamefont {J.~P.}\ \bibnamefont
  {Solovej}},\ }\bibfield  {title} {\bibinfo {title} {{Proof of the Wehrl-type
  Entropy Conjecture for Symmetric ${SU(N)}$ Coherent States}},\ }\href
  {https://doi.org/10.1007/s00220-016-2596-9} {\bibfield  {journal} {\bibinfo
  {journal} {Commun. Math. Phys.}\ }\textbf {\bibinfo {volume} {348}},\
  \bibinfo {pages} {567} (\bibinfo {year} {2016})}\BibitemShut {NoStop}%
\bibitem [{\citenamefont {Lieb}\ and\ \citenamefont
  {Solovej}(2021)}]{Lieb2021}%
  \BibitemOpen
  \bibfield  {author} {\bibinfo {author} {\bibfnamefont {E.~H.}\ \bibnamefont
  {Lieb}}\ and\ \bibinfo {author} {\bibfnamefont {J.~P.}\ \bibnamefont
  {Solovej}},\ }\bibinfo {title} {{Wehrl-type coherent state entropy
  inequalities for SU(1,1) and its AX+B subgroup}},\ in\ \href
  {https://doi.org/10.4171/ECR/18-1/18} {\emph {\bibinfo {booktitle} {Partial
  Differential Equations, Spectral Theory, and Mathematical Physics}}}\
  (\bibinfo {year} {2021})\ p.\ \bibinfo {pages} {301–314}\BibitemShut
  {NoStop}%
\bibitem [{\citenamefont {Kulikov}(2022)}]{Kulikov2022}%
  \BibitemOpen
  \bibfield  {author} {\bibinfo {author} {\bibfnamefont {A.}~\bibnamefont
  {Kulikov}},\ }\bibfield  {title} {\bibinfo {title} {{Functionals with extrema
  at reproducing kernels}},\ }\href
  {https://doi.org/10.1007/s00039-022-00608-5} {\bibfield  {journal} {\bibinfo
  {journal} {Geom. Funct. Ana.}\ }\textbf {\bibinfo {volume} {32}},\ \bibinfo
  {pages} {938} (\bibinfo {year} {2022})}\BibitemShut {NoStop}%
\bibitem [{\citenamefont {Nielsen}(1999)}]{Nielsen1999}%
  \BibitemOpen
  \bibfield  {author} {\bibinfo {author} {\bibfnamefont {M.~A.}\ \bibnamefont
  {Nielsen}},\ }\bibfield  {title} {\bibinfo {title} {{Conditions for a Class
  of Entanglement Transformations}},\ }\href
  {https://doi.org/10.1103/PhysRevLett.83.436} {\bibfield  {journal} {\bibinfo
  {journal} {Phys. Rev. Lett.}\ }\textbf {\bibinfo {volume} {83}},\ \bibinfo
  {pages} {436} (\bibinfo {year} {1999})}\BibitemShut {NoStop}%
\bibitem [{\citenamefont {Hertz}\ and\ \citenamefont {Cerf}(2019)}]{Hertz2019}%
  \BibitemOpen
  \bibfield  {author} {\bibinfo {author} {\bibfnamefont {A.}~\bibnamefont
  {Hertz}}\ and\ \bibinfo {author} {\bibfnamefont {N.~J.}\ \bibnamefont
  {Cerf}},\ }\bibfield  {title} {\bibinfo {title} {Continuous-variable entropic
  uncertainty relations},\ }\href {https://doi.org/10.1088/1751-8121/ab03f3}
  {\bibfield  {journal} {\bibinfo  {journal} {J. Phys. A Math. Theor.}\
  }\textbf {\bibinfo {volume} {52}},\ \bibinfo {pages} {173001} (\bibinfo
  {year} {2019})}\BibitemShut {NoStop}%
\bibitem [{\citenamefont {Van~Herstraeten}\ and\ \citenamefont
  {Cerf}(2021)}]{VanHerstraeten2021a}%
  \BibitemOpen
  \bibfield  {author} {\bibinfo {author} {\bibfnamefont {Z.}~\bibnamefont
  {Van~Herstraeten}}\ and\ \bibinfo {author} {\bibfnamefont {N.~J.}\
  \bibnamefont {Cerf}},\ }\bibfield  {title} {\bibinfo {title} {{Quantum Wigner
  entropy}},\ }\href {https://doi.org/10.1103/PhysRevA.104.042211} {\bibfield
  {journal} {\bibinfo  {journal} {Phys. Rev. A}\ }\textbf {\bibinfo {volume}
  {104}},\ \bibinfo {pages} {042211} (\bibinfo {year} {2021})}\BibitemShut
  {NoStop}%
\bibitem [{\citenamefont {Herstraeten}\ \emph {et~al.}(2021)\citenamefont
  {Herstraeten}, \citenamefont {Jabbour},\ and\ \citenamefont
  {Cerf}}]{VanHerstraeten2021b}%
  \BibitemOpen
  \bibfield  {author} {\bibinfo {author} {\bibfnamefont {Z.~V.}\ \bibnamefont
  {Herstraeten}}, \bibinfo {author} {\bibfnamefont {M.~G.}\ \bibnamefont
  {Jabbour}},\ and\ \bibinfo {author} {\bibfnamefont {N.~J.}\ \bibnamefont
  {Cerf}},\ }\bibfield  {title} {\bibinfo {title} {{Continuous majorization in
  quantum phase space}},\ }\href@noop {} {\bibfield  {journal} {\bibinfo
  {journal} {\href{https://arxiv.org/abs/2108.09167}{arXiv:2108.09167}}\ ,\
  \bibinfo {pages} {1}} (\bibinfo {year} {2021})}\BibitemShut {NoStop}%
\bibitem [{\citenamefont {Van~Herstraeten}(2021)}]{VanHerstraeten2021c}%
  \BibitemOpen
  \bibfield  {author} {\bibinfo {author} {\bibfnamefont {Z.}~\bibnamefont
  {Van~Herstraeten}},\ }\emph {\bibinfo {title} {{Majorization theoretical
  approach to quantum uncertainty}}},\ \href
  {http://quic.ulb.ac.be/_media/publications/2021_phd_thesis_zacharie_van_herstraeten.pdf}
  {Ph.D. thesis},\ \bibinfo  {school} {Université libre de Bruxelles}
  (\bibinfo {year} {2021})\BibitemShut {NoStop}%
\bibitem [{\citenamefont {Leonhardt}\ and\ \citenamefont
  {Paul}(1995)}]{Leonhardt1995}%
  \BibitemOpen
  \bibfield  {author} {\bibinfo {author} {\bibfnamefont {U.}~\bibnamefont
  {Leonhardt}}\ and\ \bibinfo {author} {\bibfnamefont {H.}~\bibnamefont
  {Paul}},\ }\bibfield  {title} {\bibinfo {title} {{Measuring the quantum state
  of light}},\ }\href
  {https://doi.org/https://doi.org/10.1016/0079-6727(94)00007-L} {\bibfield
  {journal} {\bibinfo  {journal} {Prog. Quantum. Electron.}\ }\textbf {\bibinfo
  {volume} {19}},\ \bibinfo {pages} {89} (\bibinfo {year} {1995})}\BibitemShut
  {NoStop}%
\bibitem [{\citenamefont {Opatrn\'y}\ and\ \citenamefont
  {Welsch}(1997)}]{Welsch1997}%
  \BibitemOpen
  \bibfield  {author} {\bibinfo {author} {\bibfnamefont {T.}~\bibnamefont
  {Opatrn\'y}}\ and\ \bibinfo {author} {\bibfnamefont {D.-G.}\ \bibnamefont
  {Welsch}},\ }\bibfield  {title} {\bibinfo {title} {{Density-matrix
  reconstruction by unbalanced homodyning}},\ }\href
  {https://doi.org/10.1103/PhysRevA.55.1462} {\bibfield  {journal} {\bibinfo
  {journal} {Phys. Rev. A}\ }\textbf {\bibinfo {volume} {55}},\ \bibinfo
  {pages} {1462} (\bibinfo {year} {1997})}\BibitemShut {NoStop}%
\bibitem [{\citenamefont {Mancini}\ \emph {et~al.}(1997)\citenamefont
  {Mancini}, \citenamefont {Tombesi},\ and\ \citenamefont
  {Man'ko}}]{Mancini1997}%
  \BibitemOpen
  \bibfield  {author} {\bibinfo {author} {\bibfnamefont {S.}~\bibnamefont
  {Mancini}}, \bibinfo {author} {\bibfnamefont {P.}~\bibnamefont {Tombesi}},\
  and\ \bibinfo {author} {\bibfnamefont {V.~I.}\ \bibnamefont {Man'ko}},\
  }\bibfield  {title} {\bibinfo {title} {{Density matrix from photon number
  tomography}},\ }\href {https://doi.org/10.1209/epl/i1997-00115-8} {\bibfield
  {journal} {\bibinfo  {journal} {EPL}\ }\textbf {\bibinfo {volume} {37}},\
  \bibinfo {pages} {79} (\bibinfo {year} {1997})}\BibitemShut {NoStop}%
\bibitem [{\citenamefont {Shen}\ \emph {et~al.}(2016)\citenamefont {Shen},
  \citenamefont {Heeres}, \citenamefont {Reinhold}, \citenamefont {Jiang},
  \citenamefont {Liu}, \citenamefont {Schoelkopf},\ and\ \citenamefont
  {Jiang}}]{Shen2016}%
  \BibitemOpen
  \bibfield  {author} {\bibinfo {author} {\bibfnamefont {C.}~\bibnamefont
  {Shen}}, \bibinfo {author} {\bibfnamefont {R.~W.}\ \bibnamefont {Heeres}},
  \bibinfo {author} {\bibfnamefont {P.}~\bibnamefont {Reinhold}}, \bibinfo
  {author} {\bibfnamefont {L.}~\bibnamefont {Jiang}}, \bibinfo {author}
  {\bibfnamefont {Y.-K.}\ \bibnamefont {Liu}}, \bibinfo {author} {\bibfnamefont
  {R.~J.}\ \bibnamefont {Schoelkopf}},\ and\ \bibinfo {author} {\bibfnamefont
  {L.}~\bibnamefont {Jiang}},\ }\bibfield  {title} {\bibinfo {title}
  {{Optimized tomography of continuous variable systems using excitation
  counting}},\ }\href {https://doi.org/10.1103/PhysRevA.94.052327} {\bibfield
  {journal} {\bibinfo  {journal} {Phys. Rev. A}\ }\textbf {\bibinfo {volume}
  {94}},\ \bibinfo {pages} {052327} (\bibinfo {year} {2016})}\BibitemShut
  {NoStop}%
\bibitem [{\citenamefont {Kirchmair}\ \emph {et~al.}(2013)\citenamefont
  {Kirchmair}, \citenamefont {Vlastakis}, \citenamefont {Leghtas},
  \citenamefont {Nigg}, \citenamefont {Paik}, \citenamefont {Ginossar},
  \citenamefont {Mirrahimi}, \citenamefont {Frunzio}, \citenamefont {Girvin},\
  and\ \citenamefont {Schoelkopf}}]{Kirchmair2013}%
  \BibitemOpen
  \bibfield  {author} {\bibinfo {author} {\bibfnamefont {G.}~\bibnamefont
  {Kirchmair}}, \bibinfo {author} {\bibfnamefont {B.}~\bibnamefont
  {Vlastakis}}, \bibinfo {author} {\bibfnamefont {Z.}~\bibnamefont {Leghtas}},
  \bibinfo {author} {\bibfnamefont {S.~E.}\ \bibnamefont {Nigg}}, \bibinfo
  {author} {\bibfnamefont {H.}~\bibnamefont {Paik}}, \bibinfo {author}
  {\bibfnamefont {E.}~\bibnamefont {Ginossar}}, \bibinfo {author}
  {\bibfnamefont {M.}~\bibnamefont {Mirrahimi}}, \bibinfo {author}
  {\bibfnamefont {L.}~\bibnamefont {Frunzio}}, \bibinfo {author} {\bibfnamefont
  {S.~M.}\ \bibnamefont {Girvin}},\ and\ \bibinfo {author} {\bibfnamefont
  {R.~J.}\ \bibnamefont {Schoelkopf}},\ }\bibfield  {title} {\bibinfo {title}
  {{Observation of quantum state collapse and revival due to the single-photon
  Kerr effect}},\ }\href {https://doi.org/10.1038/nature11902} {\bibfield
  {journal} {\bibinfo  {journal} {Nature}\ }\textbf {\bibinfo {volume} {495}},\
  \bibinfo {pages} {205} (\bibinfo {year} {2013})}\BibitemShut {NoStop}%
\bibitem [{\citenamefont {Wang}\ \emph {et~al.}(2016)\citenamefont {Wang},
  \citenamefont {Gao}, \citenamefont {Reinhold}, \citenamefont {Heeres},
  \citenamefont {Ofek}, \citenamefont {Chou}, \citenamefont {Axline},
  \citenamefont {Reagor}, \citenamefont {Blumoff}, \citenamefont {Sliwa},
  \citenamefont {Frunzio}, \citenamefont {Girvin}, \citenamefont {Jiang},
  \citenamefont {Mirrahimi}, \citenamefont {Devoret},\ and\ \citenamefont
  {Schoelkopf}}]{Wang2016}%
  \BibitemOpen
  \bibfield  {author} {\bibinfo {author} {\bibfnamefont {C.}~\bibnamefont
  {Wang}}, \bibinfo {author} {\bibfnamefont {Y.~Y.}\ \bibnamefont {Gao}},
  \bibinfo {author} {\bibfnamefont {P.}~\bibnamefont {Reinhold}}, \bibinfo
  {author} {\bibfnamefont {R.~W.}\ \bibnamefont {Heeres}}, \bibinfo {author}
  {\bibfnamefont {N.}~\bibnamefont {Ofek}}, \bibinfo {author} {\bibfnamefont
  {K.}~\bibnamefont {Chou}}, \bibinfo {author} {\bibfnamefont {C.}~\bibnamefont
  {Axline}}, \bibinfo {author} {\bibfnamefont {M.}~\bibnamefont {Reagor}},
  \bibinfo {author} {\bibfnamefont {J.}~\bibnamefont {Blumoff}}, \bibinfo
  {author} {\bibfnamefont {K.~M.}\ \bibnamefont {Sliwa}}, \bibinfo {author}
  {\bibfnamefont {L.}~\bibnamefont {Frunzio}}, \bibinfo {author} {\bibfnamefont
  {S.~M.}\ \bibnamefont {Girvin}}, \bibinfo {author} {\bibfnamefont
  {L.}~\bibnamefont {Jiang}}, \bibinfo {author} {\bibfnamefont
  {M.}~\bibnamefont {Mirrahimi}}, \bibinfo {author} {\bibfnamefont {M.~H.}\
  \bibnamefont {Devoret}},\ and\ \bibinfo {author} {\bibfnamefont {R.~J.}\
  \bibnamefont {Schoelkopf}},\ }\bibfield  {title} {\bibinfo {title} {{A
  Schrödinger cat living in two boxes}},\ }\href
  {https://doi.org/10.1126/science.aaf2941} {\bibfield  {journal} {\bibinfo
  {journal} {Science}\ }\textbf {\bibinfo {volume} {352}},\ \bibinfo {pages}
  {1087} (\bibinfo {year} {2016})}\BibitemShut {NoStop}%
\bibitem [{\citenamefont {Haas}\ \emph {et~al.}(2014)\citenamefont {Haas},
  \citenamefont {Volz}, \citenamefont {Gehr}, \citenamefont {Reichel},\ and\
  \citenamefont {Est{\`e}ve}}]{Haas2014}%
  \BibitemOpen
  \bibfield  {author} {\bibinfo {author} {\bibfnamefont {F.}~\bibnamefont
  {Haas}}, \bibinfo {author} {\bibfnamefont {J.}~\bibnamefont {Volz}}, \bibinfo
  {author} {\bibfnamefont {R.}~\bibnamefont {Gehr}}, \bibinfo {author}
  {\bibfnamefont {J.}~\bibnamefont {Reichel}},\ and\ \bibinfo {author}
  {\bibfnamefont {J.}~\bibnamefont {Est{\`e}ve}},\ }\bibfield  {title}
  {\bibinfo {title} {{Entangled states of more than 40 atoms in an optical
  fiber cavity}},\ }\href {https://doi.org/10.1126/science.1248905} {\bibfield
  {journal} {\bibinfo  {journal} {Science}\ }\textbf {\bibinfo {volume}
  {344}},\ \bibinfo {pages} {180} (\bibinfo {year} {2014})}\BibitemShut
  {NoStop}%
\bibitem [{\citenamefont {Barontini}\ \emph {et~al.}(2015)\citenamefont
  {Barontini}, \citenamefont {Hohmann}, \citenamefont {Haas}, \citenamefont
  {Est{\`e}ve},\ and\ \citenamefont {Reichel}}]{Barontini2015}%
  \BibitemOpen
  \bibfield  {author} {\bibinfo {author} {\bibfnamefont {G.}~\bibnamefont
  {Barontini}}, \bibinfo {author} {\bibfnamefont {L.}~\bibnamefont {Hohmann}},
  \bibinfo {author} {\bibfnamefont {F.}~\bibnamefont {Haas}}, \bibinfo {author}
  {\bibfnamefont {J.}~\bibnamefont {Est{\`e}ve}},\ and\ \bibinfo {author}
  {\bibfnamefont {J.}~\bibnamefont {Reichel}},\ }\bibfield  {title} {\bibinfo
  {title} {{Deterministic generation of multiparticle entanglement by quantum
  Zeno dynamics}},\ }\href {https://doi.org/10.1126/science.aaa0754} {\bibfield
   {journal} {\bibinfo  {journal} {Science}\ }\textbf {\bibinfo {volume}
  {349}},\ \bibinfo {pages} {1317} (\bibinfo {year} {2015})}\BibitemShut
  {NoStop}%
\bibitem [{\citenamefont {Leibfried}\ \emph {et~al.}(1996)\citenamefont
  {Leibfried}, \citenamefont {Meekhof}, \citenamefont {King}, \citenamefont
  {Monroe}, \citenamefont {Itano},\ and\ \citenamefont
  {Wineland}}]{Leibfried1996}%
  \BibitemOpen
  \bibfield  {author} {\bibinfo {author} {\bibfnamefont {D.}~\bibnamefont
  {Leibfried}}, \bibinfo {author} {\bibfnamefont {D.~M.}\ \bibnamefont
  {Meekhof}}, \bibinfo {author} {\bibfnamefont {B.~E.}\ \bibnamefont {King}},
  \bibinfo {author} {\bibfnamefont {C.}~\bibnamefont {Monroe}}, \bibinfo
  {author} {\bibfnamefont {W.~M.}\ \bibnamefont {Itano}},\ and\ \bibinfo
  {author} {\bibfnamefont {D.~J.}\ \bibnamefont {Wineland}},\ }\bibfield
  {title} {\bibinfo {title} {{Experimental Determination of the Motional
  Quantum State of a Trapped Atom}},\ }\href
  {https://doi.org/10.1103/PhysRevLett.77.4281} {\bibfield  {journal} {\bibinfo
   {journal} {Phys. Rev. Lett.}\ }\textbf {\bibinfo {volume} {77}},\ \bibinfo
  {pages} {4281} (\bibinfo {year} {1996})}\BibitemShut {NoStop}%
\bibitem [{\citenamefont {G{\"a}rttner}\ \emph {et~al.}(2017)\citenamefont
  {G{\"a}rttner}, \citenamefont {Bohnet}, \citenamefont {Safavi-Naini},
  \citenamefont {Wall}, \citenamefont {Bollinger},\ and\ \citenamefont
  {Rey}}]{Gaerttner2017}%
  \BibitemOpen
  \bibfield  {author} {\bibinfo {author} {\bibfnamefont {M.}~\bibnamefont
  {G{\"a}rttner}}, \bibinfo {author} {\bibfnamefont {J.~G.}\ \bibnamefont
  {Bohnet}}, \bibinfo {author} {\bibfnamefont {A.}~\bibnamefont
  {Safavi-Naini}}, \bibinfo {author} {\bibfnamefont {M.~L.}\ \bibnamefont
  {Wall}}, \bibinfo {author} {\bibfnamefont {J.~J.}\ \bibnamefont
  {Bollinger}},\ and\ \bibinfo {author} {\bibfnamefont {A.~M.}\ \bibnamefont
  {Rey}},\ }\bibfield  {title} {\bibinfo {title} {{Measuring out-of-time-order
  correlations and multiple quantum spectra in a trapped-ion quantum magnet}},\
  }\href {https://doi.org/10.1038/nphys4119} {\bibfield  {journal} {\bibinfo
  {journal} {Nat. Phys.}\ }\textbf {\bibinfo {volume} {13}},\ \bibinfo {pages}
  {781} (\bibinfo {year} {2017})}\BibitemShut {NoStop}%
\bibitem [{\citenamefont {Landon-Cardinal}\ \emph {et~al.}(2018)\citenamefont
  {Landon-Cardinal}, \citenamefont {Govia},\ and\ \citenamefont
  {Clerk}}]{Landon2018}%
  \BibitemOpen
  \bibfield  {author} {\bibinfo {author} {\bibfnamefont {O.}~\bibnamefont
  {Landon-Cardinal}}, \bibinfo {author} {\bibfnamefont {L.~C.~G.}\ \bibnamefont
  {Govia}},\ and\ \bibinfo {author} {\bibfnamefont {A.~A.}\ \bibnamefont
  {Clerk}},\ }\bibfield  {title} {\bibinfo {title} {{Quantitative Tomography
  for Continuous Variable Quantum Systems}},\ }\href
  {https://doi.org/10.1103/PhysRevLett.120.090501} {\bibfield  {journal}
  {\bibinfo  {journal} {Phys. Rev. Lett.}\ }\textbf {\bibinfo {volume} {120}},\
  \bibinfo {pages} {090501} (\bibinfo {year} {2018})}\BibitemShut {NoStop}%
\bibitem [{\citenamefont {Mandel}\ and\ \citenamefont
  {Wolf}(2013)}]{Mandel2013}%
  \BibitemOpen
  \bibfield  {author} {\bibinfo {author} {\bibfnamefont {L.}~\bibnamefont
  {Mandel}}\ and\ \bibinfo {author} {\bibfnamefont {E.}~\bibnamefont {Wolf}},\
  }\href {https://doi.org/10.1017/CBO9781139644105} {\emph {\bibinfo {title}
  {{Optical Coherence and Quantum Optics}}}}\ (\bibinfo  {publisher} {Cambridge
  University Press},\ \bibinfo {year} {2013})\BibitemShut {NoStop}%
\bibitem [{\citenamefont {Stenholm}(1992)}]{Stenholm1992}%
  \BibitemOpen
  \bibfield  {author} {\bibinfo {author} {\bibfnamefont {S.}~\bibnamefont
  {Stenholm}},\ }\bibfield  {title} {\bibinfo {title} {{Simultaneous
  measurement of conjugate variables}},\ }\href
  {https://doi.org/https://doi.org/10.1016/0003-4916(92)90086-2} {\bibfield
  {journal} {\bibinfo  {journal} {Ann. Phys.}\ }\textbf {\bibinfo {volume}
  {218}},\ \bibinfo {pages} {233} (\bibinfo {year} {1992})}\BibitemShut
  {NoStop}%
\bibitem [{\citenamefont {Noh}\ \emph {et~al.}(1991)\citenamefont {Noh},
  \citenamefont {Foug\`eres},\ and\ \citenamefont {Mandel}}]{Noh1991}%
  \BibitemOpen
  \bibfield  {author} {\bibinfo {author} {\bibfnamefont {J.~W.}\ \bibnamefont
  {Noh}}, \bibinfo {author} {\bibfnamefont {A.}~\bibnamefont {Foug\`eres}},\
  and\ \bibinfo {author} {\bibfnamefont {L.}~\bibnamefont {Mandel}},\
  }\bibfield  {title} {\bibinfo {title} {{Measurement of the quantum phase by
  photon counting}},\ }\href {https://doi.org/10.1103/PhysRevLett.67.1426}
  {\bibfield  {journal} {\bibinfo  {journal} {Phys. Rev. Lett.}\ }\textbf
  {\bibinfo {volume} {67}},\ \bibinfo {pages} {1426} (\bibinfo {year}
  {1991})}\BibitemShut {NoStop}%
\bibitem [{\citenamefont {Noh}\ \emph {et~al.}(1992)\citenamefont {Noh},
  \citenamefont {Foug\`eres},\ and\ \citenamefont {Mandel}}]{Noh1992}%
  \BibitemOpen
  \bibfield  {author} {\bibinfo {author} {\bibfnamefont {J.~W.}\ \bibnamefont
  {Noh}}, \bibinfo {author} {\bibfnamefont {A.}~\bibnamefont {Foug\`eres}},\
  and\ \bibinfo {author} {\bibfnamefont {L.}~\bibnamefont {Mandel}},\
  }\bibfield  {title} {\bibinfo {title} {{Operational approach to the phase of
  a quantum field}},\ }\href {https://doi.org/10.1103/PhysRevA.45.424}
  {\bibfield  {journal} {\bibinfo  {journal} {Phys. Rev. A}\ }\textbf {\bibinfo
  {volume} {45}},\ \bibinfo {pages} {424} (\bibinfo {year} {1992})}\BibitemShut
  {NoStop}%
\bibitem [{\citenamefont {Leonhardt}\ and\ \citenamefont
  {Paul}(1993)}]{Leonhardt1993}%
  \BibitemOpen
  \bibfield  {author} {\bibinfo {author} {\bibfnamefont {U.}~\bibnamefont
  {Leonhardt}}\ and\ \bibinfo {author} {\bibfnamefont {H.}~\bibnamefont
  {Paul}},\ }\bibfield  {title} {\bibinfo {title} {{Phase measurement and Q
  function}},\ }\href {https://doi.org/10.1103/PhysRevA.47.R2460} {\bibfield
  {journal} {\bibinfo  {journal} {Phys. Rev. A}\ }\textbf {\bibinfo {volume}
  {47}},\ \bibinfo {pages} {R2460} (\bibinfo {year} {1993})}\BibitemShut
  {NoStop}%
\bibitem [{\citenamefont {M\"uller}\ \emph {et~al.}(2016)\citenamefont
  {M\"uller}, \citenamefont {Peuntinger}, \citenamefont {Dirmeier},
  \citenamefont {Khan}, \citenamefont {Vogl}, \citenamefont {Marquardt},
  \citenamefont {Leuchs}, \citenamefont {S\'anchez-Soto}, \citenamefont {Teo},
  \citenamefont {Hradil},\ and\ \citenamefont {\ifmmode \check{R}\else
  \v{R}\fi{}eh\'a\ifmmode~\check{c}\else \v{c}\fi{}ek}}]{Mueller2016}%
  \BibitemOpen
  \bibfield  {author} {\bibinfo {author} {\bibfnamefont {C.~R.}\ \bibnamefont
  {M\"uller}}, \bibinfo {author} {\bibfnamefont {C.}~\bibnamefont
  {Peuntinger}}, \bibinfo {author} {\bibfnamefont {T.}~\bibnamefont
  {Dirmeier}}, \bibinfo {author} {\bibfnamefont {I.}~\bibnamefont {Khan}},
  \bibinfo {author} {\bibfnamefont {U.}~\bibnamefont {Vogl}}, \bibinfo {author}
  {\bibfnamefont {C.}~\bibnamefont {Marquardt}}, \bibinfo {author}
  {\bibfnamefont {G.}~\bibnamefont {Leuchs}}, \bibinfo {author} {\bibfnamefont
  {L.~L.}\ \bibnamefont {S\'anchez-Soto}}, \bibinfo {author} {\bibfnamefont
  {Y.~S.}\ \bibnamefont {Teo}}, \bibinfo {author} {\bibfnamefont
  {Z.}~\bibnamefont {Hradil}},\ and\ \bibinfo {author} {\bibfnamefont
  {J.}~\bibnamefont {\ifmmode \check{R}\else
  \v{R}\fi{}eh\'a\ifmmode~\check{c}\else \v{c}\fi{}ek}},\ }\bibfield  {title}
  {\bibinfo {title} {{Evading Vacuum Noise: Wigner Projections or Husimi
  Samples?}},\ }\href {https://doi.org/10.1103/PhysRevLett.117.070801}
  {\bibfield  {journal} {\bibinfo  {journal} {Phys. Rev. Lett.}\ }\textbf
  {\bibinfo {volume} {117}},\ \bibinfo {pages} {070801} (\bibinfo {year}
  {2016})}\BibitemShut {NoStop}%
\bibitem [{\citenamefont {Kunkel}\ \emph {et~al.}(2019)\citenamefont {Kunkel},
  \citenamefont {Pr\"ufer}, \citenamefont {Lannig}, \citenamefont
  {Rosa-Medina}, \citenamefont {Bonnin}, \citenamefont {G\"arttner},
  \citenamefont {Strobel},\ and\ \citenamefont {Oberthaler}}]{Kunkel2019}%
  \BibitemOpen
  \bibfield  {author} {\bibinfo {author} {\bibfnamefont {P.}~\bibnamefont
  {Kunkel}}, \bibinfo {author} {\bibfnamefont {M.}~\bibnamefont {Pr\"ufer}},
  \bibinfo {author} {\bibfnamefont {S.}~\bibnamefont {Lannig}}, \bibinfo
  {author} {\bibfnamefont {R.}~\bibnamefont {Rosa-Medina}}, \bibinfo {author}
  {\bibfnamefont {A.}~\bibnamefont {Bonnin}}, \bibinfo {author} {\bibfnamefont
  {M.}~\bibnamefont {G\"arttner}}, \bibinfo {author} {\bibfnamefont
  {H.}~\bibnamefont {Strobel}},\ and\ \bibinfo {author} {\bibfnamefont {M.~K.}\
  \bibnamefont {Oberthaler}},\ }\bibfield  {title} {\bibinfo {title}
  {{Simultaneous Readout of Noncommuting Collective Spin Observables beyond the
  Standard Quantum Limit}},\ }\href
  {https://doi.org/10.1103/PhysRevLett.123.063603} {\bibfield  {journal}
  {\bibinfo  {journal} {Phys. Rev. Lett.}\ }\textbf {\bibinfo {volume} {123}},\
  \bibinfo {pages} {063603} (\bibinfo {year} {2019})}\BibitemShut {NoStop}%
\bibitem [{\citenamefont {Kunkel}\ \emph {et~al.}(2022)\citenamefont {Kunkel},
  \citenamefont {Pr\"ufer}, \citenamefont {Lannig}, \citenamefont {Strohmaier},
  \citenamefont {G\"arttner}, \citenamefont {Strobel},\ and\ \citenamefont
  {Oberthaler}}]{Kunkel2021}%
  \BibitemOpen
  \bibfield  {author} {\bibinfo {author} {\bibfnamefont {P.}~\bibnamefont
  {Kunkel}}, \bibinfo {author} {\bibfnamefont {M.}~\bibnamefont {Pr\"ufer}},
  \bibinfo {author} {\bibfnamefont {S.}~\bibnamefont {Lannig}}, \bibinfo
  {author} {\bibfnamefont {R.}~\bibnamefont {Strohmaier}}, \bibinfo {author}
  {\bibfnamefont {M.}~\bibnamefont {G\"arttner}}, \bibinfo {author}
  {\bibfnamefont {H.}~\bibnamefont {Strobel}},\ and\ \bibinfo {author}
  {\bibfnamefont {M.~K.}\ \bibnamefont {Oberthaler}},\ }\bibfield  {title}
  {\bibinfo {title} {{Detecting Entanglement Structure in Continuous Many-Body
  Quantum Systems}},\ }\href {https://doi.org/10.1103/PhysRevLett.128.020402}
  {\bibfield  {journal} {\bibinfo  {journal} {Phys. Rev. Lett.}\ }\textbf
  {\bibinfo {volume} {128}},\ \bibinfo {pages} {020402} (\bibinfo {year}
  {2022})}\BibitemShut {NoStop}%
\bibitem [{\citenamefont {G\"arttner}\ \emph {et~al.}(2023)\citenamefont
  {G\"arttner}, \citenamefont {Haas},\ and\ \citenamefont {Noll}}]{PRL}%
  \BibitemOpen
  \bibfield  {author} {\bibinfo {author} {\bibfnamefont {M.}~\bibnamefont
  {G\"arttner}}, \bibinfo {author} {\bibfnamefont {T.}~\bibnamefont {Haas}},\
  and\ \bibinfo {author} {\bibfnamefont {J.}~\bibnamefont {Noll}},\ }\bibfield
  {title} {\bibinfo {title} {{General Class of Continuous Variable Entanglement
  Criteria}},\ }\href {https://doi.org/10.1103/PhysRevLett.131.150201}
  {\bibfield  {journal} {\bibinfo  {journal} {Phys. Rev. Lett.}\ }\textbf
  {\bibinfo {volume} {131}},\ \bibinfo {pages} {150201} (\bibinfo {year}
  {2023})}\BibitemShut {NoStop}%
\bibitem [{\citenamefont {Gerving}\ \emph {et~al.}(2012)\citenamefont
  {Gerving}, \citenamefont {Hoang}, \citenamefont {Land}, \citenamefont
  {Anquez}, \citenamefont {Hamley},\ and\ \citenamefont
  {Chapman}}]{Gerving2012}%
  \BibitemOpen
  \bibfield  {author} {\bibinfo {author} {\bibfnamefont {C.}~\bibnamefont
  {Gerving}}, \bibinfo {author} {\bibfnamefont {T.}~\bibnamefont {Hoang}},
  \bibinfo {author} {\bibfnamefont {B.}~\bibnamefont {Land}}, \bibinfo {author}
  {\bibfnamefont {M.}~\bibnamefont {Anquez}}, \bibinfo {author} {\bibfnamefont
  {C.}~\bibnamefont {Hamley}},\ and\ \bibinfo {author} {\bibfnamefont
  {M.}~\bibnamefont {Chapman}},\ }\bibfield  {title} {\bibinfo {title}
  {{Non-equilibrium dynamics of an unstable quantum pendulum explored in a
  spin-1 Bose–Einstein condensate}},\ }\href
  {https://doi.org/10.1038/ncomms2179} {\bibfield  {journal} {\bibinfo
  {journal} {Nat. Commun.}\ }\textbf {\bibinfo {volume} {3}},\ \bibinfo {pages}
  {1169} (\bibinfo {year} {2012})}\BibitemShut {NoStop}%
\bibitem [{\citenamefont {Hamley}\ \emph {et~al.}(2012)\citenamefont {Hamley},
  \citenamefont {Gerving}, \citenamefont {Hoang}, \citenamefont {Bookjans},\
  and\ \citenamefont {Chapman}}]{Hamley2012}%
  \BibitemOpen
  \bibfield  {author} {\bibinfo {author} {\bibfnamefont {C.~D.}\ \bibnamefont
  {Hamley}}, \bibinfo {author} {\bibfnamefont {C.~S.}\ \bibnamefont {Gerving}},
  \bibinfo {author} {\bibfnamefont {T.~M.}\ \bibnamefont {Hoang}}, \bibinfo
  {author} {\bibfnamefont {E.~M.}\ \bibnamefont {Bookjans}},\ and\ \bibinfo
  {author} {\bibfnamefont {M.~S.}\ \bibnamefont {Chapman}},\ }\bibfield
  {title} {\bibinfo {title} {{Spin-nematic squeezed vacuum in a quantum gas}},\
  }\href {https://doi.org/10.1038/nphys2245} {\bibfield  {journal} {\bibinfo
  {journal} {Nat. Phys.}\ }\textbf {\bibinfo {volume} {8}},\ \bibinfo {pages}
  {305} (\bibinfo {year} {2012})}\BibitemShut {NoStop}%
\bibitem [{\citenamefont {Kunkel}\ \emph {et~al.}(2018)\citenamefont {Kunkel},
  \citenamefont {Prüfer}, \citenamefont {Strobel}, \citenamefont {Linnemann},
  \citenamefont {Fr{\"o}lian}, \citenamefont {Gasenzer}, \citenamefont
  {Gärttner},\ and\ \citenamefont {Oberthaler}}]{Kunkel2018}%
  \BibitemOpen
  \bibfield  {author} {\bibinfo {author} {\bibfnamefont {P.}~\bibnamefont
  {Kunkel}}, \bibinfo {author} {\bibfnamefont {M.}~\bibnamefont {Prüfer}},
  \bibinfo {author} {\bibfnamefont {H.}~\bibnamefont {Strobel}}, \bibinfo
  {author} {\bibfnamefont {D.}~\bibnamefont {Linnemann}}, \bibinfo {author}
  {\bibfnamefont {A.}~\bibnamefont {Fr{\"o}lian}}, \bibinfo {author}
  {\bibfnamefont {T.}~\bibnamefont {Gasenzer}}, \bibinfo {author}
  {\bibfnamefont {M.}~\bibnamefont {Gärttner}},\ and\ \bibinfo {author}
  {\bibfnamefont {M.~K.}\ \bibnamefont {Oberthaler}},\ }\bibfield  {title}
  {\bibinfo {title} {{Spatially distributed multipartite entanglement enables
  EPR steering of atomic clouds}},\ }\href
  {https://doi.org/10.1126/science.aao2254} {\bibfield  {journal} {\bibinfo
  {journal} {Science}\ }\textbf {\bibinfo {volume} {360}},\ \bibinfo {pages}
  {413} (\bibinfo {year} {2018})}\BibitemShut {NoStop}%
\bibitem [{\citenamefont {Zhang}\ \emph {et~al.}(1990)\citenamefont {Zhang},
  \citenamefont {Feng},\ and\ \citenamefont {Gilmore}}]{Zhang1990}%
  \BibitemOpen
  \bibfield  {author} {\bibinfo {author} {\bibfnamefont {W.-M.}\ \bibnamefont
  {Zhang}}, \bibinfo {author} {\bibfnamefont {D.~H.}\ \bibnamefont {Feng}},\
  and\ \bibinfo {author} {\bibfnamefont {R.}~\bibnamefont {Gilmore}},\
  }\bibfield  {title} {\bibinfo {title} {{Coherent states: Theory and some
  applications}},\ }\href {https://doi.org/10.1103/RevModPhys.62.867}
  {\bibfield  {journal} {\bibinfo  {journal} {Rev. Mod. Phys.}\ }\textbf
  {\bibinfo {volume} {62}},\ \bibinfo {pages} {867} (\bibinfo {year}
  {1990})}\BibitemShut {NoStop}%
\bibitem [{\citenamefont {Radcliffe}(1971)}]{Radcliffe1971}%
  \BibitemOpen
  \bibfield  {author} {\bibinfo {author} {\bibfnamefont {J.~M.}\ \bibnamefont
  {Radcliffe}},\ }\bibfield  {title} {\bibinfo {title} {{Some properties of
  coherent spin states}},\ }\href {https://doi.org/10.1088/0305-4470/4/3/009}
  {\bibfield  {journal} {\bibinfo  {journal} {J. Phys. A}\ }\textbf {\bibinfo
  {volume} {4}},\ \bibinfo {pages} {313} (\bibinfo {year} {1971})}\BibitemShut
  {NoStop}%
\bibitem [{\citenamefont {Gilmore}(1974)}]{Gilmore1974}%
  \BibitemOpen
  \bibfield  {author} {\bibinfo {author} {\bibfnamefont {R.}~\bibnamefont
  {Gilmore}},\ }\bibfield  {title} {\bibinfo {title} {{On Properties Of
  Coherent States}},\ }\href
  {http://www.physics.drexel.edu/~bob/GroupTheory/Prop_Coh_States.pdf}
  {\bibfield  {journal} {\bibinfo  {journal} {Rev. Mex. de Fis.}\ }\textbf
  {\bibinfo {volume} {23}},\ \bibinfo {pages} {143} (\bibinfo {year}
  {1974})}\BibitemShut {NoStop}%
\bibitem [{\citenamefont {Schleich}(2001)}]{Schleich2001}%
  \BibitemOpen
  \bibfield  {author} {\bibinfo {author} {\bibfnamefont {W.~P.}\ \bibnamefont
  {Schleich}},\ }\href {https://doi.org/10.1002/3527602976} {\emph {\bibinfo
  {title} {{Quantum Optics in Phase Space}}}}\ (\bibinfo  {publisher}
  {Wiley‐VCH Verlag Berlin},\ \bibinfo {year} {2001})\BibitemShut {NoStop}%
\bibitem [{\citenamefont {Heisenberg}(1927)}]{Heisenberg1927}%
  \BibitemOpen
  \bibfield  {author} {\bibinfo {author} {\bibfnamefont {W.}~\bibnamefont
  {Heisenberg}},\ }\bibfield  {title} {\bibinfo {title} {{{\"U}ber den
  anschaulichen Inhalt der quantentheoretischen Kinematik und Mechanik}},\
  }\href {https://doi.org/10.1007/BF01397280} {\bibfield  {journal} {\bibinfo
  {journal} {Z. Phys.}\ }\textbf {\bibinfo {volume} {43}},\ \bibinfo {pages}
  {172} (\bibinfo {year} {1927})}\BibitemShut {NoStop}%
\bibitem [{\citenamefont {Kennard}(1927)}]{Kennard1927}%
  \BibitemOpen
  \bibfield  {author} {\bibinfo {author} {\bibfnamefont {E.~H.}\ \bibnamefont
  {Kennard}},\ }\bibfield  {title} {\bibinfo {title} {{Zur Quantenmechanik
  einfacher Bewegungs-typen}},\ }\href {https://doi.org/10.1007/BF01391200}
  {\bibfield  {journal} {\bibinfo  {journal} {Z. Phys.}\ }\textbf {\bibinfo
  {volume} {44}},\ \bibinfo {pages} {326} (\bibinfo {year} {1927})}\BibitemShut
  {NoStop}%
\bibitem [{\citenamefont {Schrödinger}(1930)}]{Schroedinger1930}%
  \BibitemOpen
  \bibfield  {author} {\bibinfo {author} {\bibfnamefont {E.}~\bibnamefont
  {Schrödinger}},\ }\bibfield  {title} {\bibinfo {title} {{Zum Heisenbergschen
  Unschärfeprinzip}},\ }\href@noop {} {\bibfield  {journal} {\bibinfo
  {journal} {Sitzungsberichte der Preußischen Akademie der Wissenschaften.
  Physikalisch-mathematische Klasse}\ }\textbf {\bibinfo {volume} {14}},\
  \bibinfo {pages} {296} (\bibinfo {year} {1930})}\BibitemShut {NoStop}%
\bibitem [{\citenamefont {Robertson}(1929)}]{Robertson1929}%
  \BibitemOpen
  \bibfield  {author} {\bibinfo {author} {\bibfnamefont {H.~P.}\ \bibnamefont
  {Robertson}},\ }\bibfield  {title} {\bibinfo {title} {{The Uncertainty
  Principle}},\ }\href {https://doi.org/10.1103/PhysRev.34.163} {\bibfield
  {journal} {\bibinfo  {journal} {Phys. Rev.}\ }\textbf {\bibinfo {volume}
  {34}},\ \bibinfo {pages} {163} (\bibinfo {year} {1929})}\BibitemShut
  {NoStop}%
\bibitem [{\citenamefont {Robertson}(1930)}]{Robertson1930}%
  \BibitemOpen
  \bibfield  {author} {\bibinfo {author} {\bibfnamefont {H.~P.}\ \bibnamefont
  {Robertson}},\ }\bibfield  {title} {\bibinfo {title} {A general formulation
  of the uncertainty principle and its classical interpretation},\ }\href@noop
  {} {\bibfield  {journal} {\bibinfo  {journal} {Phys. Rev.}\ }\textbf
  {\bibinfo {volume} {35}},\ \bibinfo {pages} {667} (\bibinfo {year}
  {1930})}\BibitemShut {NoStop}%
\bibitem [{\citenamefont {Maassen}\ and\ \citenamefont
  {Uffink}(1988)}]{Maassen1988}%
  \BibitemOpen
  \bibfield  {author} {\bibinfo {author} {\bibfnamefont {H.}~\bibnamefont
  {Maassen}}\ and\ \bibinfo {author} {\bibfnamefont {J.~B.~M.}\ \bibnamefont
  {Uffink}},\ }\bibfield  {title} {\bibinfo {title} {{Generalized entropic
  uncertainty relations}},\ }\href
  {https://doi.org/10.1103/PhysRevLett.60.1103} {\bibfield  {journal} {\bibinfo
   {journal} {Phys. Rev. Lett.}\ }\textbf {\bibinfo {volume} {60}},\ \bibinfo
  {pages} {1103} (\bibinfo {year} {1988})}\BibitemShut {NoStop}%
\bibitem [{\citenamefont {Bia{\l}ynicki-Birula}\ and\ \citenamefont
  {Mycielski}(1975)}]{Bialynicki-Birula1975}%
  \BibitemOpen
  \bibfield  {author} {\bibinfo {author} {\bibfnamefont {I.}~\bibnamefont
  {Bia{\l}ynicki-Birula}}\ and\ \bibinfo {author} {\bibfnamefont
  {J.}~\bibnamefont {Mycielski}},\ }\bibfield  {title} {\bibinfo {title}
  {{Uncertainty relations for information entropy in wave mechanics}},\ }\href
  {https://doi.org/10.1007/BF01608825} {\bibfield  {journal} {\bibinfo
  {journal} {Commun. Math. Phys.}\ }\textbf {\bibinfo {volume} {44}},\ \bibinfo
  {pages} {129} (\bibinfo {year} {1975})}\BibitemShut {NoStop}%
\bibitem [{\citenamefont {Coles}\ \emph {et~al.}(2017)\citenamefont {Coles},
  \citenamefont {Berta}, \citenamefont {Tomamichel},\ and\ \citenamefont
  {Wehner}}]{Coles2017}%
  \BibitemOpen
  \bibfield  {author} {\bibinfo {author} {\bibfnamefont {P.~J.}\ \bibnamefont
  {Coles}}, \bibinfo {author} {\bibfnamefont {M.}~\bibnamefont {Berta}},
  \bibinfo {author} {\bibfnamefont {M.}~\bibnamefont {Tomamichel}},\ and\
  \bibinfo {author} {\bibfnamefont {S.}~\bibnamefont {Wehner}},\ }\bibfield
  {title} {\bibinfo {title} {{Entropic uncertainty relations and their
  applications}},\ }\href {https://doi.org/10.1103/RevModPhys.89.015002}
  {\bibfield  {journal} {\bibinfo  {journal} {Rev. Mod. Phys.}\ }\textbf
  {\bibinfo {volume} {89}},\ \bibinfo {pages} {015002} (\bibinfo {year}
  {2017})}\BibitemShut {NoStop}%
\bibitem [{\citenamefont {Floerchinger}\ \emph
  {et~al.}(2021{\natexlab{b}})\citenamefont {Floerchinger}, \citenamefont
  {Haas},\ and\ \citenamefont {Hoeber}}]{Haas2021a}%
  \BibitemOpen
  \bibfield  {author} {\bibinfo {author} {\bibfnamefont {S.}~\bibnamefont
  {Floerchinger}}, \bibinfo {author} {\bibfnamefont {T.}~\bibnamefont {Haas}},\
  and\ \bibinfo {author} {\bibfnamefont {B.}~\bibnamefont {Hoeber}},\
  }\bibfield  {title} {\bibinfo {title} {{Relative entropic uncertainty
  relation}},\ }\href {https://doi.org/10.1103/PhysRevA.103.062209} {\bibfield
  {journal} {\bibinfo  {journal} {Phys. Rev. A}\ }\textbf {\bibinfo {volume}
  {103}},\ \bibinfo {pages} {062209} (\bibinfo {year}
  {2021}{\natexlab{b}})}\BibitemShut {NoStop}%
\bibitem [{\citenamefont {Floerchinger}\ \emph
  {et~al.}(2022{\natexlab{b}})\citenamefont {Floerchinger}, \citenamefont
  {Haas},\ and\ \citenamefont {Schröfl}}]{Haas2022b}%
  \BibitemOpen
  \bibfield  {author} {\bibinfo {author} {\bibfnamefont {S.}~\bibnamefont
  {Floerchinger}}, \bibinfo {author} {\bibfnamefont {T.}~\bibnamefont {Haas}},\
  and\ \bibinfo {author} {\bibfnamefont {M.}~\bibnamefont {Schröfl}},\
  }\bibfield  {title} {\bibinfo {title} {{Relative entropic uncertainty
  relation for scalar quantum fields}},\ }\href
  {https://doi.org/10.21468/SciPostPhys.12.3.089} {\bibfield  {journal}
  {\bibinfo  {journal} {SciPost Phys.}\ }\textbf {\bibinfo {volume} {12}},\
  \bibinfo {pages} {089} (\bibinfo {year} {2022}{\natexlab{b}})}\BibitemShut
  {NoStop}%
\bibitem [{\citenamefont {Marshall}\ \emph {et~al.}(2011)\citenamefont
  {Marshall}, \citenamefont {Olkin},\ and\ \citenamefont
  {Arnold}}]{Marshall2011}%
  \BibitemOpen
  \bibfield  {author} {\bibinfo {author} {\bibfnamefont {A.~W.}\ \bibnamefont
  {Marshall}}, \bibinfo {author} {\bibfnamefont {I.}~\bibnamefont {Olkin}},\
  and\ \bibinfo {author} {\bibfnamefont {B.~C.}\ \bibnamefont {Arnold}},\
  }\href {https://doi.org/10.1007/978-0-387-68276-1} {\emph {\bibinfo {title}
  {{Inequalities: Theory of Majorization and Its Applications}}}}\ (\bibinfo
  {publisher} {Springer New York, NY},\ \bibinfo {year} {2011})\BibitemShut
  {NoStop}%
\bibitem [{\citenamefont {G\"uhne}(2004)}]{Guehne2004}%
  \BibitemOpen
  \bibfield  {author} {\bibinfo {author} {\bibfnamefont {O.}~\bibnamefont
  {G\"uhne}},\ }\bibfield  {title} {\bibinfo {title} {{Characterizing
  Entanglement via Uncertainty Relations}},\ }\href
  {https://doi.org/10.1103/PhysRevLett.92.117903} {\bibfield  {journal}
  {\bibinfo  {journal} {Phys. Rev. Lett.}\ }\textbf {\bibinfo {volume} {92}},\
  \bibinfo {pages} {117903} (\bibinfo {year} {2004})}\BibitemShut {NoStop}%
\bibitem [{\citenamefont {G\"uhne}\ and\ \citenamefont
  {Lewenstein}(2004)}]{Guehne2004b}%
  \BibitemOpen
  \bibfield  {author} {\bibinfo {author} {\bibfnamefont {O.}~\bibnamefont
  {G\"uhne}}\ and\ \bibinfo {author} {\bibfnamefont {M.}~\bibnamefont
  {Lewenstein}},\ }\bibfield  {title} {\bibinfo {title} {{Entropic uncertainty
  relations and entanglement}},\ }\href
  {https://doi.org/10.1103/PhysRevA.70.022316} {\bibfield  {journal} {\bibinfo
  {journal} {Phys. Rev. A}\ }\textbf {\bibinfo {volume} {70}},\ \bibinfo
  {pages} {022316} (\bibinfo {year} {2004})}\BibitemShut {NoStop}%
\bibitem [{\citenamefont {Schneeloch}\ \emph {et~al.}(2019)\citenamefont
  {Schneeloch}, \citenamefont {Tison}, \citenamefont {Fanto}, \citenamefont
  {Alsing},\ and\ \citenamefont {Howland}}]{Schneeloch2019}%
  \BibitemOpen
  \bibfield  {author} {\bibinfo {author} {\bibfnamefont {J.}~\bibnamefont
  {Schneeloch}}, \bibinfo {author} {\bibfnamefont {C.~C.}\ \bibnamefont
  {Tison}}, \bibinfo {author} {\bibfnamefont {M.~L.}\ \bibnamefont {Fanto}},
  \bibinfo {author} {\bibfnamefont {P.~M.}\ \bibnamefont {Alsing}},\ and\
  \bibinfo {author} {\bibfnamefont {G.~A.}\ \bibnamefont {Howland}},\
  }\bibfield  {title} {\bibinfo {title} {{Quantifying entanglement in a
  68-billion-dimensional quantum state space}},\ }\href
  {https://doi.org/10.1038/s41467-019-10810-z} {\bibfield  {journal} {\bibinfo
  {journal} {Nat. Commun.}\ }\textbf {\bibinfo {volume} {10}},\ \bibinfo
  {pages} {1} (\bibinfo {year} {2019})}\BibitemShut {NoStop}%
\bibitem [{\citenamefont {Bergh}\ and\ \citenamefont
  {G\"arttner}(2021)}]{Bergh2021a}%
  \BibitemOpen
  \bibfield  {author} {\bibinfo {author} {\bibfnamefont {B.}~\bibnamefont
  {Bergh}}\ and\ \bibinfo {author} {\bibfnamefont {M.}~\bibnamefont
  {G\"arttner}},\ }\bibfield  {title} {\bibinfo {title} {Entanglement detection
  in quantum many-body systems using entropic uncertainty relations},\ }\href
  {https://doi.org/10.1103/PhysRevA.103.052412} {\bibfield  {journal} {\bibinfo
   {journal} {Phys. Rev. A}\ }\textbf {\bibinfo {volume} {103}},\ \bibinfo
  {pages} {052412} (\bibinfo {year} {2021})}\BibitemShut {NoStop}%
\bibitem [{\citenamefont {Bergh}\ and\ \citenamefont
  {G{\"a}rttner}(2021)}]{Bergh2021b}%
  \BibitemOpen
  \bibfield  {author} {\bibinfo {author} {\bibfnamefont {B.}~\bibnamefont
  {Bergh}}\ and\ \bibinfo {author} {\bibfnamefont {M.}~\bibnamefont
  {G{\"a}rttner}},\ }\bibfield  {title} {\bibinfo {title} {{Experimentally
  Accessible Bounds on Distillable Entanglement from Entropic Uncertainty
  Relations}},\ }\href {https://doi.org/10.1103/PhysRevLett.126.190503}
  {\bibfield  {journal} {\bibinfo  {journal} {Phys. Rev. Lett.}\ }\textbf
  {\bibinfo {volume} {126}},\ \bibinfo {pages} {190503} (\bibinfo {year}
  {2021})}\BibitemShut {NoStop}%
\bibitem [{\citenamefont {Vignat}\ \emph {et~al.}(2006)\citenamefont {Vignat},
  \citenamefont {Hero},\ and\ \citenamefont {Costa}}]{Vignat2006}%
  \BibitemOpen
  \bibfield  {author} {\bibinfo {author} {\bibfnamefont {C.}~\bibnamefont
  {Vignat}}, \bibinfo {author} {\bibfnamefont {A.~O.}\ \bibnamefont {Hero}},\
  and\ \bibinfo {author} {\bibfnamefont {J.~A.}\ \bibnamefont {Costa}},\
  }\bibfield  {title} {\bibinfo {title} {{A Geometric Characterization of
  Maximum Rényi Entropy Distributions}},\ }in\ \href
  {https://doi.org/10.1109/ISIT.2006.261749} {\emph {\bibinfo {booktitle} {2006
  IEEE International Symposium on Information Theory}}}\ (\bibinfo {year}
  {2006})\ pp.\ \bibinfo {pages} {1822--1826}\BibitemShut {NoStop}%
\bibitem [{\citenamefont {Johnson}\ and\ \citenamefont
  {Vignat}(2007)}]{Johnson2007}%
  \BibitemOpen
  \bibfield  {author} {\bibinfo {author} {\bibfnamefont {O.}~\bibnamefont
  {Johnson}}\ and\ \bibinfo {author} {\bibfnamefont {C.}~\bibnamefont
  {Vignat}},\ }\bibfield  {title} {\bibinfo {title} {{Some results concerning
  maximum Rényi entropy distributions}},\ }\href
  {https://doi.org/https://doi.org/10.1016/j.anihpb.2006.05.001} {\bibfield
  {journal} {\bibinfo  {journal} {Ann. I. H. Poincare-PR}\ }\textbf {\bibinfo
  {volume} {43}},\ \bibinfo {pages} {339} (\bibinfo {year} {2007})}\BibitemShut
  {NoStop}%
\bibitem [{\citenamefont {Beckner}(1975)}]{Beckner1975}%
  \BibitemOpen
  \bibfield  {author} {\bibinfo {author} {\bibfnamefont {W.}~\bibnamefont
  {Beckner}},\ }\bibfield  {title} {\bibinfo {title} {{Inequalities in Fourier
  Analysis}},\ }\href {https://doi.org/10.2307/1970980} {\bibfield  {journal}
  {\bibinfo  {journal} {Ann. Math.}\ }\textbf {\bibinfo {volume} {102}},\
  \bibinfo {pages} {159} (\bibinfo {year} {1975})}\BibitemShut {NoStop}%
\bibitem [{\citenamefont {Bia{\l}ynicki-Birula}\ and\ \citenamefont
  {Rudnicki}(2011)}]{Bialynicki-Birula2011}%
  \BibitemOpen
  \bibfield  {author} {\bibinfo {author} {\bibfnamefont {I.}~\bibnamefont
  {Bia{\l}ynicki-Birula}}\ and\ \bibinfo {author} {\bibfnamefont
  {{\L}.}~\bibnamefont {Rudnicki}},\ }\bibinfo {title} {{Entropic Uncertainty
  Relations in Quantum Physics}},\ in\ \href
  {https://doi.org/10.1007/978-90-481-3890-6_1} {\emph {\bibinfo {booktitle}
  {{Statistical Complexity}}}}\ (\bibinfo  {publisher} {Springer, Dordrecht},\
  \bibinfo {year} {2011})\BibitemShut {NoStop}%
\bibitem [{\citenamefont {Gomes}\ \emph
  {et~al.}(2009{\natexlab{a}})\citenamefont {Gomes}, \citenamefont {Salles},
  \citenamefont {Toscano}, \citenamefont {Ribeiro},\ and\ \citenamefont
  {Walborn}}]{Gomes2009a}%
  \BibitemOpen
  \bibfield  {author} {\bibinfo {author} {\bibfnamefont {R.~M.}\ \bibnamefont
  {Gomes}}, \bibinfo {author} {\bibfnamefont {A.}~\bibnamefont {Salles}},
  \bibinfo {author} {\bibfnamefont {F.}~\bibnamefont {Toscano}}, \bibinfo
  {author} {\bibfnamefont {P.~H.~S.}\ \bibnamefont {Ribeiro}},\ and\ \bibinfo
  {author} {\bibfnamefont {S.~P.}\ \bibnamefont {Walborn}},\ }\bibfield
  {title} {\bibinfo {title} {{Quantum entanglement beyond Gaussian criteria}},\
  }\href {https://doi.org/10.1073/pnas.0908329106} {\bibfield  {journal}
  {\bibinfo  {journal} {{PNAS}}\ }\textbf {\bibinfo {volume} {106}},\ \bibinfo
  {pages} {21517} (\bibinfo {year} {2009}{\natexlab{a}})}\BibitemShut {NoStop}%
\bibitem [{\citenamefont {Gomes}\ \emph
  {et~al.}(2009{\natexlab{b}})\citenamefont {Gomes}, \citenamefont {Salles},
  \citenamefont {Toscano}, \citenamefont {Ribeiro},\ and\ \citenamefont
  {Walborn}}]{Gomes2009b}%
  \BibitemOpen
  \bibfield  {author} {\bibinfo {author} {\bibfnamefont {R.~M.}\ \bibnamefont
  {Gomes}}, \bibinfo {author} {\bibfnamefont {A.}~\bibnamefont {Salles}},
  \bibinfo {author} {\bibfnamefont {F.}~\bibnamefont {Toscano}}, \bibinfo
  {author} {\bibfnamefont {P.~H.~S.}\ \bibnamefont {Ribeiro}},\ and\ \bibinfo
  {author} {\bibfnamefont {S.~P.}\ \bibnamefont {Walborn}},\ }\bibfield
  {title} {\bibinfo {title} {{Observation of a Nonlocal Optical Vortex}},\
  }\href {https://doi.org/10.1103/PhysRevLett.103.033602} {\bibfield  {journal}
  {\bibinfo  {journal} {Phys. Rev. Lett.}\ }\textbf {\bibinfo {volume} {103}},\
  \bibinfo {pages} {033602} (\bibinfo {year} {2009}{\natexlab{b}})}\BibitemShut
  {NoStop}%
\bibitem [{\citenamefont {Nogueira}\ \emph {et~al.}(2004)\citenamefont
  {Nogueira}, \citenamefont {Walborn}, \citenamefont {P\'adua},\ and\
  \citenamefont {Monken}}]{Nogueira2004}%
  \BibitemOpen
  \bibfield  {author} {\bibinfo {author} {\bibfnamefont {W.~A.~T.}\
  \bibnamefont {Nogueira}}, \bibinfo {author} {\bibfnamefont {S.~P.}\
  \bibnamefont {Walborn}}, \bibinfo {author} {\bibfnamefont {S.}~\bibnamefont
  {P\'adua}},\ and\ \bibinfo {author} {\bibfnamefont {C.~H.}\ \bibnamefont
  {Monken}},\ }\bibfield  {title} {\bibinfo {title} {{Generation of a
  Two-Photon Singlet Beam}},\ }\href
  {https://doi.org/10.1103/PhysRevLett.92.043602} {\bibfield  {journal}
  {\bibinfo  {journal} {Phys. Rev. Lett.}\ }\textbf {\bibinfo {volume} {92}},\
  \bibinfo {pages} {043602} (\bibinfo {year} {2004})}\BibitemShut {NoStop}%
\bibitem [{\citenamefont {Rudnicki}\ \emph {et~al.}(2012)\citenamefont
  {Rudnicki}, \citenamefont {Walborn},\ and\ \citenamefont
  {Toscano}}]{Rudnicki2012}%
  \BibitemOpen
  \bibfield  {author} {\bibinfo {author} {\bibfnamefont {{\L}.}~\bibnamefont
  {Rudnicki}}, \bibinfo {author} {\bibfnamefont {S.~P.}\ \bibnamefont
  {Walborn}},\ and\ \bibinfo {author} {\bibfnamefont {F.}~\bibnamefont
  {Toscano}},\ }\bibfield  {title} {\bibinfo {title} {{Heisenberg uncertainty
  relation for coarse-grained observables}},\ }\href
  {https://doi.org/10.1209/0295-5075/97/38003} {\bibfield  {journal} {\bibinfo
  {journal} {{EPL}}\ }\textbf {\bibinfo {volume} {97}},\ \bibinfo {pages}
  {38003} (\bibinfo {year} {2012})}\BibitemShut {NoStop}%
\bibitem [{\citenamefont {Tasca}\ \emph {et~al.}(2013)\citenamefont {Tasca},
  \citenamefont {Rudnicki}, \citenamefont {Gomes}, \citenamefont {Toscano},\
  and\ \citenamefont {Walborn}}]{Tasca2013}%
  \BibitemOpen
  \bibfield  {author} {\bibinfo {author} {\bibfnamefont {D.~S.}\ \bibnamefont
  {Tasca}}, \bibinfo {author} {\bibfnamefont {{\L}.}~\bibnamefont {Rudnicki}},
  \bibinfo {author} {\bibfnamefont {R.~M.}\ \bibnamefont {Gomes}}, \bibinfo
  {author} {\bibfnamefont {F.}~\bibnamefont {Toscano}},\ and\ \bibinfo {author}
  {\bibfnamefont {S.~P.}\ \bibnamefont {Walborn}},\ }\bibfield  {title}
  {\bibinfo {title} {{Reliable Entanglement Detection under Coarse-Grained
  Measurements}},\ }\href {https://doi.org/10.1103/PhysRevLett.110.210502}
  {\bibfield  {journal} {\bibinfo  {journal} {Phys. Rev. Lett.}\ }\textbf
  {\bibinfo {volume} {110}},\ \bibinfo {pages} {210502} (\bibinfo {year}
  {2013})}\BibitemShut {NoStop}%
\bibitem [{\citenamefont {Jones}\ \emph {et~al.}(1996)\citenamefont {Jones},
  \citenamefont {Marron},\ and\ \citenamefont {Sheather}}]{Jones1996}%
  \BibitemOpen
  \bibfield  {author} {\bibinfo {author} {\bibfnamefont {M.~C.}\ \bibnamefont
  {Jones}}, \bibinfo {author} {\bibfnamefont {J.~S.}\ \bibnamefont {Marron}},\
  and\ \bibinfo {author} {\bibfnamefont {S.~J.}\ \bibnamefont {Sheather}},\
  }\bibfield  {title} {\bibinfo {title} {{A Brief Survey of Bandwidth Selection
  for Density Estimation}},\ }\href
  {https://doi.org/10.1080/01621459.1996.10476701} {\bibfield  {journal}
  {\bibinfo  {journal} {J. Am. Stat. Assoc.}\ }\textbf {\bibinfo {volume}
  {91}},\ \bibinfo {pages} {401} (\bibinfo {year} {1996})}\BibitemShut
  {NoStop}%
\bibitem [{\citenamefont {Armstrong}\ \emph {et~al.}(2019)\citenamefont
  {Armstrong}, \citenamefont {Sutton},\ and\ \citenamefont
  {Hibbert}}]{Armstrong2019}%
  \BibitemOpen
  \bibfield  {author} {\bibinfo {author} {\bibfnamefont {N.}~\bibnamefont
  {Armstrong}}, \bibinfo {author} {\bibfnamefont {G.~J.}\ \bibnamefont
  {Sutton}},\ and\ \bibinfo {author} {\bibfnamefont {D.~B.}\ \bibnamefont
  {Hibbert}},\ }\bibfield  {title} {\bibinfo {title} {{Estimating probability
  density functions using a combined maximum entropy moments and Bayesian
  method. Theory and numerical examples}},\ }\href
  {https://doi.org/10.1088/1681-7575/aaf7d1} {\bibfield  {journal} {\bibinfo
  {journal} {Metrologia}\ }\textbf {\bibinfo {volume} {56}},\ \bibinfo {pages}
  {015019} (\bibinfo {year} {2019})}\BibitemShut {NoStop}%
\bibitem [{\citenamefont {Uria}\ \emph {et~al.}(2016)\citenamefont {Uria},
  \citenamefont {C{{\^o}}t{{\'e}}}, \citenamefont {Gregor}, \citenamefont
  {Murray},\ and\ \citenamefont {Larochelle}}]{Uria2016}%
  \BibitemOpen
  \bibfield  {author} {\bibinfo {author} {\bibfnamefont {B.}~\bibnamefont
  {Uria}}, \bibinfo {author} {\bibfnamefont {M.-A.}\ \bibnamefont
  {C{{\^o}}t{{\'e}}}}, \bibinfo {author} {\bibfnamefont {K.}~\bibnamefont
  {Gregor}}, \bibinfo {author} {\bibfnamefont {I.}~\bibnamefont {Murray}},\
  and\ \bibinfo {author} {\bibfnamefont {H.}~\bibnamefont {Larochelle}},\
  }\bibfield  {title} {\bibinfo {title} {{Neural Autoregressive Distribution
  Estimation}},\ }\href {http://jmlr.org/papers/v17/16-272.html} {\bibfield
  {journal} {\bibinfo  {journal} {J. Mach. Learn. Res.}\ }\textbf {\bibinfo
  {volume} {17}},\ \bibinfo {pages} {1} (\bibinfo {year} {2016})}\BibitemShut
  {NoStop}%
\bibitem [{\citenamefont {Reed}\ \emph {et~al.}(2017)\citenamefont {Reed},
  \citenamefont {van~den Oord}, \citenamefont {Kalchbrenner}, \citenamefont
  {Colmenarejo}, \citenamefont {Wang}, \citenamefont {Belov},\ and\
  \citenamefont {de~Freitas}}]{Reed2017}%
  \BibitemOpen
  \bibfield  {author} {\bibinfo {author} {\bibfnamefont {S.~E.}\ \bibnamefont
  {Reed}}, \bibinfo {author} {\bibfnamefont {A.}~\bibnamefont {van~den Oord}},
  \bibinfo {author} {\bibfnamefont {N.}~\bibnamefont {Kalchbrenner}}, \bibinfo
  {author} {\bibfnamefont {S.~G.}\ \bibnamefont {Colmenarejo}}, \bibinfo
  {author} {\bibfnamefont {Z.}~\bibnamefont {Wang}}, \bibinfo {author}
  {\bibfnamefont {D.}~\bibnamefont {Belov}},\ and\ \bibinfo {author}
  {\bibfnamefont {N.}~\bibnamefont {de~Freitas}},\ }\bibfield  {title}
  {\bibinfo {title} {{Parallel Multiscale Autoregressive Density Estimation}},\
  }in\ \href@noop {} {\emph {\bibinfo {booktitle} {{Proceedings of the 34th
  International Conference on Machine Learning - Volume 70}}}}\ (\bibinfo
  {year} {2017})\ p.\ \bibinfo {pages} {2912–2921}\BibitemShut {NoStop}%
\bibitem [{\citenamefont {Paninski}(2003)}]{Paninski2003}%
  \BibitemOpen
  \bibfield  {author} {\bibinfo {author} {\bibfnamefont {L.}~\bibnamefont
  {Paninski}},\ }\bibfield  {title} {\bibinfo {title} {{Estimation of Entropy
  and Mutual Information}},\ }\href
  {https://doi.org/10.1162/089976603321780272} {\bibfield  {journal} {\bibinfo
  {journal} {Neural Comput.}\ }\textbf {\bibinfo {volume} {15}},\ \bibinfo
  {pages} {1191} (\bibinfo {year} {2003})}\BibitemShut {NoStop}%
\bibitem [{\citenamefont {Kraskov}\ \emph {et~al.}(2004)\citenamefont
  {Kraskov}, \citenamefont {St\"ogbauer},\ and\ \citenamefont
  {Grassberger}}]{Kraskov2004}%
  \BibitemOpen
  \bibfield  {author} {\bibinfo {author} {\bibfnamefont {A.}~\bibnamefont
  {Kraskov}}, \bibinfo {author} {\bibfnamefont {H.}~\bibnamefont
  {St\"ogbauer}},\ and\ \bibinfo {author} {\bibfnamefont {P.}~\bibnamefont
  {Grassberger}},\ }\bibfield  {title} {\bibinfo {title} {{Estimating mutual
  information}},\ }\href {https://doi.org/10.1103/PhysRevE.69.066138}
  {\bibfield  {journal} {\bibinfo  {journal} {Phys. Rev. E}\ }\textbf {\bibinfo
  {volume} {69}},\ \bibinfo {pages} {066138} (\bibinfo {year}
  {2004})}\BibitemShut {NoStop}%
\bibitem [{\citenamefont {Palma}(2017)}]{dePalma2018c}%
  \BibitemOpen
  \bibfield  {author} {\bibinfo {author} {\bibfnamefont {G.~D.}\ \bibnamefont
  {Palma}},\ }\bibfield  {title} {\bibinfo {title} {{Uncertainty relations with
  quantum memory for the Wehrl entropy}},\ }\href
  {https://doi.org/10.1007/s11005-018-1067-y} {\bibfield  {journal} {\bibinfo
  {journal} {Commun. Math. Phys.}\ }\textbf {\bibinfo {volume} {108}},\
  \bibinfo {pages} {2139} (\bibinfo {year} {2017})}\BibitemShut {NoStop}%
\end{thebibliography}%

\end{document}